\documentclass[10pt]{article}
\usepackage{authblk}

\usepackage[T1]{fontenc} 

\usepackage[normalem]{ulem} 

\usepackage[T1]{fontenc} 

\usepackage[normalem]{ulem} 

\usepackage[usenames,dvipsnames,svgnames,x11names]{xcolor}
\usepackage{fontawesome5}

\definecolor{modebeige}{rgb}{0.59, 0.44, 0.09}
\newcommand{\cfaCheck}[0]{\textcolor{ForestGreen}{\faArrowCircleUp}}
\newcommand{\cfaTimes}[0]{\textcolor{WildStrawberry}{\faArrowCircleDown}}
\newcommand{\cfaNeutral}[0]{\textcolor{modebeige}{\faBullseye}}

\newcommand{\modc}[1]{{\textcolor{blue}{#1}}}
\newcommand{\addc}[1]{{\textcolor{teal}{#1}}}

\newcommand{\delc}[1]{ {\textcolor{gray} {\sout{#1}} }}

\renewcommand{\modc}[1]{#1} 
\renewcommand{\addc}[1]{#1} 
\renewcommand{\delc}[1]{} 

\newcommand{\addcrtwo}[1]{{\textcolor{teal}{#1}}}

\newcommand{\delcrtwo}[1]{ {\textcolor{gray} {\sout{#1}} }}

\renewcommand{\addcrtwo}[1]{#1} 
\renewcommand{\delcrtwo}[1]{} 

\usepackage{balance}

\usepackage{listings}

\lstdefinelanguage{XML}
{
basicstyle=\ttfamily\footnotesize,
  morestring=[b]",
  moredelim=[s][\bfseries\color{Maroon}]{<}{\ },
  moredelim=[s][\bfseries\color{Maroon}]{</}{>},
  moredelim=[l][\bfseries\color{Maroon}]{/>},
  moredelim=[l][\bfseries\color{Maroon}]{>},
  morecomment=[s]{<?}{?>},
  morecomment=[s]{<!--}{-->},
  commentstyle=\color{gray},
  stringstyle=\color{blue},
  identifierstyle=\color{red}
}

\lstdefinestyle{yaml}{
     basicstyle=\color{blue}\small\tt,
     rulecolor=\color{black},
     string=[s]{'}{'},
     stringstyle=\color{blue},
     comment=[l]{:},
     commentstyle=\color{black},
     morecomment=[l]{-}
 }
\usepackage{moreverb}

\usepackage[nounderscore]{syntax}

\usepackage[cmex10]{amsmath}
\usepackage{amssymb}
\usepackage{mathtools}
\usepackage{amsthm}
\usepackage{amsfonts}
\usepackage{gensymb}

\usepackage{subfig}

\usepackage{algorithmicx}
\usepackage{algpseudocode}
\usepackage[ruled]{algorithm}
\definecolor{light-gray}{gray}{0.75}
\algrenewcommand{\algorithmiccomment}[1]{\hskip3em{{\footnotesize \textcolor{light-gray}{$\blacktriangleright$}}} #1}

\usepackage{multirow} 
\usepackage{rotating} 
\usepackage{booktabs} 
\usepackage{colortbl} 
\usepackage{tablefootnote} 
\usepackage{longtable}
\usepackage{tabularx}

\usepackage{array}
\newcolumntype{L}[1]{>{\raggedright\let\newline\\\arraybackslash\hspace{0pt}}m{#1}}
\newcolumntype{C}[1]{>{\centering\let\newline\\\arraybackslash\hspace{0pt}}m{#1}}
\newcolumntype{R}[1]{>{\raggedleft\let\newline\\\arraybackslash\hspace{0pt}}m{#1}}

\usepackage[pdftex,colorlinks=true,urlcolor=blue,citecolor=blue]{hyperref}

\usepackage{xspace}

\usepackage{enumitem}

\usepackage{blindtext}

\usepackage{soul}

\newcommand{\FL}{\textsc{Flotilla}\xspace}

\begin{document}

\title{\textit{Flotilla:} A Scalable, Modular and Resilient Federated Learning Framework for Heterogeneous Resources\thanks{~Paper published in the Journal of Parallel and Distributed Computing, \href{https://doi.org/10.1016/j.jpdc.2025.105103}{https://doi.org/10.1016/j.jpdc.2025.105103}}}

\author[1]{Roopkatha Banerjee}
\author[1]{Prince Modi}
\author[1]{Jinal Vyas}
\author[1]{Chunduru Sri Abhijit}
\author[1]{Tejus Chandrashekar}
\author[2]{Harsha Varun Marisetty}
\author[2]{Manik Gupta}
\author[1]{Yogesh Simmhan}

\affil[1]{Department of Computational and Data Sciences (CDS), Indian Institute of Science (IISc), Bangalore 560012, India\\
\texttt{\{roopkathab, simmhan\}@iisc.ac.in}}
\affil[2]{Birla Institute of Technology and Science (BITS), Pilani, Hyderabad Campus, Hyderabad 500078, India}

\date{}
\maketitle

\begin{abstract}
With the recent improvements in mobile and edge computing and rising concerns of data privacy, \textit{Federated Learning~(FL)} has rapidly gained popularity as a privacy-preserving, distributed machine learning methodology. Several FL frameworks have been built for testing novel FL strategies. However, most focus on validating the \textit{learning} aspects of FL through pseudo-distributed simulation but not for deploying on real edge hardware in a distributed manner to meaningfully evaluate the \textit{federated} aspects from a systems perspective. Current frameworks are also inherently not designed to support asynchronous aggregation, which is gaining popularity, and have limited resilience to client and server failures.  We introduce \FL, a scalable and lightweight FL framework. It adopts a ``user-first'' modular design to help rapidly compose various synchronous and asynchronous FL strategies while being agnostic to the DNN architecture. 
It uses stateless clients and a server design that separates out the session state, which are periodically or incrementally checkpointed.
We demonstrate the modularity of \FL by evaluating five different FL strategies for training five DNN models. We also evaluate the client and server-side fault tolerance on 200+ clients, and showcase its ability to rapidly failover within seconds. Finally, we show that \FL's resource usage on Raspberry Pis and Nvidia Jetson edge accelerators are comparable to or better than three state-of-the-art FL frameworks, Flower, OpenFL and FedML. It also scales significantly better compared to Flower for 1000+ clients. This positions \FL as a competitive candidate to build novel FL strategies on, compare them uniformly, rapidly deploy them, and perform systems research and optimizations.
\end{abstract}

\section{Introduction} \label{sec:intro}

\subsection{Motivation}\label{subsec:intro-motivation}
The popularity of smartphones and Internet of Things (IoT) deployments for society~\cite{iot-smart-city-1, kumar2019internet} and science~\cite{iot-science-anl, iot-science-parashar} has unleashed a torrent of sensor data generated continuously on edge devices. Machine Learning (ML) and Deep Learning (DL) models have been developed to gain value from such pervasive datasets, e.g., to optimize traffic signalling using camera feeds~\cite{ml-traffic}, to detect wildfires using field instruments~\cite{ml-science-field} and for low-cost diagnostics using medical devices~\cite{ml-medical}. In a traditional ML setting, data from these edge devices are pushed to a central server, typically on the cloud, to train a model~\cite{liu2020client}. However, this raises concerns about data privacy if the server is not fully trusted (e.g., honest but curious~\cite{Nguyen2024preservingprivacy}), is not permitted due to regulatory restrictions (e.g., health~\cite{ml-health-regulatory} or financial data~\cite{fintech-health-regulatory}), and/or can have a high resource cost for network data movement (e.g., at remote locations with low bandwidth~\cite{ml-science-field} or moving large video feeds over a Wide Area Network (WAN)~\cite{ml-traffic}).

\begin{figure}[t]
    \centering
    \includegraphics[width=0.7\columnwidth]{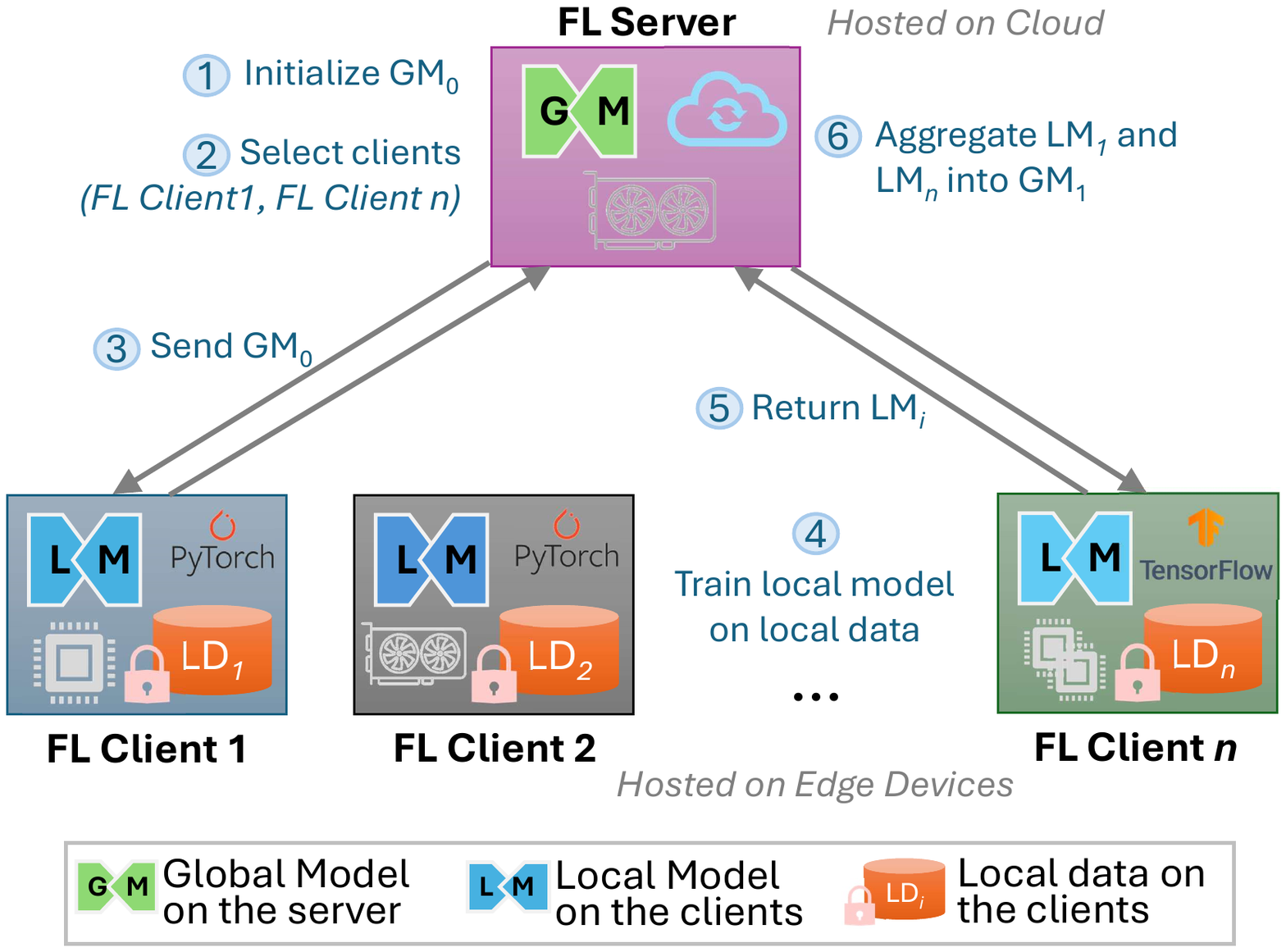}
    \caption{A typical federated learning round on a central server \modc{with} clients holding the data.
    }
    \label{fig:fl}
\end{figure}

Recently, \textit{Federated Learning~(FL)}~\cite{McMahan2016CommunicationEfficientLO} has emerged as a privacy-preserving, decentralized training paradigm for edge devices that are growing more powerful. In a typical FL setup (Fig.~\ref{fig:fl}), a \textit{global model} maintained by the \textit{central server} is iteratively trained by a collaborative set of \textit{edge devices}, typically with limited resources and on a WAN, which hold the \textit{training data} that they have collected. 
At a high level, \textbf{(1)} the server initializes the \textit{global model ($GM_0$)}, \textbf{(2)} selects a subset of edge devices/clients, and \textbf{(3)} sends the $GM_0$ to them. \textbf{(4)} Each client \modc{further trains} the global model using its local data, and \textbf{(5)} sends their updated \textit{local models ($LM_i$)} back to the server.
\textbf{(6)} The server \textit{aggregates} the local models into a new global model ($GM_1=\oplus_i( LM_i)$), and \textbf{(3$^+$)} sends it back to the same or a different subset of clients for the next \textit{round} of training. This repeats until the global model converges, or until some time or iteration budget is reached.
The privacy-sensitive local data on the edge remains localized on the clients, and the global model thus trained has shown to converge to an accuracy comparable to a centrally trained global model using all data, under ideal conditions~\cite{9252927}. Further, the size of the models exchanged can be much smaller than the sizes of the training data, thereby reducing network usage as well.

Systems research on FL has focused on reducing training time and increasing accuracy under diverse conditions of data distribution, device capabilities and network conditions~\cite{refl,ching2024totoro,khan2024float}.
The potentially large number of clients~(100--1000s of IoT devices and smartphones) and their resource heterogeneity~(from low-end Raspberry Pis to GPU-accelerated Jetson edge devices) has resulted in diverse \textit{client selection strategies}~\cite{chai2020tifl, wolfrath2022haccs}. These try to accelerate convergence over non-IID (non-independently and identically distributed) data distribution across clients~\cite{chai2020tifl} and account for stragglers in a round~\cite{refl}. \textit{Model aggregation strategies} attempt to weight local models during aggregation based on a client's data distribution and prior participation and also use asynchronous aggregation to avoid delays from slower clients slowing down the progress~\cite{xie2019asynchronous,chai2021fedat}. Failed clients or network connectivity to them over WAN also poses a challenge, especially on constrained devices or under field deployments~\cite{refl, zhang2020federated}.

\begin{figure}
    \centering
    \includegraphics[width=\linewidth, trim = {0cm 12cm 2cm 0cm}, clip]{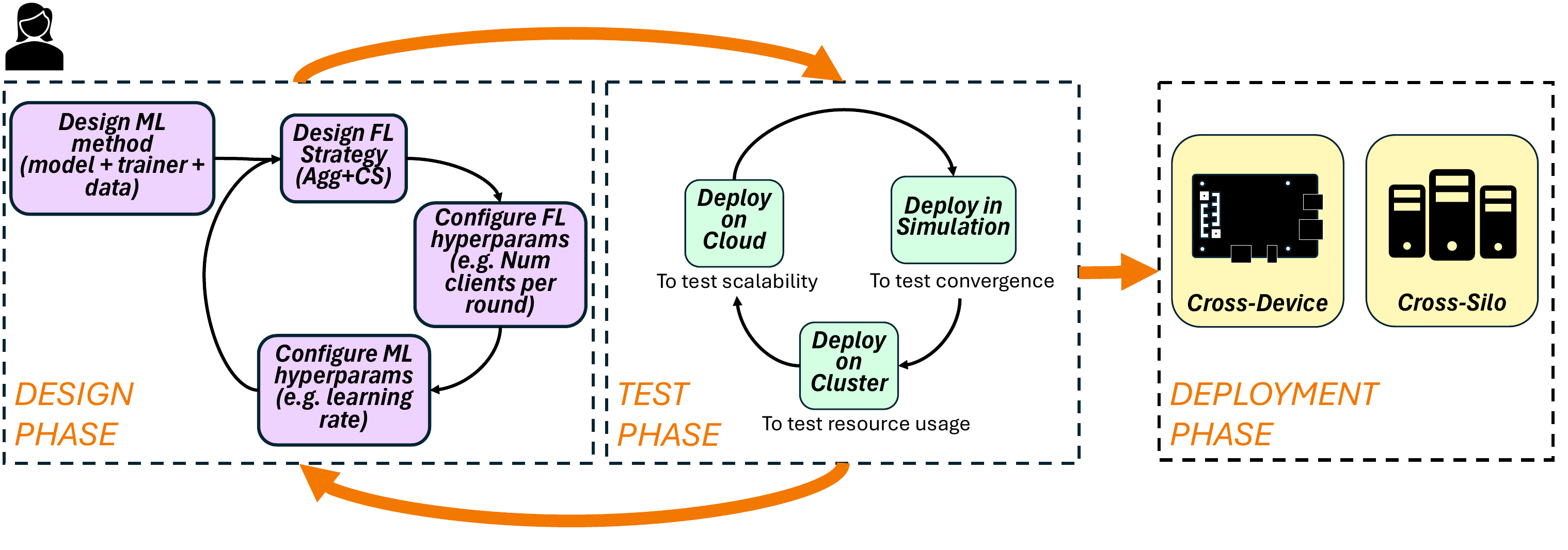}
    \caption{The lifecycle of Federated Learning application, from development to deployment.}
    \label{fig:fl_development_lifecycle}
\end{figure}

\subsection{FL Development Lifecycle}
In Fig.~\ref{fig:fl_development_lifecycle}, we illustrate the typical production lifecycle of a FL application. The process generally begins with domain experts identifying the need for FL-based training for their domain.

Initially, in the \textit{design phase}, a traditional Machine Learning (ML) or Deep Learning (DL) solution is developed for training a local model. Subsequently, this is extended to an FL configuration by selecting the client selection, aggregation, etc., strategies tailored to the deployment setting, data distribution and privacy constraints. These are validated in a pseudo-distributed or by simulating FL execution in a single-machine setting~\cite{lai2022fedscale, refl}.

Once the FL strategy is designed and implemented, the \textit{testing phase} validates its behaviour in a realistic distributed setting, e.g., in containerized environments on clusters or cloud VMs~\cite{liu2020client}. Here, the interplay of critical FL parameters, such as the number of clients selected in each round and frequency/type of aggregation, and hyper-parameters such as learning rate and batch size, with the distributed environment need to be assessed for convergence, scaling and robustness and tuned as required. 
The process of designing, testing and fine-tuning is iterative, often requiring multiple rounds of adjustment to develop an effective and convergent FL strategy under variable conditions imposed by the distributed environment.

Lastly, the \textit{deployment phase} involves rolling out the FL model and runtime on real edge devices and the server for practical orchestration.

\subsection{\modc{State-of-the-Art FL Frameworks and Limitations}}
FL frameworks enable users to design, test and deploy FL applications. \addc{A number of open-source FL platforms such as Google's Tensorflow Federated (TFF)~\cite{tff}, CMU's LEAF~\cite{caldas2018leaf}, Flower from University of Cambridge~\cite{beutel2020flower}, OpenFL from the The Linux Foundation~\cite{openfl_citation} and USC's FedML~\cite{he2020fedml} exist.}
\modc{However, they focus more on the design and pseudo-distributed validation of FL strategies, rather than seamlessly allow iterative testing, scaling and eventual deployment onto real hardware. In other words,} the FL platforms target \textit{ML researchers} who develop and validate model architectures and strategies. But they offer inadequate capabilities for \textit{systems researchers} to examine and tune the distributed FL system for efficiency and reliability on real hardware and networks; for \textit{practitioners} to evaluate FL configurations in a managed distributed setup; or for \textit{operational users} deploying the solutions on real field devices. This gives rise to several gaps summarized below and discussed exhaustively in Sec.~\ref{sec:challenges:req}.

\begin{enumerate}[leftmargin=*,noitemsep]
    \item
LEAF~\cite{caldas2018leaf} and Tensorflow Federated (TFF)~\cite{tff} simulate FL training on a single machine and mimic multiple devices as local processes. These do not have a \textit{distributed framework} to deploy and orchestrate on networked clients. Translating a pseudo-distributed FL implementation to a reliable, manageable and scalable implementation in a distributed setup takes substantial effort, and may even require revisiting the FL strategy.

\item
Popular frameworks like Flower~\cite{beutel2020flower} and OpenFL~\cite{openfl_citation} omit key features like \textit{asynchronous aggregation} that have grown popular to address data and device heterogeneity. While these have some modularity, incorporating asynchrony into these platforms requires a fundamental redesign of their client interactions and the orchestration of model aggregation\footnote{Asynchronous FL using Flower, Github Issues \href{https://github.com/adap/flower/issues/469}{\#469} and \href{https://github.com/adap/flower/pull/3932}{\#3932}}.

\item
While Flower~\cite{beutel2020flower} is \textit{resilient} to client failures, most others like FedScale~\cite{lai2022fedscale} and FedML~\cite{he2020fedml} assume that the server (or even the clients) are reliable. This curtails training in practical settings.
Concerns on reliability are real, as we report from our hours-long runs where edge failures and non-responsiveness were common due to overheating and resource constraints (Sec.~\ref{subsec: server_resilience_exp}). Having server reliability also allows researchers to snapshot and resume a previous training session that has run for hours, possibly with different hyper-parameters.

\item
Lastly, \textit{developer and operational features} such as pushing the model code to the clients at runtime (rather than assume it is pre-deployed on the edge) or having a mechanism for clients to join/leave the training pool dynamically are lacking.

\item It is cumbersome if not impossible to implement and evaluate different State-of-the-Art (SOTA) FL strategies on a single FL platform for an apples-to-apples comparison. This inability to quickly implement and reproduce FL research claims is a key concern, as we show in our evaluation of popular strategies (Sec.~\ref{subsec: scalability and edge deployment}). 

\end{enumerate}
In summary, no single FL framework meets the needs for the managing the complete lifecycle of an FL application.

\subsection{Contributions}

In this paper, we propose \FL, a novel open-source FL framework \textit{designed using a principled approach} to offer substantial flexibility in quickly implementing, testing and scaling FL strategies in a simulated environment, and subsequently translating it to a real distributed setup with minimal effort.
It meets the detailed requirements that we posit in Sec.~\ref{sec:gaps} and addresses the gaps present in popular FL frameworks. 

Some of the key contributions we make through \FL are:
\begin{enumerate}[leftmargin=*,noitemsep]
    \item \textit{Modularity:} \FL adopts a \textit{user-first design principle}. Its defining feature is the flexibility of incorporating new FL strategies for \textit{client selection} and \textit{aggregation}, including \textit{synchronous} and \textit{asynchronous}, through modular interfaces that expose states and are triggered by an event-driven training lifecycle (Sec.~\ref{sec:arch}). 
   It supports PyTorch and TensorFlow as training engines, and diverse architectures such as
   CNNs, LSTMs and transformer models, besides supervised and unsupervised methods.
    This allows for rapid development and distributed deployment. 
    We compose 5 diverse baseline and SOTA FL strategies -- FedAvg~\cite{McMahan2016CommunicationEfficientLO}, FedAsync~\cite{xie2019asynchronous}, TiFL~\cite{chai2020tifl}, HACCS~\cite{wolfrath2022haccs} and FedAT~\cite{chai2021fedat} -- in 59--246 lines of code each (Sec.~\ref{subsec: modularity_exp}). \FL also enhances reproducibility of FL research. Comparing these strategies to train several DNN models shows that even simple baselines like FedAvg can match the performance of more sophisticated SOTA ones (Sec.~\ref{sec:exp:fl-perf}).
    
    \item \textit{Resilience:} \FL is designed to be resilient not only to client dropouts, but also to  server failures (Sec.~\ref{sec:arch:reliability}). 
    Clients are stateless, with all required states sent on-demand from the server to clients in each round; client-caching reduces overheads.
    The server can externalize all its states to a durable key-value store, and make these accessible to custom user modules. This allows recovery even after a permanent loss of the server.
    The server also check-points the global model to rapidly recover from transient failures, with loss limited to a single training round.
    This resilience is demonstrated through fault-injection on 200+ clients, with a drop in accuracy of only $0.28\%$; and with fault-injection in the server, whose training session is recovered and training resumed by an alternate server within $820$ms of detection (Sec.~\ref{subsec: server_resilience_exp}).

    \item \textit{Compatibility and Scalability:} \FL can be deployable on a wide range of edge hardware, with low memory footprint of $\approx 230$MB and CPU overhead, which is comparable to or better than other FL platforms (Sec.~\ref{sec:exp:compare}). We evaluate this on Raspberry Pis and Nvidia Jetsons edge clusters spanning 7 device types and 58 hardware devices. \FL also requires minimal client configuration and dynamically delivers model training code (not just the model architecture), allowing stateless clients to join and leave during training -- vital when training on 100s of clients. \FL's Dockerized containers enables rapid deployment of edge clients for testing. \FL also scales to 1000+ clients, with a weak-scaling efficiency of $92.5\%$ (Sec.~\ref{subsec: scalability and edge deployment}). 
    
   \item Lastly, \FL provides native performance tracing and logging to assist large scale systems research and distributed debugging.
\end{enumerate}

\FL will be released under an open-source license at \url{https://github.com/dream-lab/flotilla}.

\section{\modc{Requirements and Gaps in Contemporary FL Platforms}}
\label{sec:gaps}

In this section, we first review recent literature on FL strategies and systems-research prototypes to establish the need for a flexible and extensible platform. We then propose a suite of requirements for such an FL framework to meet diverse user needs, and finally contrast with existing open-source platforms on their suitability and gaps. \addc{This requirements analysis is informed by discussions with our collaborators from industry and academia on applications such as smart power grids, traffic management and financial technology, personal experiences with FL deployments using diverse frameworks, and a review of existing literature and industry reports, enabling us to identify key limitations in SOTA FL frameworks.}


 \subsection{Related Research on FL}\label{sec:related}

While the idea for federated learning was proposed in \textit{FedAvg}~\cite{McMahan2016CommunicationEfficientLO}, it grew popular with Google's adoption of FL for their Android smartphones, e.g., to train their Gboard keyboard over millions of devices~\cite{flatscale}. However, this is designed for millions of mobile and ephemeral devices rather than 100--1000s of IoT and edge devices that are of practical interest to most domains. Further, the complex server-side orchestration service of Google's Federated Compute Platform is proprietary, with only the client-side code open-sourced~\footnote{Federated Compute Platform, https://github.com/google-parfait/federated-compute}.

The original synchronous averaging of local models over IID data has expanded into FL research for non-ideal conditions of: (1) \textit{data heterogeneity}~\cite{vahidian2023rethinkingdatahetero}~\cite{wolfrath2022haccs}, where clients have varying volumes or non-IID distribution of local data; (2) \textit{device heterogeneity}~\cite{chai2020tifl}~\cite{fl-on-hetero-dev-survey}, where clients have different computing capabilities; and (3) \textit{behavioural heterogeneity}~\cite{refl}~\cite{emperical-hetero-fl}, where device availability and network bandwidths vary across rounds. Data heterogeneity affects the accuracy of the global model~\cite{fedcav}~\cite{wolfrath2022haccs}~\cite{li2019convergence}, while device diversity impacts the time to converge to a certain accuracy~\cite{chai2020tifl}. Behavioural heterogeneity can cause longer round durations in synchronous FL strategies~\cite{AQFL}~\cite{emperical-hetero-fl}.

A number of FL strategies and optimization have been proposed to combat this~\cite{hetero-fl-survey}, such as
synchronous~\cite{McMahan2016CommunicationEfficientLO}, asynchronous~\cite{xie2019asynchronous}, semi-synchronous~\cite{Wu2019SAFAAS}, tiered~\cite{chai2020tifl}~\cite{chai2021fedat}, hierarchical~\cite{fedPEC}~\cite{feddyn} or even decentralized~\cite{Roy2019BrainTorrentAP}.

Some also include privacy preserving techniques like differential privacy into their solutions~\cite{chai2020tifl}~\cite{wolfrath2022haccs}. 

However,translating this growing body of research into working and reproducible prototypes, let alone practical deployments, is limited by the availability of reusable and programmable FL frameworks in the public domain. This forces researchers and developers to invest time and effort on custom implementations of these strategies for different environments. 
E.g., TiFL~\cite{chai2020tifl} and FedAT~\cite{chai2021fedat}, which address data and device heterogeneity, test their strategies on LEAF~\cite{caldas2018leaf}, while HACCS~\cite{wolfrath2022haccs} is built on PySyft~\cite{pysyft}. REFL~\cite{refl}, which is a semi-asynchronous algorithm, 
is evaluated using FedScale~\cite{lai2022fedscale}. 

Due to challenges in implementing such strategies, researchers often resort to simulations. However, results from simulations may not accurately reflect the complexity of real-world distributed runtime scenarios.

In this article, we do not present any new FL strategies, but rather a new modular platform for researchers and practitioners to quickly develop, test and deploy their FL strategies in a resilient and scalable manner in practical settings. \addcrtwo{We focus on a parameter-server-like centralized aggregation rather than decentralized FL.}

\begin{table}[t]
\footnotesize
\caption{Feature-matrix comparing popular FL Frameworks on the proposed requirements. \addc{\cfaCheck~indicates that the feature is fully supported by the framework; \cfaTimes~implies that the feature not supported; and \cfaNeutral~means that there is partial support for the feature or it can be enabled with modest modifications.}}
\label{tbl:framework_comparison}
\resizebox{0.99\linewidth}{!}{%
\centering
\setlength{\tabcolsep}{2pt}
\begin{tabular}{p{1.6cm}|p{2.0cm}p{1.5cm}p{1.5cm}p{1.4cm}p{1cm}p{1.2cm}|p{1.2cm}p{1.2cm}p{1.2cm}} 
\hline
\multicolumn{1}{c|}{\multirow{3}{*}{\textbf{Framework}}} & \multicolumn{6}{c|}{\cellcolor{cyan!20}\textbf{\textsc{Modularity}}} & \multicolumn{3}{c}{\cellcolor{cyan!20}\textbf{\textsc{Resilience}}} \\ 
\cline{2-10}
\multicolumn{1}{c|}{} & \multicolumn{1}{c|}{\multirow{2}{*}{\begin{tabular}[c]{@{}c@{}}\textbf{Custom}\\\textbf{Client Select.}\end{tabular}}} & \multicolumn{2}{c|}{\begin{tabular}[c]{@{}c@{}}\textbf{Custom Model Agg.}\end{tabular}} & \multicolumn{2}{c|}{\textbf{ML Engine}} & \multicolumn{1}{c|}{\multirow{2}{*}{\begin{tabular}[c]{@{}c@{}}\textbf{Plugins}\\\textbf{Avail.?}\end{tabular}}} & \multicolumn{1}{c}{\multirow{2}{*}{\begin{tabular}[c]{@{}c@{}}\textbf{Server}\\\textbf{Restart}\end{tabular}}} & \multicolumn{1}{c}{\multirow{2}{*}{\begin{tabular}[c]{@{}c@{}}\textbf{Server}\\\textbf{Failover}\end{tabular}}} & \multirow{2}{*}{\begin{tabular}[c]{@{}c@{}}\textbf{Client}\\\textbf{Failures}\end{tabular}} \\ 
\cline{3-6}
\multicolumn{1}{c|}{} & \multicolumn{1}{c|}{} & \multicolumn{1}{c}{\em 
 \textbf{Synchron.}} & \multicolumn{1}{c|}{\em 
 \textbf{Asynchron.}} & \em \textbf{PyTorch} & \multicolumn{1}{c|}{\em \textbf{TFlow}} & \multicolumn{1}{c|}{} &  \\ 
\hline\hline
\textbf{FedML AI} & \multicolumn{1}{c}{\cfaCheck} & \multicolumn{1}{c}{\cfaCheck} & \multicolumn{1}{c}{\hspace{0.1cm}\cfaTimes$^1$} & \multicolumn{1}{c}{\cfaCheck} & \multicolumn{1}{c}{\cfaCheck} & \multicolumn{1}{c|}{\cfaCheck}& \multicolumn{1}{c}{\hspace{0.1cm}\cfaNeutral$^6$} & \multicolumn{1}{c}{\cfaTimes} & \multicolumn{1}{c}{\cfaTimes} \\

 \textbf{FedScale} & \multicolumn{1}{c}{\hspace{0.1cm}\cfaNeutral$^2$} & \multicolumn{1}{c}{\hspace{0.1cm}\cfaNeutral$^2$} & \multicolumn{1}{c}{\hspace{0.1cm}\cfaTimes$^1$} & \multicolumn{1}{c}{\cfaCheck}& \multicolumn{1}{c}{\cfaCheck} & \multicolumn{1}{c|}{\cfaCheck}& \multicolumn{1}{c}{\hspace{0.1cm}\cfaNeutral$^6$} & \multicolumn{1}{c}{\cfaTimes} & \multicolumn{1}{c}{\cfaCheck} \\
 
 \textbf{Flower} & \multicolumn{1}{c}{\cfaCheck} & \multicolumn{1}{c}{\cfaCheck} & \multicolumn{1}{c}{\hspace{0.1cm}\cfaNeutral$^3$} & \multicolumn{1}{c}{\cfaCheck} & \multicolumn{1}{c}{\cfaCheck}& \multicolumn{1}{c|}{\cfaCheck} & \multicolumn{1}{c}{\hspace{0.1cm}\cfaNeutral$^6$} & \multicolumn{1}{c}{\hspace{0.1cm}\cfaNeutral$^4$} & \multicolumn{1}{c}{\cfaCheck} \\
 
 \textbf{LEAF} & \multicolumn{1}{c}{\cfaTimes} & \multicolumn{1}{c}{\cfaTimes} & \multicolumn{1}{c}{\cfaTimes} & \multicolumn{1}{c}{\cfaTimes} & \multicolumn{1}{c}{\cfaCheck}& \multicolumn{1}{c|}{\cfaTimes} & \multicolumn{1}{c}{\hspace{0.1cm}\cfaNeutral$^6$}& \multicolumn{1}{c}{\cfaTimes} & \multicolumn{1}{c}{\cfaTimes} \\
 
\textbf{OpenFL} & \multicolumn{1}{c}{\cfaCheck} & \multicolumn{1}{c}{\cfaCheck} & \multicolumn{1}{c}{\cfaTimes} & \multicolumn{1}{c}{\cfaCheck} & \multicolumn{1}{c}{\cfaCheck}& \multicolumn{1}{c|}{\cfaCheck} & \multicolumn{1}{c}{\hspace{0.1cm}\cfaNeutral$^6$} & \multicolumn{1}{c}{\cfaTimes} & \multicolumn{1}{c}{\cfaTimes} \\

\textbf{TFF} & \multicolumn{1}{c}{\cfaCheck} & \multicolumn{1}{c}{\cfaCheck} & \multicolumn{1}{c}{\cfaTimes} & \multicolumn{1}{c}{\cfaTimes} & \multicolumn{1}{c}{\cfaCheck} & \multicolumn{1}{c|}{\cfaCheck} & \multicolumn{1}{c}{\hspace{0.1cm}\cfaNeutral$^6$} & \multicolumn{1}{c}{\cfaTimes} & \multicolumn{1}{c}{\cfaTimes} \\\hline
\cellcolor{yellow!60}\textbf{\FL} & \multicolumn{1}{c}{\cfaCheck} & \multicolumn{1}{c}{\cfaCheck} & \multicolumn{1}{c}{\cfaCheck} & \multicolumn{1}{c}{\cfaCheck} & \multicolumn{1}{c}{\cfaCheck} & \multicolumn{1}{c|}{\cfaCheck} & \multicolumn{1}{c}{\cfaCheck} & \multicolumn{1}{c}{\cfaCheck} & \multicolumn{1}{c}{\cfaCheck} \\ 
\hline
\end{tabular}}
\resizebox{0.92\textwidth}{!}{%
\setlength{\tabcolsep}{2pt}
\begin{tabular}{p{1.6cm}|p{0.5cm}p{0.5cm}p{1.6cm}p{1.6cm}p{1.6cm}p{1.6cm}}
\cline{1-7}
\multicolumn{1}{c|}{\multirow{3}{*}{\textbf{Framework}}} & \multicolumn{6}{c}{\cellcolor{cyan!20} \textbf{\textsc{Deployment}}} \\ \cline{2-7} 
& 
\multicolumn{1}{l}{\textbf{On-device}} &
\multicolumn{1}{l}{\textbf{Simulation}} &
\textbf{Containers} &
\multicolumn{1}{c}{\textbf{Client Disco.}} &
\multicolumn{1}{l}{\textbf{Model Delivery}} &
\multicolumn{1}{l}{\textbf{Easy FL Config.}} \\ 

\hline\hline
\textbf{FedML AI} &\multicolumn{1}{c}{\cfaCheck}&\multicolumn{1}{c}{\cfaCheck}&\multicolumn{1}{c}{\cfaCheck}&\multicolumn{1}{c}{\cfaTimes}&\multicolumn{1}{c}{\cfaTimes} & \multicolumn{1}{c}{\textbf{\cfaTimes}} \\

\textbf{FedScale} &\multicolumn{1}{c}{\cfaCheck}&\multicolumn{1}{c}{\cfaCheck}&\multicolumn{1}{c}{\cfaCheck}&\multicolumn{1}{c}{\cfaTimes}& \multicolumn{1}{c}{\hspace{0.15cm}\cfaNeutral$^5$} & \multicolumn{1}{c}{\textbf{\cfaCheck}} \\ 

\textbf{Flower} &\multicolumn{1}{c}{\cfaCheck}&\multicolumn{1}{c}{\cfaCheck}&\multicolumn{1}{c}{\cfaCheck}&\multicolumn{1}{c}{\cfaTimes}&\multicolumn{1}{c}{\cfaTimes}  & \multicolumn{1}{c}{\textbf{\cfaCheck}} \\ 

\textbf{LEAF} &\multicolumn{1}{c}{\cfaTimes}&\multicolumn{1}{c}{\cfaCheck}&\multicolumn{1}{c}{\cfaTimes}&\multicolumn{1}{c}{\cfaTimes}&\multicolumn{1}{c}{\cfaTimes} & \multicolumn{1}{c}{\textbf{\cfaTimes}} \\ 

\textbf{OpenFL} &\multicolumn{1}{c}{\cfaCheck}&\multicolumn{1}{c}{\cfaCheck}&\multicolumn{1}{c}{\cfaCheck}&\multicolumn{1}{c}{\cfaTimes}&\multicolumn{1}{c}{\cfaTimes} & \multicolumn{1}{c}{\textbf{\cfaTimes}} \\

\textbf{TFF} &\multicolumn{1}{c}{\cfaTimes}&\multicolumn{1}{c}{\cfaCheck}&\multicolumn{1}{c}{\cfaTimes}&\multicolumn{1}{c}{\cfaTimes}&\multicolumn{1}{c}{\cfaTimes} & \multicolumn{1}{c}{\textbf{\cfaTimes}} \\\cline{1-7}

\cellcolor{yellow!60}\textbf{\FL} &\multicolumn{1}{c}{\cfaCheck}&\multicolumn{1}{c}{\cfaCheck}&\multicolumn{1}{c}{\cfaCheck}&\multicolumn{1}{c}{\cfaCheck}&\multicolumn{1}{c}{\cfaCheck} & \multicolumn{1}{c}{\textbf{\cfaCheck}} \\ \cline{1-7}
\multicolumn{7}{p{13cm}}{$^1$~Asynchronous Strategies are supported in simulation.}\\ 
\multicolumn{7}{p{13cm}}{$^2$~Claims custom strategy is possible by modifying the framework code.} \\
\multicolumn{7}{p{13cm}}{$^3$~Async is an experimental feature and does not have documentation to test it.}\\
\multicolumn{7}{p{13cm}}{$^4$~Claims to allow persisting the server state to a SQLite~Database, But no documentation to verify.}\\
\multicolumn{7}{p{13cm}}{$^5$~Some updates to the model are layers possible using a config file.}\\
\multicolumn{7}{p{13cm}}{$^6$~Not natively supported but can be added programmatically by user.}
\end{tabular}
}
\\

\end{table}


\subsection{Synthesis of FL Framework Requirements and Limitations}\label{sec:challenges:req}
\label{subsec: requirements-good-framework}

We consider the requirements of three primary user groups of FL frameworks -- \textit{FL researchers} focus on designing and refining novel FL algorithms in order to achieve faster model convergence; \textit{Systems researchers} work on optimizing the efficiency and scalability of FL systems, addressing challenges like communication overheads in distributed training; and \textit{FL practitioners} focus on adapting existing FL algorithms for real-world deployments. We do not consider \textit{Enterprise users}, who may have additional requirements beyond practitioners (e.g., authentication, governance, etc.).

We synthesize their requirements below and compare the feature-matrix of popular FL frameworks: \textit{FedML AI}~\cite{he2020fedml}, \textit{FedScale}~\cite{lai2022fedscale}, \textit{Flower}~\cite{beutel2020flower}, \textit{LEAF}~\cite{caldas2018leaf}, \textit{Intel's OpenFL}~\cite{openfl_citation} and \textit{Google's TensorFlow Federated~(TFF)}~\cite{tff}, in Table~\ref{tbl:framework_comparison}.

\subsubsection{Modularity and Extensibility}

Planning an FL application is an iterative process (Fig.~\ref{fig:fl_development_lifecycle}).
This necessitates frameworks with easily accessible and modifiable ML and FL modules. We take a \textit{user-first design principle} in \FL, which focuses on two key aspects: (1) identifying the necessary modules and interfaces to design novel FL strategies, and (2) enabling rapid assembly and prototyping of these components.

\paragraph{Modularity}~We identify \textit{client selection} and \textit{aggregation} as modular logic-blocks that need to be customized as part of an FL strategy rather than as a monolith.
Client selection picks the optimal subset of clients in each training round, while model aggregation combines the trained local models with the global model in novel ways to achieve high accuracy in the presence of data, device and behavioural heterogeneity. 
Designing \textit{custom client selection strategies} needs a holistic view of the available clients, their participation history and data distribution to help select a subset, using flat or hierarchical approaches. The aggregation module should support both \textit{synchronous and asynchronous strategies} over local models, and be able to decide when to perform aggregation. Several FL strategies focus on either custom selection or aggregation, while some jointly develop both. Both must be possible.

\paragraph{Extensibility and Rapid Design} \label{para:Extensibility and Rapid Design}~It should be possible to mix and match a library of pre-defined popular modules with custom ones to rapidly design, evaluate, adopt and extend new strategies.
FL frameworks must support interfacing with \textit{standard training engines} like PyTorch and TensorFlow to offload the actual training and model aggregation. 
In addition, higher-level ML libraries and abstractions like HuggingFace also need to be natively supported for rapid development. This will give researchers a benchmark platform to compare against uniformly and also engender reproducibility. Further, the test setup should be translatable to deployment with minimal to no extra configuration.

\paragraph{Limitations of SOTA Platforms}
As shown in the \textit{Modularity} columns of Table~\ref{tbl:framework_comparison}, 
most frameworks support both PyTorch and TensorFlow as external \textit{ML training engines}, and offer several FL strategies as \textit{built-in plugins}.
But these frameworks are limited in their extensibility, particularly in their support for complex FL algorithms. 
Many frameworks either \textit{do not maintain sufficient information} about client characteristics and performance, and/or \textit{lack an interface} for the custom client selection and aggregation modules to access these values. This lack of a flexible interface hinders the implementation and modular extension of more advanced FL strategies.

A critical limitation is a \textit{lack of support for asynchronous aggregation}, which has demonstrated significant potential in recent FL research\cite{chai2021fedat, xu2023asynchronous}. Extending existing platforms for asynchronous aggregation requires substantial modifications and a fundamental rethinking of their orchestration. E.g., asynchronous aggregation strategies such as FedAsync~\cite{xie2019asynchronous} need to adapt to 
client updates as they are received, running aggregation and client selection modules concurrently, while managing potential race conditions.

Further, frameworks like FedML and OpenFL maintain a static list of clients throughout the training session, offering \textit{no mechanism to detect client failures} and prevent their selection in subsequent rounds, let alone dynamically include newly joining clients. While OpenFL includes a timeout mechanism based on estimated round-trip times to avoid indefinite waits for unresponsive clients, this can prolong training rounds when inactive clients are repeatedly selected.

Complex FL strategies like TiFL~\cite{chai2020tifl} rely on \textit{dynamic client characteristics} such as performance, latency, availability, etc. This requires modifications to the core components of the frameworks to maintain these client statistics. OpenFL, Flower and FedML require client selection and aggregation modules to be implemented together for such strategies, further complicating their extensibility.

\paragraph{\FL capabilities} As described later, \FL exposes interfaces to define custom client selection and aggregation strategies. This user logic has visibility into a variety of system information 

across rounds, e.g., the model accuracy and runtime training history, and the performance, availability and hardware specs of the clients.

These are required by popular strategies like TiFL~\cite{chai2020tifl}.
Further, \FL maintains a shared global session state with read and write restrictions that allow client selection and aggregation logic to coordinate without interfering with each other, if they are being co-designed.

\FL also allows custom trainers, dataloaders and DNN training frameworks to be plugged-in with support for high-level ML libraries like HuggingFace~\cite{wolf2020transformers} 
and DGL~\cite{wang2019deep}. E.g., \FL could rapidly implement a Federated GNN framework~\cite{naman2024optimizing} using DGL.
We also offer a number of built-in strategies that can be combined with custom ones, such as
client selection based on device performance and client data skews, and 

weighted averaging and staleness-aware asynchronous averaging for model aggregation. Further, we provide a declarative model to define the FL training session, using just a YAML config file to specify the pre-defined or custom training strategies and their parameters. 
This avoiding users having to write any code if existing modules suffice.

\subsubsection{Resilience}
\paragraph{Handling Client and Server Failures}~FL training can be long-running, operating for hours or even days. FL clients can be unreliable, fail in challenging field environments, or become unresponsive due to their resource-constraints (e.g., we see Pi and Jetson devices freeze up and need rebooting during sessions that run for $1$--$5$h) or when operatin over unreliable networks. There should be native ability to \textit{handle client failures} during a training session.
Server failures are less frequent but have a catastrophic impact, causing loss of hours of training. Simple checkpointing of the global model can help \textit{restart} the FL training using this prior global model as a bootstrap. But it will not be able to \textit{resume} the FL session, losing any historical session information used by the strategy.
More granular session recovery can ensure that the FL session behaviour can resume from the last checkpointed round, or even better, midway through a round.

\paragraph{Limitations of SOTA platforms and \FL capabilities}~ None of the SOTA frameworks natively support checkpointing the global model to disk and restarting an FL session from it. However, Flower offers sample user-code to checkpoint the global model from its aggregation strategy,
which can be extended to restart with the checkpointed model, adapted by other frameworks as well.
\FL has inherent support both for server restart with the checkpointed model and for  seamless server failover, just by enabling a flag, allowing resumption of the FL session from the last checkpointed session state. It also allows mid-round resumption of an FL session by externalizing session states to a Redis store,
all done in a declarative no-code manner by the user. 
Flower initiates a gRPC connection at the start of an FL session and retains it through training. Client failures are detected when this connection terminates. But maintaining a persistent network connection to 100-1000s of clients can be costly.

OpenFL lacks a failure detection mechanism but can impose timeout limits on a training round, ensuring the training progresses. However, this may result in unavailable client still being selected in subsequent rounds.
FedML lacks both client failure detection and round timeouts, causing the session to hang if a client fails.
\FL uses periodic heartbeats from clients to detect their liveliness, allowing the framework to differentiate between failed and slow clients. It also allows incorporating client behavior into FL strategies, giving users and their strategies the control to manage such failures effectively.

\subsubsection{Deployment Models}
FL frameworks should support seamless deployment on single-machine pseudo-distributed  \textit{simulation} setting for rapid testing, deployment \textit{on-device} using edge hardware for operations, and \textit{containerized deployment} for large-scale validation of strategies~(research) or training models/datasets~(practice), \addc{and emulating large edge clusters and IoT networks~\cite{violet}.} It is non-trivial to translate a simulation or pseudo-distributed setup to a real-world operational deployment, with its complexity and challenges that can require the FL strategy itself to be revisited. 

Given the $100$--$1000$s of client devices that may be present, scaling from a single-machine prototype to a large cluster setup should be \textit{easy to configure}. Their instantiation and configuration should not have manual overheads, e.g., copying model files, when launching training sessions. 

The framework should support \textit{heterogeneous edge resources}, from Pis to GPU workstations, given the need for both cross-device and cross-silo FL~\cite{lai2022fedscale,openfl_citation}. 
There should be managed ways for clients to dynamically join and leave the training pool during a session. The server should be capable of dynamic client discovery to maintain the \textit{clients' availability} for use during client selection.

\paragraph{Limitations of SOTA platforms}
Most FL frameworks either do not support deployment on edge devices or offer only a limited set of features (\textit{Deployment} columns of Table~\ref{tbl:framework_comparison}). 
Early FL frameworks like TFF and LEAF designed for FL researchers allow only \textit{simulation} of clients training on a single machine, which also hosts the server. Recent frameworks like FedML, Flower and OpenFL support \textit{on-device} deployment. However, 
\textit{model delivery} is not automated and requires manual copying of model definitions and file dependencies to each client.

Further, FedML and OpenFL require the client list to be statically defined, and each client's configuration manually edited. This lack of \textit{client discovery} imposes a massive overhead when running at scale and makes translating from single-machine prototype to deployment on large clusters cumbersome.

\paragraph{\FL capabilities}
\FL offers a containerized client package that can be used for \textit{initial bootstrapping} of the client stub. This helps deploy clients at scale on VMs or edge devices.
Further, during an FL session, \FL delivers models, trainers, dataloaders and related dependencies from the server to the clients at runtime to \textit{initialize the session}, if they are not already cached. Clients locally cache these components, identified using a unique package name, for reuse across training sessions. These significantly eases the deployment of \FL and execution of FL sessions across 1000s of clients.
Configuring FL sessions using a declarative YAML file also allows easy hyper-parameter tuning and deployment of different FL runs.
As mentioned before, we track the liveliness of clients and also allow registration of new client using an advertisement mechanism that client selection and aggregation modules can leverage. 
Although not the core focus of our work, \FL also allows users to rapidly prototype and test models and strategies on a single machine by launching multiple client processes, and requires minimal configuration for deploying training to \textit{cross-device} setting, as we evaluate in this article, or in a \textit{cross-silo} setting~\cite{naman2024optimizing}.

\subsubsection{Security and Privacy Non-goals}

 \addc{Addressing security issues in \FL is a non-goal in this article. Understanding and identifying security threats, attack vectors and vulnerabilities is a whole research dimension in itself and requires substantial investigation of the system design, deployment model and FL strategy design. That said, \FL's design makes it easy to support such capabilities in future.}

Some FL frameworks support specialized security and privacy features required by some FL applications or deployments, which we enumerate but not yet support. 
While \FL assumes a \textit{fully trusted server}, several FL techniques operate with  \textit{honest but curious} servers.
Here, cryptographic and Multi-Party Computational (MPC) methods have been proposed for secure aggregation~\cite{Bonawitz2017}. Homomorphic encryption has also been used for aggregation over encrypted client models~\cite{BatchCrypt}.
Differential privacy introduces noise into the local models that cancel each other out during aggregation, to prevent the server from deciphering the trained model~\cite{MaoqiangWu2021}. 
Further, there are server/client authentication mechanisms e.g., using certificates,
that provide additional security \addc{and mutual trust}.
\FL does not support these yet, but are easy to extend as an orthogonal layer.

\addc{\FL also assumes that \textit{clients are not malicious}. However, \FL offers inherent modularity to implement existing FL strategies to account for adversarial clients.
Statistical tests of the client models using euclidean distance~\cite{cao2019understanding} or Pearson correlation coefficient~\cite{liu2021privacy} have been used to detect data or model poisoning attacks. This can be implemented as part of the Validation phase of FL strategies in \FL. Other defense strategies~\cite{Wang2020model} evaluate the trained local or global model against a held-out dataset in the leader to detect outliers. This can be incorporated as part of the Aggregation phase. We leave the implementation of such FL strategies using \FL to be robust to client attacks as part of future work. 

}

\begin{figure}[t!]
    \centering
    \includegraphics[width=0.95\columnwidth]{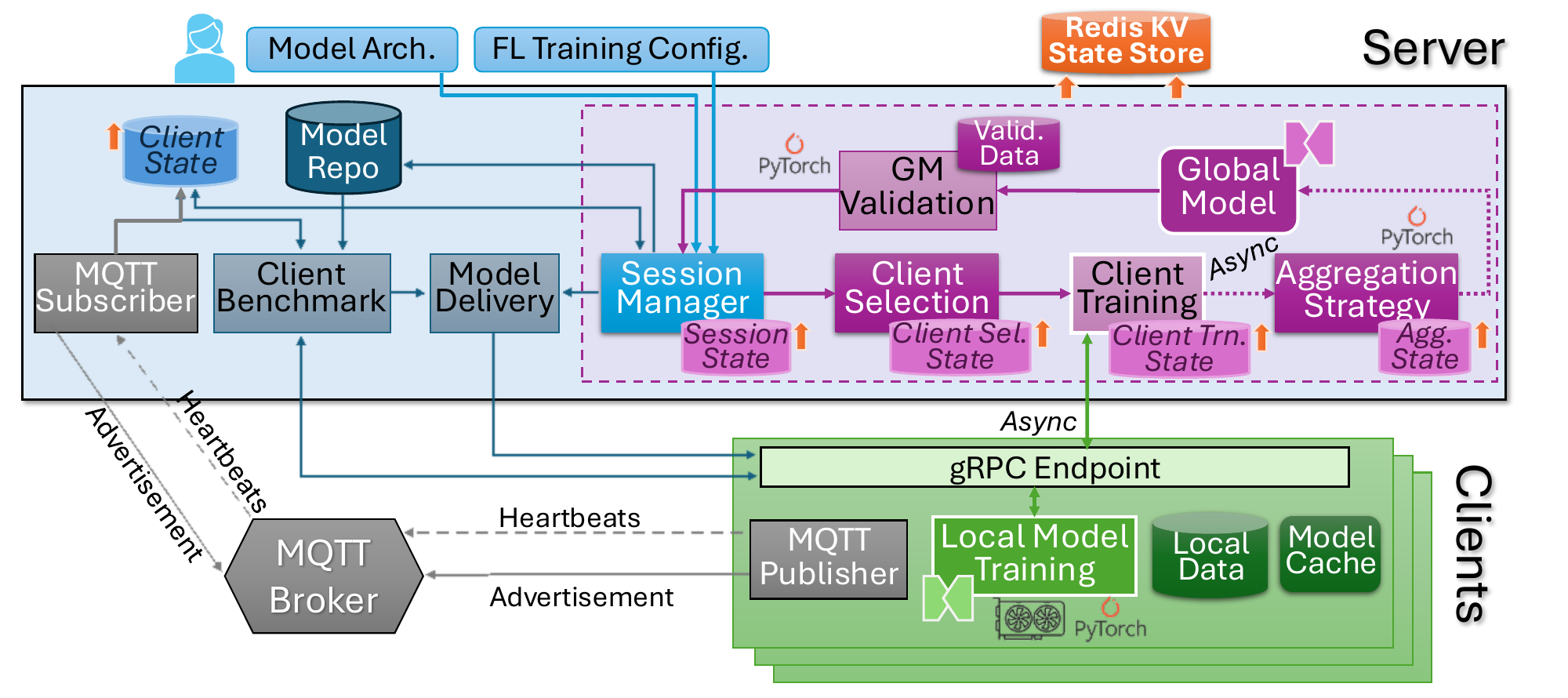}
    \caption{\FL Architecture}
    
    \label{fig:arch}
    \vspace{-0.15in}
\end{figure}

\section{\FL Architecture}
\label{sec:arch}

Fig.~\ref{fig:arch} shows the architecture diagram for \FL and its various components.
\FL has a \textit{stateful Leader Service} running on a central server machine or VM, and \textit{stateless clients} running on each potential device that will participate in the training. 
The Leader Service has three components: \textit{client discovery} coordinated using MQTT, the \textit{Server Manager} that manages the model repository, benchmarking clients, etc. across sessions, and the \textit{Session Manager} which \delc{is} orchestrates training of a single FL model. 
A \FL training session operates over a set of \textit{states}, and well-defined \textit{module interfaces} that are triggered based on \textit{events} and can access and/or modify these states as part of the training lifecycle. These are described below.


\subsection{FL Strategy Composition and Session Launch}
\label{subsec:compose}

Users who wish to train a FL model can specify the model architecture and configure key FL parameters, including client selection and aggregation strategies, within a training YAML file (sample in ~\ref{appendix: training-config}). These strategies can be either built-in, requiring only changes to the YAML configuration file, or custom ones, where users implement their own client selection and/or aggregation strategies as detailed in Sec.~\ref{sec:interfaces}. 
The configuration file also has hyperparamters for the session (e.g., \# of training rounds, if validation should be run, etc.), the client training (e.g., \# of epochs per round, batch sizes, optimizer, etc.), and the specific model architecture (e.g., custom dataloader). Such a \textit{declarative approach} makes it simple for users to design and launch new FL sessions rapidly.

The submitted model and configuration files are used to create a \textit{FL training session} in the Leader Service, which is executed as part of the training lifecycle. The \textit{session} captures all state of the training. This design allows the same deployment of \FL leader and clients to perform multiple concurrent FL training sessions. \addc{This is helpful for parallel training on the same set of client of the same model with different hyper-parameters, different models on the same data, or across different data present in the same set of client. This can improve resource efficiency and throughput of FL systems~\cite{zhuang2022mufl, bhuyan2022multi}.}
However, currently, only one session is allowed at a time.


\subsection{Event-driven Training Lifecycle}
\label{subsec: training_lifecycle}

\begin{figure}[t!]
    \centering
    \includegraphics[width=0.85\columnwidth, trim={0cm 5.5cm 0cm 0cm}]{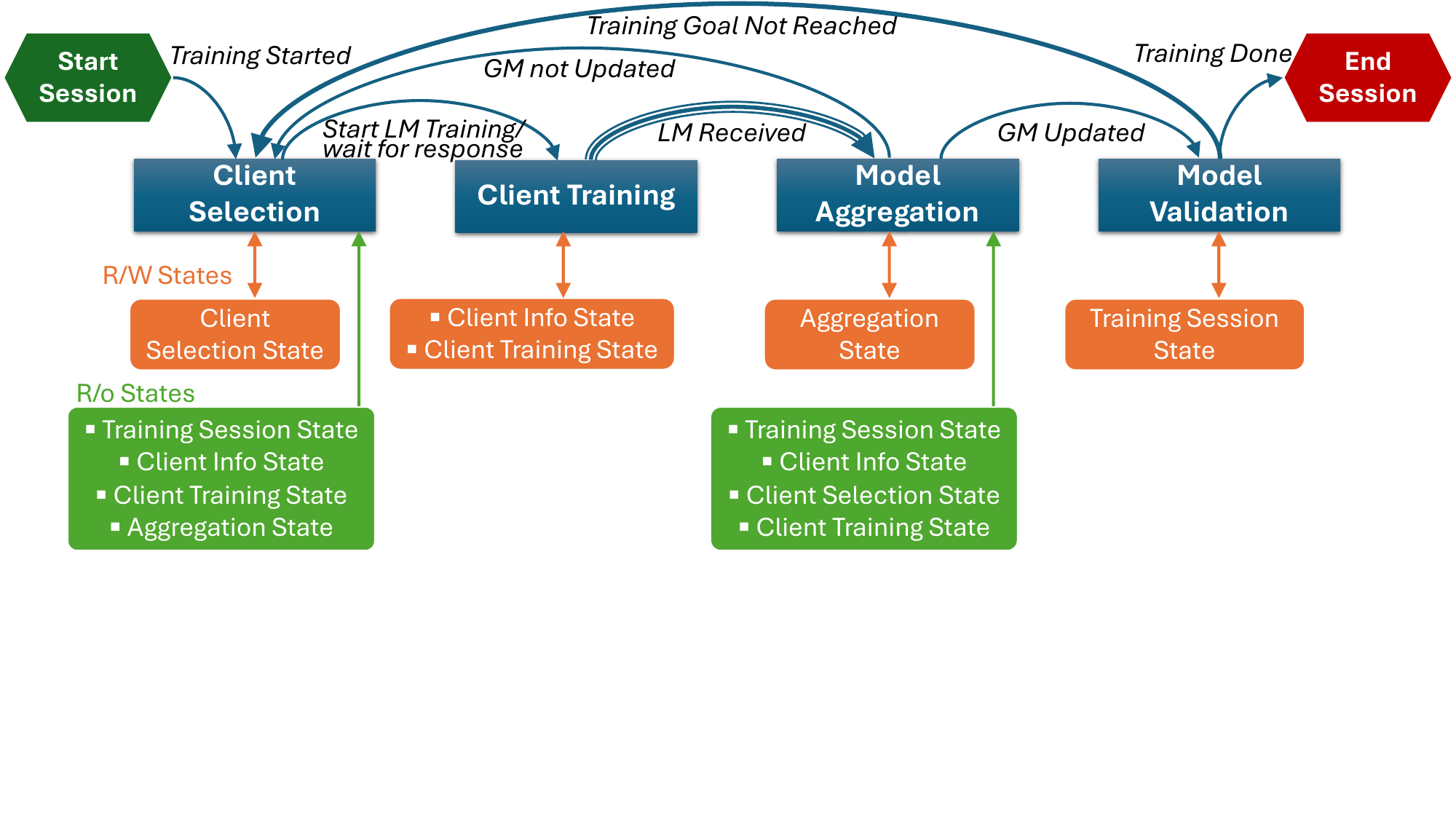}
    \caption{\FL Training Lifecycle showing event triggers between phases and read-only (green) and read/write (orange) states accessible to each module.}
    
    \label{fig:lifecycle}
\end{figure}

\paragraph{\FL Training Lifecycle}
The \FL leader maintains one deployment-specific and four session-specific persistent states (Sec.~\ref{subsec: server_resilience_theory}), which encapsulate all information related to a specific training session. 
Each state object has a set of key-value pairs, where keys may be pre-defined or custom entries, and values can be any value or Python object.
When a \FL leader starts, it populates the \textit{Client Info State} based on client discovery advertisements, and continuously maintains this in the background across sessions (Sec.~\ref{sec:arch:discovery}). 
Once a training session starts, the lifecycle for a single round consists of four \textit{phases}: Client Selection $\rightarrow$ Client Training $\rightarrow$ Model Aggregation $\rightarrow$ Model Validation, which are each triggered by the \textit{Session Manager} using an event-driven approach. We consider one training round as all the steps that are performed to update the global model once. This repeats for multiple rounds until a user-defined termination condition (e.g., accuracy, \# of rounds) is reached. Fig.~\ref{fig:lifecycle} shows the sequence of modules the \FL server calls during the training phases and the access permissions of the modules to the states. We discuss these modules next. The set of pre-defined states are given in \ref{appendix:System States}.

\subsubsection{Client Selection Phase}
\label{subsubsec:client_sel}
At the start of the training session, the Session Manager calls the \textit{Client Selection Strategy (CS) module} specified in the training configuration. Subsequently, every response from the client will result in a call to this module, after the Model Aggregation and optionally the Model Validation modules are invoked (Fig.~\ref{fig:lifecycle}). 

The CS module has read-write access to the \textit{Client Selection State}, where it maintains the details required across multiple rounds to make decisions in this session. The Session Manager also passes it read-only access to the \textit{Session, Client Info, Client Training} and \textit{Aggregation States}.  
Such read access to states from other parts of the lifecycle allows CS to make intelligent local decisions based on global session knowledge. Leveraging this, CS can optionally select a set of clients, whom the Session Manager will then request to run a model training on in the Client Training phase. 
E.g., the CS of FedAT~\cite{chai2021fedat} uses the list of clients and their performance benchmarks in the Client Info State to pick ones with balanced performance while the CS of TiFL~\cite{chai2020tifl} uses the accuracy of the local model training, maintained in the Client Training State, to select the clients~(pseudocode in \ref{appendix:psudocode}).

\begin{figure}[t!]
\centering
    \subfloat[FedAvg Strategy\label{fig: strategy-seq-diagram-fedavg}]{\includegraphics[width=0.99\textwidth]{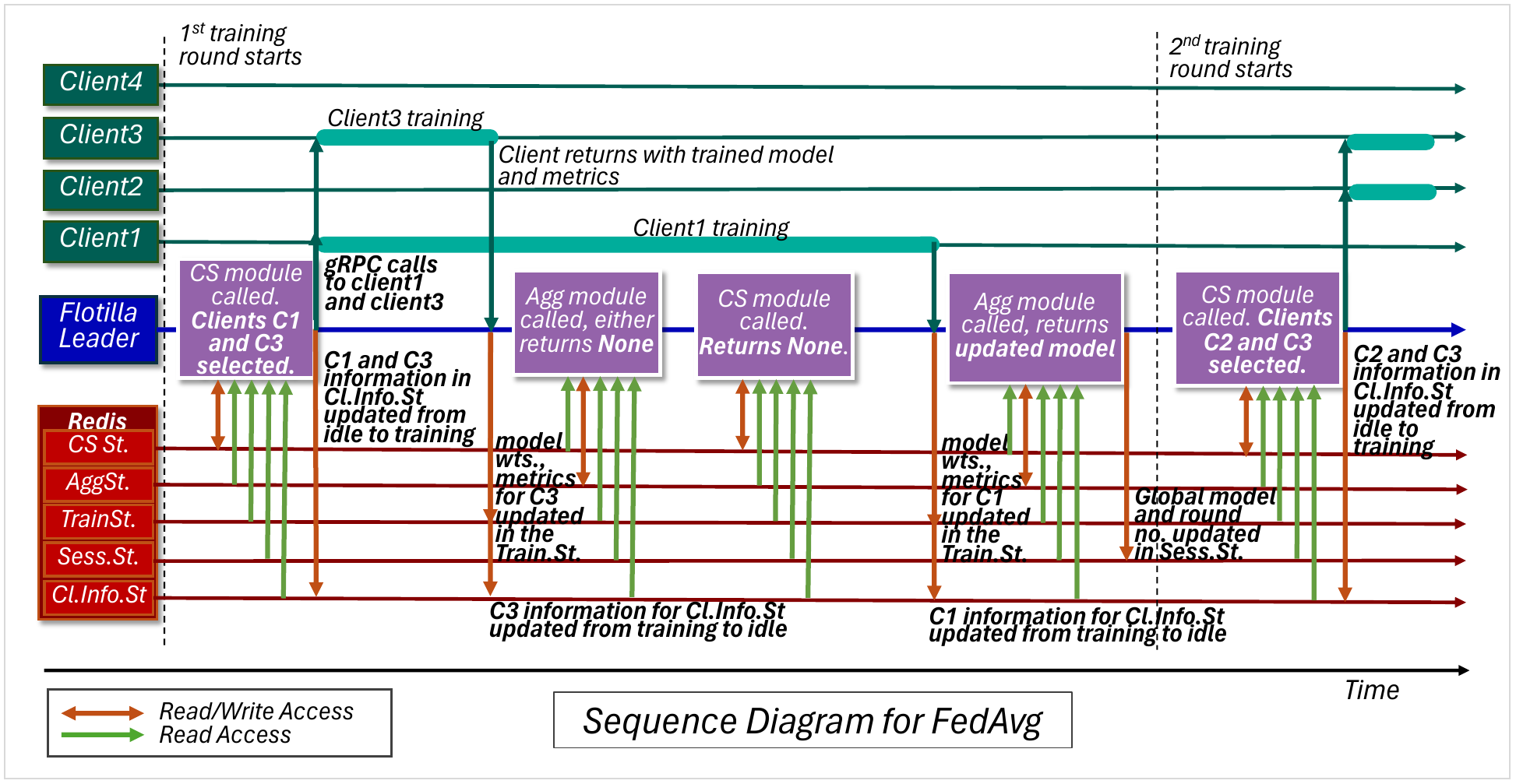}}\\
    \subfloat[FedAsync Strategy\label{fig: strategy-seq-diagram-fedasync}]{\includegraphics[width=0.99\textwidth]{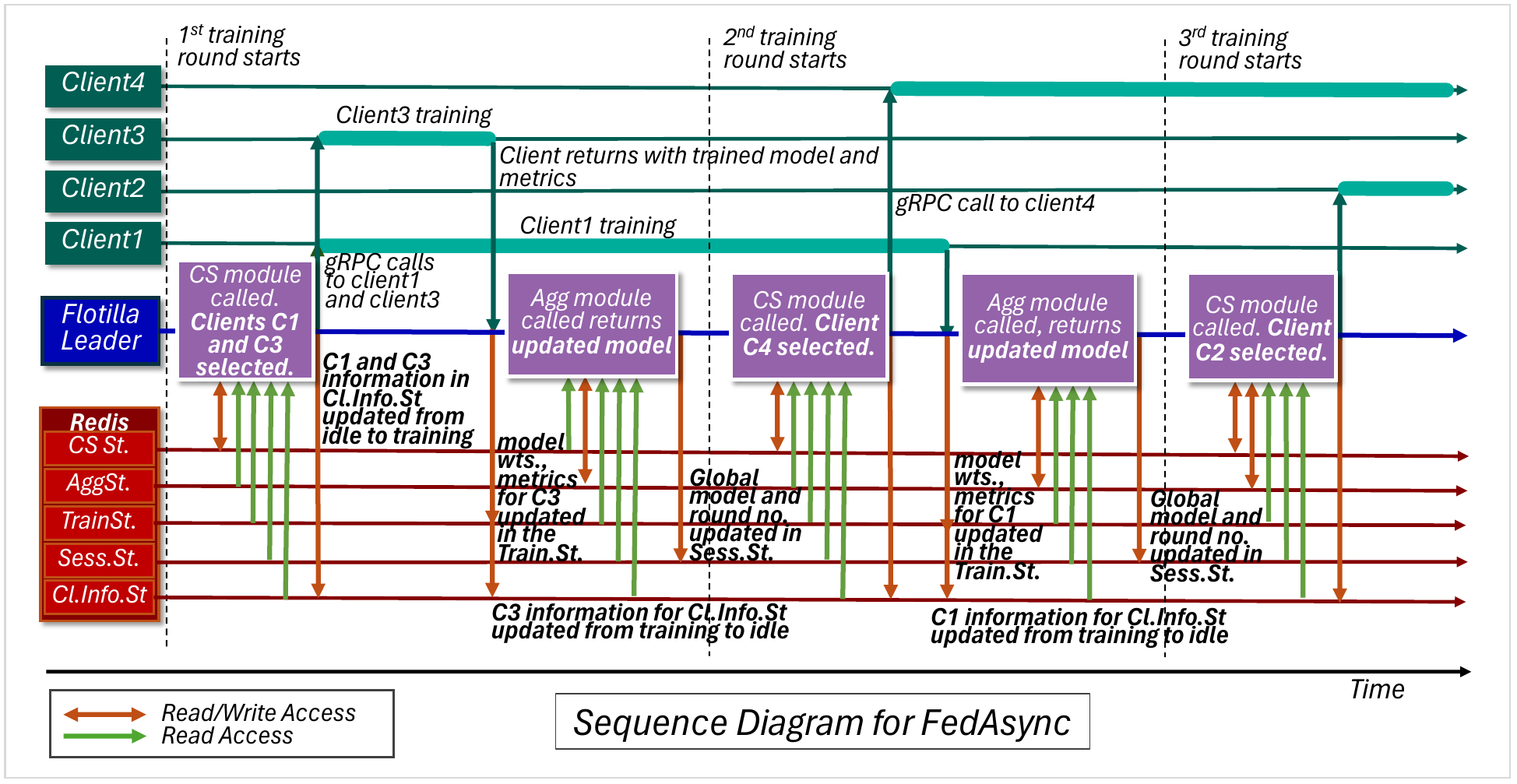}}
    \caption{Sequence diagram of interactions between Leader modules and clients in \FL and access to the states, for example execution of FedAvg (top) and FedAsync (bottom).}
    \label{fig: strategy-seq-diagram}
    
\end{figure}

Fig.~\ref{fig: strategy-seq-diagram} shows the sequence diagram of the execution flow for the FedAvg and FedAsync strategies, for 4 clients across a couple of rounds.

Both FedAvg and FedAsync select \texttt{Client1} and \texttt{Client3} to train in the first round. For FedAvg (Fig.~\ref{fig: strategy-seq-diagram-fedavg}), when \texttt{Client3} finishes training and returns the local model, 
the Model Aggregation decides not to perform aggregation since the other client is pending. Model validation also is skipped since there is no new global model. The CS module is called and, since it does not detect an updated global model version in the Training Session State, it does not select any new client for training. 
Later, when Aggregation happens on receipt of the local model from \texttt{Client1} and a new round is initiated, the call to CS returns two new clients, \texttt{Client2} and \texttt{Client3}, selected for training in the second round. So, the states form a flexible means for coordination and parameter-passing between modules.
FedAsync (Fig.~\ref{fig: strategy-seq-diagram-fedasync}) causes each local model returned by a client to trigger Aggregation of a new global model and its Validation, and starts the next round. Here, the CS module selects \texttt{Client4} to train a new local model for round 2 after \texttt{Client3}'s model is received and aggregated.

\subsubsection{Client Training Phase}
If the CS module selects a list of clients for local training in a round, the Session Manager updates its \textit{Client Info State} with this.
The Manager calls the \textit{Client Training module}, which reads the \textit{Client Training State} to determine if client training is required, and makes asynchronous gRPC calls to the relevant clients in the Client Info State to start their local training. 
The gRPC request sends the current global model from the Training Session State,
and any user-defined hyper-parameters such as the number of epochs, batch size, etc. to the client.
The module marks the clients as being in a training status in the Client Info State.
Depending on the FL strategy, CS can request clients to perform client-side validation. E.g., TiFL complements the server-side validation accuracy using client-side validation on all clients at every $n^{th}$ round (configurable) to help assign tier probabilities. In Figs.\ref{fig: strategy-seq-diagram-fedavg} and \ref{fig: strategy-seq-diagram-fedasync}, we see such gRPC calls being made to the clients selected by the Client Selection module.

\subsubsection{Model Aggregation Phase}

The Session Manager receives an asynchronous gRPC response from each client once its local model training is completed, or if 
the gRPC connection with the client is lost. It updates the Client Training State with the client's response, which contains the client's updated local model and training statistics. 
The Client Info state is also updated to indicate that the client is no longer training. 

The Manager then triggers the \textit{Aggregation (Agg) module}, which has read-only access to the \textit{Session, Client Info, Client Selection and Client Training States}, while it is granted read-write access to its own \textit{Aggregation State}.  
When each local model update~(or a ``failure flag'' if the client fails) is passed to the Agg module, it can either defer aggregation by adding the local model to its state, or proceed to aggregate all previously stashed models and return a new global model to the Session Manager. 
The Manager updates the Session state with the new global model and increments the round number.
If creating a new global model, Agg clears the Aggregation State.
Other than monitoring the global model version number in the Session state, the CS module can also use a cleared Aggregation state to determine that a new round has started.

In Fig.~\ref{fig: strategy-seq-diagram}, Agg is executed each time a client finishes training. For FedAvg, when \texttt{Client3} returns from its training, the Agg module adds its ID and its model to the Aggregation state, and checks if both the training clients listed in the Client Selection state have returned. Since \texttt{Client1} is still pending, it waits for it to return its local model, and then aggregates it with the stashed model of \texttt{Client3} to generate a new global model.
It also uses the data-count for clients present in the Client Info state to weight the local models during aggregation.
For FedAsync, a new global model is aggregated each time a local model is received from a client, and no Aggregation state is maintained.

\subsubsection{Model Validation Phase}
\addc{Validation is a key phase that allows the users to determine if the FL training is making adequate progress and if further rounds of training are required.} If the user has configured the FL session to perform model validation, the \textit{Validation module} is executed by the Session Manager\addc{, typically on some held-out dataset,} to estimate the accuracy of the newly aggregated global model \modc{that} is created. \addc{This can be used by the Manager to decide if a particular accuracy threshold set by the user has been met, upon which the FL training session can be terminated.} \addc{This leader-side validation logic can also be complemented by client-side logic that the users can provide. The loss reported by the clients or seen for the global model can also be a metric for convergence.} \modc{Besides accuracy, other termination conditions that can be specified by the user include a fixed number of rounds or a fixed time budget.} If starting the next round, the CS module is once again called.
\addc{The accuracy or other scores calculated in the validation phase also comes in handy during client selection. E.g., TiFL~\cite{chai2020tifl} uses these to adjust client selection probabilities, favoring clients that contribute more effectively to model quality. It can also be used to improve the robustness of the training in semi-trusted environments, e.g., by comparing the validation data from different clients to detect malicious ones or to prevent against poisoning attacks~\cite{kabir2023flshield}.}
For brevity, the Validation module is not shown in Fig.~\ref{fig: strategy-seq-diagram}.


\subsection{State Management}
\label{subsec: server_resilience_theory}

While \FL clients are stateless, the \FL leader is built in a \textit{state-centric manner}.
For each of the five state objects -- Client Info, Training Session, Client Selection State, Client Training State and Aggregation State -- the Manager has two wrapper objects: one that exposes a \textit{read-write interface} to the state and is passed only to the ``owner'' module for the state, e.g., Aggregation State is accessed and updated by the Aggregation module, and a \textit{read-only interface} that is passed to other modules to access and take decisions, e.g., Client Info State read by the Client Selection logic.  
There are a set of pre-defined state entries that the \FL platform uses, populates and interprets, which we describe next and whose entries are listed in \ref{appendix:System States}. In addition, user-logic for CS and Aggregation can also define their own custom state entries to access across different rounds of the session, or to coordinate with each other.

\subsubsection{Client Info State}

The Client Info state 
is persisted across training sessions with details about every client seen by the server as part of discovery (Sec.~\ref{sec:arch:discovery}). 
It maintains the client's gRPC endpoint, hardware specs, the dataset tags and their distribution (which can be revealed), and optional soft-state information such as prior models, benchmarks, etc. 
This also has the uptime history for this client, updated by the Discovery module using heartbeats. In future, we plan to also list the network visibility of the client in case it is behind a firewall. 
This state is only updated by the Client Discovery component, Session Manager and Client Trainer module, but read by CS and Agg modules. E.g., benchmarks of clients are used to cluster them into tiers by TiFL and FedAT~(pseudocode in \ref{appendix:psudocode}). The Client Info state also keeps track of clients that are supposed to be currently training, allowing the Client Training module to know which set of clients have been selected to train by the CS module.

\subsubsection{Training Session State}
This and the next three states are initialized at the start of a training session. The Training Session state contains the configuration provided by the user for that training: the ML model architecture, dataset name to train it on, global training rounds or duration of training for termination, CS and Agg strategies and their config parameters, and hyper-parameters for client training.
The Training Session also has the latest global model and training round number, updated by the Session Manager. 
The Training Session serves as a bootstrap, and maintains the logical ID to the other three states for this session.
Contents of this state are used to restore/revive the server from a failure (Sec.~\ref{sec:arch:reliability}).

\subsubsection{Client Training State}
The Client Training State tracks the training state of each individual client for the session, maintaining details of the last round in which each client participated, along with the training and validation metrics (accuracy, loss, training time, etc.) reported for that round. The Client Training module updates this information, and it is accessed by the CS, Aggregation, and Manager modules.

\subsubsection{Client Selection and Aggregation States}
The final two states are provided to the client selection and aggregation modules to track any custom information and metadata not already captured by the above states, and necessary for these FL strategy to make decisions within or across a round. \addc{The writable entries in these state are all defined by the user logic specific to an FL strategy, e.g., FedAvg maintains the list of selected client IDs per round in the Client Selection State while FedAT tracks the selected clients in each tier (\ref{appendix:System States}).} These are treated as black-boxes by the other modules as their entries are specific to the CS and Agg logic; only CS and Agg modules access each others' states to make coordinated decisions within and across rounds.


\subsection{Custom Client Selection and Aggregation Modules}
\label{sec:interfaces}
\label{subsec: cs-agg-interface}

\FL provides simple interfaces for users to separately develop and plug-in their custom modules for Client Selection and Aggregation strategies. These are the two key parts of the FL lifecycle with substantial prior research. 

The Client Selection module's interface is:
\begin{center}
\texttt{\textbf{clientSelect}(sessionID, availableClients, \textit{client\-Sel\-StateRW, aggStateRO, clientTrainStateRO, client\-Info\-StateRO, trainSessionStateRO,} clientSelUserConfig)} $\rightarrow$ \texttt{List<clientID> | NULL}
\end{center}
and it returns a list of selected clients which need to be trained on, or a null object in case the selection is deferred till pending async client training completes; an \texttt{RO} suffix indicates read-only while \texttt{RW} means a read-write state object. 

The Aggregation module's interface is:
\begin{center}
\texttt{\textbf{aggregate}(sessionID, clientID, localModel, \textit{aggStateRW, client\-Sel\-StateRO, clientTrainStateRO, client\-Info\-StateRO, trainSessionStateRO,} aggUserConfig)} $\rightarrow$ \texttt{globalModel | NULL}
\end{center}
and it returns either the updated \texttt{globalModel} or \texttt{NULL} if the aggregation was not triggered for this model update. We provide pseudo-codes for using these interfaces to implement the popular SOTA synchronous strategy, TiFL~\cite{chai2020tifl}, and asynchronous strategy, FedAT~\cite{chai2021fedat} in \ref{appendix:psudocode}. These illustrate the interaction between various states and the client selection and aggregation modules within the framework, showcasing its flexibility.


\subsection{Client and Server Resilience}\label{sec:arch:reliability}

\begin{figure}
    \centering
    \includegraphics[width=0.9\columnwidth]{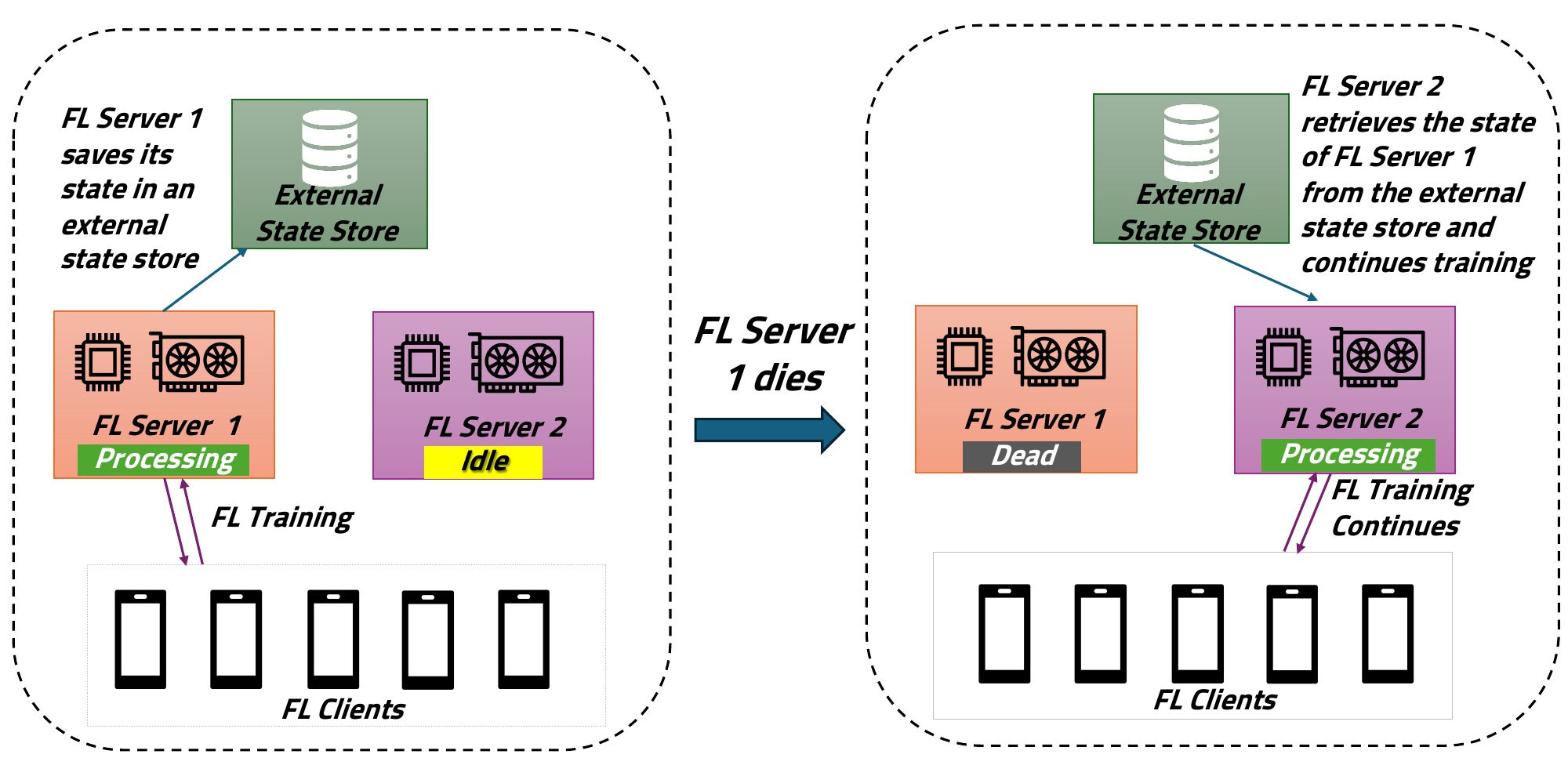}
    \caption{
    Using \FL's external state store to \textit{resume} a training session on server failure. 
    }
    \label{fig: server resilience}
    \vspace{-0.15in}
\end{figure}

\textit{Client failures} are common when $100$--$1000$s of clients are in the field or when intense training causes them to fail. The Client Discovery module (Sec.~\ref{sec:arch:discovery}) maintains the availability of active clients using periodic heartbeats. This ensures that when calling the CS module, the devices selected by it from the ones that are available will mostly likely be able to participate in the round.

A client failure can be detected at several points of the FL lifecycle: missing a certain number of heartbeats before or during participation in a training round, failure when invoking a gRPC call during benchmarking, training or validation, invoking the gRPC request successfully but the client fails in-between or fails to respond 
within the user-defined timeout. 

If the client misses a user-specified number of heartbeats, the client is marked not active in the Client Info State and the CS module would exclude it from being selected. The status of the client is reinstated as active after regular heartbeats resume. 
If a client's gRPC endpoint is not accessible when initiating training, or if the training timeout is reached without receiving the local model, the gRPC call will trigger a fault in its callback to the Session Manager.
The Client Info state is updated to reflect that the client is not training and the training round in which the failure occurred is also captured. The Aggregation module is triggered with a failure flag for this client and it can use this detail when deciding on aggregation for the round. E.g., we provide variants of FedAvg that performs global model aggregation once $m$ out of $n$ clients have returned their local model updates, thus tolerating up to $(n-m)$ client failures. 

However, a gRPC failure does not remove the client from the available pool, but just marks its status in the Client Info state and lets the CS module decide on its inclusion.
This design allows the framework to differentiate between failed clients that are unavailable
and over-worked clients ones
that still be available to train in subsequent rounds. 
Other than this, the loss of a client does not disrupt the training session as it is stateless.

\textit{Server failures} are less common but still possible, e.g., due to VM failures, power outages, adversarial requests etc.. This can cause the catastrophic loss of hours of training. Recovering from server failure is more involved. Here, we offer two complementary strategies. A discrete \textit{checkpointing} strategy allows users to specify a flag for the Session Manager to save the five session states to disk (potentially mounted on a remote server) after every $k$ rounds of training. This ensures that if the server fails, the \textit{session can be resumed} in future on a different server by restoring these five states into the new Leader's Session Manager and resuming from the last checkpointed round. This can cause up to the last $k$ rounds~(including the current round) to be lost. But this is better than starting a clean FL session without the states and only using the global model as a bootstrap.

An added feature in \FL optionally \textit{externalizes the state objects} in realtime. Here, the states are persisted and continuously updated at a remote and reliable Key-Value store. Specifically, we provide implementations of the state object where all \texttt{put} and \texttt{get} operations pass through to a durable Redis service, e.g., offered by a Cloud provider. This ensures that at any point in time, the KV store has the current state of the session. This complements the default implementation of the state object that maintains the values in an in-memory \texttt{dictionary}.
This externalized state, however, needs to be used with care. Since the server might fail in the midst of a round, the state that is persisted may be a partial state. 
So the relevant modules that access these states when the session is restored need to estimate the consistency of the keys they wish to use. Stronger built-in models of consistency will be explored as future work.

When a server fails, we can rapidly restore the session on a secondary server (but connected to the same MQTT broker for Client Discovery, to access the same client pool), and have the session \textit{resume} from the relevant checkpoint or the externalized state. The overheads of checkpointing and state persistence, and their impact on lost rounds are discussed in Sec.~\ref{subsec: server_resilience_exp}.


\subsection{Client Discovery and Initialization}
\label{sec:arch:discovery}

Clients and server(s) interested in participating in FL training need a means to discover each other. We use an advertisement-based approach~\cite{frey2002condor}, where knowledge of a well-known \textit{MQTT publish-subscribe broker} endpoint serves as the bootstrapping mechanism. The clients advertise their availability for training on the \texttt{clientAdvert} topic of this broker and publish basic details about themselves such as their gRPC endpoint, etc., which is a subset of the details present in the Client Info state.
The server subscribes to this topic to discover the clients, and creates an entry in its Client Info state.

Subsequently, each client periodically sends a \textit{heartbeat message} to the \texttt{client\-heartbeat} topic, e.g., every 5~secs or 60~secs (configurable). The server subscribes to this topic as well, and updates the Client Info state with the availability of the clients. It also maintains a history of their availability, if a CS strategy wishes to use this to pick reliable clients~\cite{Hu2022incentive}. 
The server can be configured to mark a client as inactive if more than a certain number, or a certain time period of heartbeats, are missed; such clients can be skipped by CS.

As part of the discovery process, the client may report its \textit{performance benchmark}, if specified in the client configuration, and this is included in the Client Info state. Otherwise, by default, a Session Manager wishing to use an available client will initiate a benchmark of a model training on it for a few mini-batches, using a canonical model (or the current model being trained) for which the client has relevant training data. This training time taken gives a measure of its relative performance, which can be used by CS strategies. The benchmark result is cached in the client and reported in future advertisements.

Clients are stateless. The only prerequisites for the client to participate in training are to have a \FL agent with a gRPC endpoint running, 
PyTorch or TensorFlow training framework installed, and host the relevant local data to train over. The Python code to train the individual model architecture in an FL session is dynamically pushed by the Session Manager to the client when it is first selected for training. This model is cached, and reused if the same model is used for local training in future; we use a SHA256 hash
over the model file to ensure that they are identical. Even if this cache is deleted, the Manager can push the model again on-demand.
This limits the client's software dependencies and management overheads. A client can even set a configuration flag to wipe all session related files after a training round for enhanced privacy.

We currently assume that the server may be behind a firewall and the clients are directly accessible from the server to perform synchronous or asynchronous gRPC calls. But this can be relaxed. It is possible that some or all clients may be on a private network or behind a firewall. Here, we can either use NAT or port forwarding to send requests to the clients, or introduce a pub-sub broker visible to both clients and servers to act as an message-bus to send requests and receive responses from such clients~\cite{10.1049/iet-sen.2017.0251}.


\subsection{Implementation}
\label{sec:impl}

All code for \FL is written from the ground-up in Python 3.6+. We use the \texttt{AsyncIO} library of Python to implement all the asynchronous event-driven functions. This was a conscious design choice as our experiments indicate that the server is primarily I/O bound rather than CPU bound. Python's \texttt{threading} package is used for parallel execution of the various modules of \FL, e.g., heartbeats, Session Manager, asynchronous response from the clients, etc. 
We use \textit{asynchronous gRPC} and \textit{ProtoBuf} for remote procedural calls from the server to the client.
The \texttt{paho} library is used for MQTT invocations and \textit{Mosquitto} serves as the MQTT Broker; the broker can run on a separate VM or, for a simple setup, on the same VM as the Leader Service.

The default in-memory state store is implemented as Python nested dictionaries, while the externalized state store uses Redis key-value store accessed using a Python client. The interfaces to both stores are identical, allowing drop-in replacement. 
Redis can be replaced with other NoSQL stores like Memcached.
Besides downloading and running the client agent and Leader service from the commandline, we also offer their Docker containers for quick deployment.

\FL supports \textit{PyTorch} and \textit{TensorFlow} as ML training engines. We provide PyTorch dataloaders for common datasets, and also support seamless access to artifacts from \textit{Huggingface}.
We natively provide 8 client selection strategies and 3 aggregation strategies as plugins for FL sessions. Several of these can be mixed to define new training configurations; the ones evaluated are described in Sec.~\ref{sec:eval}. Others can be easily implemented. \addc{While \FL is designed for real-world deployments, it also supports a \textit{simulation} and \textit{emulation} based FL execution. During a simulation run, \FL runs each client training sequentially on a single machine, typically a GPU server or workstation, colocated with the leader. For an emulation based deployment, our container-based deployment allows each edge client to be run within its own container, with the vCPUs for the container configured to match the performance of an equivalent edge hardware. For more complex emulation of device failures, network topologies, bandwidth/latency, etc. we can leverage our prior work on VioLET~\cite{violet} that can help deploy such IoT/edge containerized topologies within which \FL can be executed.}
\FL will be open-sourced at: \url{https://github.com/dream-lab/flotilla}.


\section{Experimental Evaluation 
}\label{sec:eval}

In this section, we evaluate the modular and flexible composition capabilities of \FL (Sec.~\ref{subsec: modularity_exp}) and its benefit in comparing baseline and State-of-the-Art (SOTA) FL strategies (Sec.~\ref{sec:exp:fl-perf}), its resilience to client and server failure (Sec.~\ref{subsec: server_resilience_exp}), its scalability to 1000+ clients (Sec.~\ref{subsec: scalability and edge deployment}), and its low resource overheads, comparable to other frameworks, which allows it to be deployed on diverse edge devices (Sec.~\ref{sec:exp:compare}).


\subsection{Setup}

\subsubsection{Hardware Setup}
\label{subsec: hardware setup}

\begin{table}[t]
\setlength{\tabcolsep}{3pt}
\centering
\caption{Hardware configuration of edge device types used in the clusters.}
\label{subtab: device specs}
\resizebox{1\textwidth}{!}{%
\begin{tabular}{p{1cm}|p{1.7cm}|p{1.7cm}|p{1.7cm}|p{2.05cm}|p{2.15cm}|p{2.05cm}|p{2.15cm}|p{1.5cm}}
\hline
 & \textbf{RPi 3B+} & \textbf{RPi 4B/2} & \textbf{RPi 4B/8} & \textbf{JXNX} & \textbf{JONA} & \textbf{JAGX} & \textbf{JORA} & \textbf{Container} \\ \hline\hline
\textbf{CPU} & ARM A53 \newline 4c@1.4GHz & ARM A72 \newline 4c@1.8GHz & ARM A72 \newline 4c@1.8GHz & ARM Carmel\newline 6c@1.9GHz & ARM A78AE \newline 6c@1.4GHz & ARM Carmel \newline 8c@2.2GHz & ARM A78AE \newline 12c@2.2GHz & 1 vCPU \\ \hline
\textbf{Mem.} & 1GB RAM \newline 1GB swap & 2GB RAM \newline 2GB swap & 8GB RAM \newline 4GB swap & 8GB RAM \newline 4GB swap & 8GB RAM \newline 4GB swap & 32GB RAM \newline 16GB swap & 32GB RAM \newline 16GB swap & 4GB RAM \\ \hline
\textbf{GPU} & None & None & None & Volta 384c & Ampere 1024c & Volta 512c & Ampere 2048c & None \\ \hline
\end{tabular}%
}
\end{table}

\begin{table}[t]
\centering
\setlength{\tabcolsep}{2pt}
\caption{Composition of device types and counts per cluster.}
\label{substab: cluster details}
\resizebox{1\textwidth}{!}{%
\begin{tabular}{l||c|c|c||c|c|c|c||c||c}
\hline
\textbf{\em Cluster} & \multicolumn{3}{c||}{\textbf{\em Raspberry Pi}} & \multicolumn{4}{c||}{\textbf{\em Jetson}} & \textbf{\em Local} & \textbf{\em EC2} \\ \hline\hline
\textbf{Device}  & \textbf{RPi 3B+} & \textbf{RPi 4B/2} & \textbf{RPi 4B/8} & \textbf{JXNX} & \textbf{JONA} & \textbf{JAGX} & \textbf{JORA} & \textbf{Container} & \textbf{Container} \\ \hline
\textbf{Count}   & 16                & 18                & 12                & 4            & 4            & 2            & 2            & 208                & 1080               \\ \hline
\end{tabular}
}
\end{table}

\begin{table}[t]
\centering
\setlength{\tabcolsep}{2pt}
\caption{Training models and datasets evaluated.}
\label{tab:model_datatset_table}
\resizebox{1\textwidth}{!}{%
\begin{tabular}{l||c|c|p{1.6cm}|c|p{2.1cm}|p{1.5cm}|p{2.2cm}}
\hline
\textbf{Models}    & \textbf{\#Layers} & \textbf{\#Params} & \textbf{Dataset}      & \textbf{\#Classes} & \textbf{\#Train. Samples}   & \textbf{\#Test Samples} & \textbf{Hyper-params}            \\ \hline\hline
\textbf{LeNet5}             & 5                  & 0.06M              & EMNIST               & 10                  & Docker: 240k\newline Pi: 110.4k     & 10k                       & bs=16, lr=5e-5                   \\ \hline
\textbf{LSTM}               & 2                  & 0.07M              & OpenEIA              & NA                  & 46 buildings                & 100 buildings             & bs=16, lr=1e-5                   \\ \hline
\multirow{2}{*}{\textbf{FACNN}} & 5 & 0.12M & CIFAR10 & 10 & 50k & 10k & bs=4, lr=5e-5 \\ \cline{2-8} 
 & 5 & 0.12M & CIFAR100 & 100 & 500k\newline(50k$\times$10) & 10k & bs=4, lr=5e-4 \\ \hline
\textbf{MobileNet2}         & 2                  & 2.35M              & CIFAR10              & 10                  & 50k                         & 10k                       & bs=8, lr=1e-4, dropout=0.2       \\ \hline
\textbf{ResNet18}           & 10                 & 11.28M             & ImageNet Subset      & 200                 & 200k                        & 10k                       & bs=8, lr=5e-4, dropout=0.1       \\ \hline
\end{tabular}%
}
\end{table}

\addc{We evaluate the performance of \FL on real-world baremetal and containerized distributed systems with substantial heterogeneity.}
We conduct our experiments on four diverse clusters (Tables~\ref{subtab: device specs} and~\ref{substab: cluster details}): a \textit{Raspberry Pi cluster} with 46 ARM-based devices of 3 types; an \textit{Nvidia Jetson cluster} with 12 GPU-accelerated edge devices of 4 types; and two larger \textit{Docker clusters} formed from client containers deployed on commodity servers/VMs. 

The Pi and Jetson setups run the \FL \textit{client agent} directly on the devices.
The containerized clusters run each client within a Docker container, and have two variants: (1) \textit{Docker-208} runs 208 client containers on 13 servers of our local commodity cluster, each with an Intel Xeon Gold 6208U 16-Core CPU at 2.90GHz, 64GB RAM and 1~Gigabit Ethernet, and with each container pinned to 1~CPU core and 4~GB RAM; and (2) \textit{Docker-1080} runs 1080 client containers on 6 AWS EC2 m6a.48XL VMs in the US West (Oregon) region, with each VM having an AMD EPYC Milan CPU with 192 cores at 3.9GHz, 768GB RAM and 50Gbps Ethernet, and hosting 180 containers with 1~vCPU and 4~GB RAM.

The \FL \textit{Leader Service} for the Pi and Jetson experiments is hosted on an AMD Ryzen 9 3900X 12-Core CPU workstation with a GeForce RTX 3080 GPU and 32GB RAM, and connected over Gigabit Ethernet.
For the 208-client Local cluster, it is on a larger Docker container hosted on a separate baremetal server, while for the 1080-client EC2 cluster, it is on a separate m6a.4xlarge VM in the same region with 16 vCPUs, 64GB RAM and 12.5Gbps Ethernet.
The \textit{MQTT broker} is hosted on the same system as the Leader Service for convenience. 
The external \textit{Redis key-value store} is enabled only for for the server reliability experiments (Sec.~\ref{subsec: server_resilience_exp}), where it runs on a separate server with a Intel i9-10850K 10-core CPU and 32GB RAM connected over Gigabit Ethernet.

These client devices have varying computing architectures (ARM/Intel/AMD CPU, Nvidia GPU) and RAM (1--32GB), which capture the resource heterogeneity seen in edge deployments and showcase \FL's light-weight footprint and scaling despite its modularity and features.

\subsubsection{\FL Setup}
\label{subsubsec: Flotilla Setup}

Unless stated otherwise, we use these default configurations for \FL. All clients send heartbeats every 5 seconds to the server. Client missing 5 consecutive heartbeats are marked inactive and do not participate in subsequent training rounds.
The gRPC timeout for training is set to $1.5\times$ the round time of the slowest client, calculated using the initial benchmark round reported by the clients. 
The disk checkpointing of state is done after every 5 rounds and we use the in-memory state store by default.

\begin{table}[t]
\caption{\addc{Non-IID datasets used and their metrics of Non-IIDness of the data partitions on clients. $\delta$ is the number of labels per client.}}
\label{tab:dataset-summary-table}
\centering
\footnotesize
\begin{tabular}{l|l|c|r|r}
\hline
\textbf{Dataset} & \textbf{Non-IID Type} & \textbf{\begin{tabular}[c]{@{}l@{}}Partitioner \\ Parameters\end{tabular}} & \textbf{\begin{tabular}[c]{@{}l@{}}Coeff. of \\ Variation\end{tabular}} & \textbf{\begin{tabular}[c]{@{}l@{}}Jenson-\\ Shannon Score\end{tabular}} \\ \hline\hline
EMNIST & Label Skew & $\delta=3$ & 1.61 & 0.342 \\ \hline
OpenEIA & Seasonal Variability & N/A & 0.74 & 0.282 \\ \hline
CIFAR100 & Label Skew & $\delta=10$ & 3.01 & 0.525 \\ \hline
\multirow{2}{*}{CIFAR10} & Label Skew & $\delta=3$ & 1.64 & 0.342 \\ \cline{2-5} 
 & Label and Volume Skew & $Dir~(\alpha=0.05)$ & 2.67 & 0.443 \\ \hline
ImageNet (Subset) & Label Skew & $\delta=10$ & 1.56 & 0.343 \\ \hline
\end{tabular}%
\end{table}

\subsubsection{Models and Datasets}
\label{subsec: models_and_datasets}

We report experiments for five training model--dataset pairs (Table~\ref{tab:model_datatset_table}), which are widely used in FL research and offer diversity in their parameters (60k--11.3M), architectures~(CNN, LSTM) and data types (images, time-series). Three of them -- \textit{LeNet5}~\cite{lecun1998gradient}, \textit{LSTM} and a \textit{Custom CNN~(CCNN)}~\cite{chai2021fedat} -- are compact enough to run on the Pi and Docker clusters, while the two larger ones -- \textit{MobileNetv2}~\cite{sandler2018mobilenetv2} and \textit{ResNet18}~\cite{he2016deep} -- run only on the Jetson cluster. We train models with parameter sizes ranging from a tiny $0.06M$ LeNet5 model to a moderate-sized $11.3M$ ResNet18 model. 
Besides image classification, we also include an LSTM model for time-series demand prediction in a microgrid, using training data for 46 buildings' energy usage for 1 year from US DOE's Energy Information Authority (EIA).
In each round, the models are trained for multiple epochs, decided through hyperparamter tuning: 3 for CCNN, 1 each for LSTM and LeNet5, 2 for ResNet18, and 5 for MobileNetv2.
Separately, we have also run ALBERT transformer, as well as GraphConv~\cite{yao2024fedgcn} and GraphSAGE~\cite{hamilton2017inductive} GNN models using \FL~\cite{naman2024optimizing}, but do not report their results for brevity.

We use standard image datasets with varying complexity and sizes for training: EMNIST, CIFAR10, CIFAR100 and a subset of ImageNet~\cite{deng2009imagenet} that fit in the Jetsons (200 classes $\times$ 1000 images = 200k samples, FP-16). We create both IID and non-IID partitioning of all datasets for evaluation. 
For the \textit{IID partitioning}, data for each class label is partitioned evenly among all clients. 
For the \textit{non-IID setting}, the data for each label is partitioned uniformly into $\left\lceil\frac{c\times\delta}{l} \right\rceil$ shards, where $c$ is the number of clients, $\delta$ is the number of labels assigned per client
and $l$ is the number of labels in the dataset. We use $\delta=3$ for EMNIST and CIFAR10, and $\delta=60$ for ImageNet-Subset. For Docker-1080, we replicate the CIFAR100 dataset $\approx 10$ times and perform a non-IID sharding among the 1080 clients.
This results in them collectively hosting 540k image samples, with some images from CIFAR100 repeated 10 times and some 11 times. \addc{These partitioned datasets exhibit \textit{label distribution skew}. To further demonstrate \textit{data quantity and label skew}, we perform a Dirichlet-based partitioning~\cite{yurochkin2019bayesian} of the CIFAR10 data (Dir-NIID) with 12 clients and the Dirichlet parameter set to $\alpha=0.05$, and evaluate this for training MobileNet.}

\addc{Table~\ref{tab:dataset-summary-table} summarizes the partitioning hyperparameters and metrics of label/count heterogeneity for the different non-IID partitioning used. We report the \textit{Coefficient of Variation~(CV)} and \textit{Jenson-Shannon~(JS) divergence} for each dataset, relative to IID partitioning for categorical datasets and relative to global test dataset for OpenEIA timeseries dataset. CV quantifies the relative dispersion of label proportions within a client while JS captures the divergence between each client's label distribution and the ideal distribution.}

\FL has a built-in dataloader and trainer for torchvision datasets for classification, which we use for all CNN models. It also allows users to write a custom dataloader and trainer, which we use for the LSTM model.


\begin{table}[t!]
\centering
\captionof{table}{FL strategies implemented using \FL and developer overhead in Lines of Code~(LOC), which is used to quantify ease of use from the user's perspective.}
\label{tab:FL Strategies}
\resizebox{1\textwidth}{!}{%
\begin{tabular}{l | p{4cm} | p{4cm} | p{1cm} | p{1cm} | p{1cm}}
\hline
\textbf{Strategy} & \textbf{Client Selection (CS)} & \textbf{Aggregation Strategy (Agg)} & \textbf{Sync/ Async} & \textbf{\#LOC (CS)} & \textbf{\#LOC (Agg)} \\ \hline\hline
\textbf{FedAvg ~\cite{McMahan2016CommunicationEfficientLO}} & Selects a user-provided fraction of active clients. & 
Average of the local models of all selected clients, weighted to the number of data samples. & Sync & 32 & 67 \\ \hline
\textbf{TiFL ~\cite{chai2020tifl}} & Tiers clients based on response latencies. A tier is sampled based on its validation loss, with random clients sampled from a chosen tier. & 
FedAvg. \newline \textit{Config:} \# of tiers is set to 3 for the Pi cluster and 4 for the Jetson cluster. Agglomerative clustering is used to tier clients. & Sync & 179 & 67 \\ \hline
\textbf{HACCS ~\cite{wolfrath2022haccs}} & Clusters clients based on their data histograms and assigns weights to each cluster based on cluster average training loss and cluster max. latency. Samples clusters with replacement and picks the fastest client from each. & 
FedAvg.\newline \textit{Config:} Agglomerative clustering is performed with the \# of tiers set to 10 for the Pi cluster and 4 for the Jetson Cluster. The loss-latency tradeoff parameter is set to 0.5. & Sync & 141 & 67 \\ \hline
\textbf{FedAsync ~\cite{xie2019asynchronous}} & Selects a fraction of active clients in the first round and one client randomly at each aggregation thereafter. & 

Aggregates model updates from every client with the global model on receipt, weighted by the staleness of the updated model.\newline \textit{Config:} Mixing hyper-parameter is set to 0.9. & Async & 32 & 37 \\ \hline
\textbf{FedAT ~\cite{chai2021fedat}} & Tiers clients based on response latencies and selects a fraction of clients from each tier randomly to train at each aggregation. & 
Model updates from clients in a tier are aggregated into a tier-model using FedAvg. Tier-models are averaged, weighted by the \# of updates from each tier, to form global model. \newline \textit{Config:} \# of tiers is set to 3 for the Pi cluster and 4 for Jetson Cluster. Agglomerative clustering is used to tier clients. & Async & 101 & 77 \\ \hline

\end{tabular}
}
\end{table}

\subsection{Evaluation of Composition Modularity for FL Strategies}
\label{subsec: modularity_exp}

We demonstrate the ease of developing new FL strategies in \FL by implementing five baseline and SOTA FL strategies using the client selection and aggregation module interfaces in \FL (Table~\ref{tab:FL Strategies}).

Briefly, \textit{FedAvg}~\cite{McMahan2016CommunicationEfficientLO} is the \textit{de facto} FL baseline strategy, and does simple synchronous model averaging, while \textit{FedAsync}~\cite{xie2019asynchronous} is the default asynchronous counterpart of FedAvg. \textit{TiFL}~\cite{chai2020tifl} performs client tiering based on performance for client selection, for which we use agglomerative clustering.
\textit{HACCS}~\cite{wolfrath2022haccs} leverages knowledge of local data distribution on clients for tiering, and here again we use agglomerative clustering.
\textit{FedAT}~\cite{chai2021fedat} is a sophisticated approach that blends synchronous training within tiers and asynchronous training across tiers, which makes it particularly challenging to implement.
The client fraction per round is set to $5$ of $46$ ($\approx 11\%$) for the Pi cluster runs, $6$ of $12$ ($50\%$) for the Jetson cluster runs, floor of $10\%$ for the Docker-208 cluster runs, and 100 out of 1080 ($\approx 9\%$) for the Docker-1080 cluster runs. 

These modules take $32$--$179$ lines of Python code (LOC) for the custom CS logic and $37$--$77$ lines of code for the aggregation logic. LOC serves as a proxy to measure the ease of implementing FL strategies. As can be seen, these are concise even for complex strategies since only the core logic needs to be provided with the rest of the state management, event triggering and orchestration performed by \FL.
Once defined, individual CS or Agg modules can also be reused in other FL strategies. E.g., TiFL and HACCS have custom CS logic but share the same logic as FedAvg for Agg. Here, 
no Python code needs to be written and only the YAML config files need to be changed. This makes it easy to develop new strategies or reuse existing ones.

\addc{In addition to these, we also examine the feasibility of incorporating more recent FL approaches such as Personalized FL, where we learn personalized local models on the clients to combat statistical and system heterogeneity~\cite{tan2023towards}. Specifically, we implement FedPer~\cite{arivazhagan2019federated}, which uses parameter decoupling where some layers of the model being trained are private to each client while the other base layers are common to all. 
In \FL's \textit{Client Training phase}, we send the model to the selected clients with the base and personalized layers separately marked. Once the local models are trained on each client, they only send the weights of the base layers to the leader rather than the entire model. A custom aggregator averages only these base-layer weights and sends the updated global model back to the clients in the next round of training.}

\begin{figure}[t!]
\centering
    \subfloat[MobileNet / IID / Acc. vs Time\label{subfig: jetson accuracy mobnet iid avg5}]{
    \quad\quad\quad\includegraphics[width=0.28\linewidth]{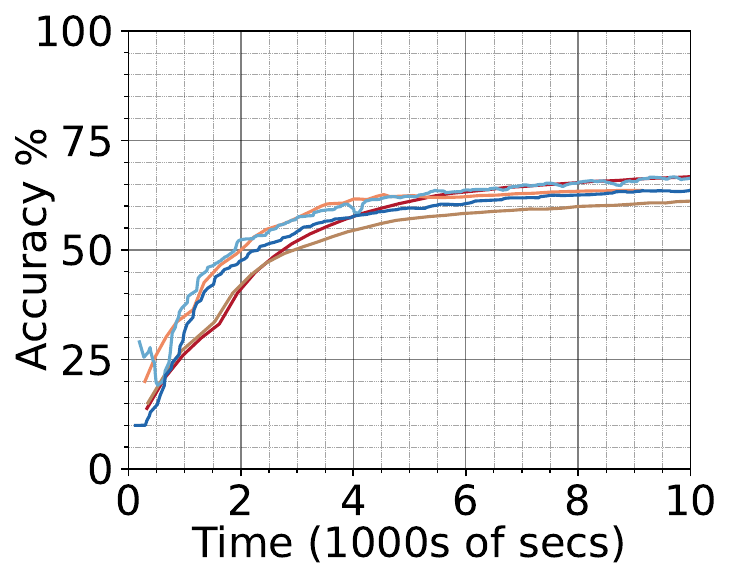}\quad\quad\quad
    }~~
    \hspace{-0.12in}
    \subfloat[MobileNet / IID / Acc. vs Round]{
    \quad\quad\quad\includegraphics[width=0.28\linewidth]{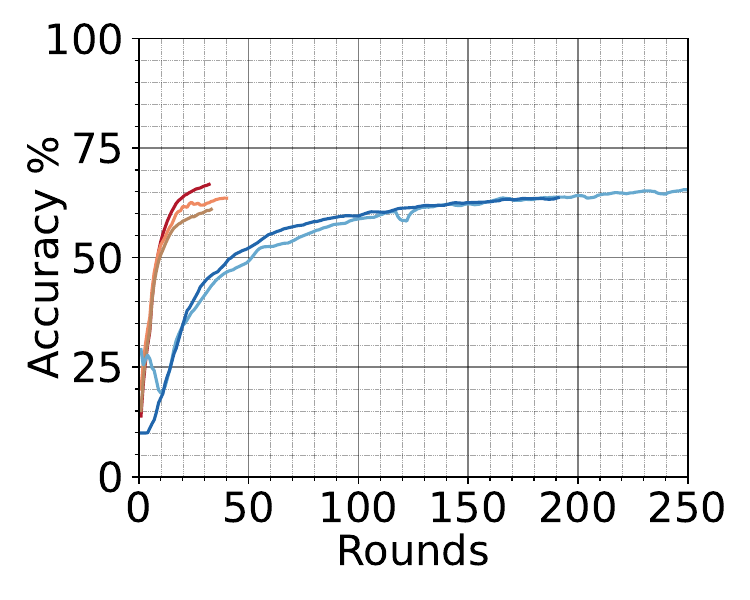}\quad\quad\quad
    }\\
    \vspace{-0.13in}
    \subfloat[MobileNet / NIID / Acc. vs Time\label{subfig: jetson accuracy mobnet noniid avg5}]{
    \quad\quad\quad\includegraphics[width=0.28\linewidth]{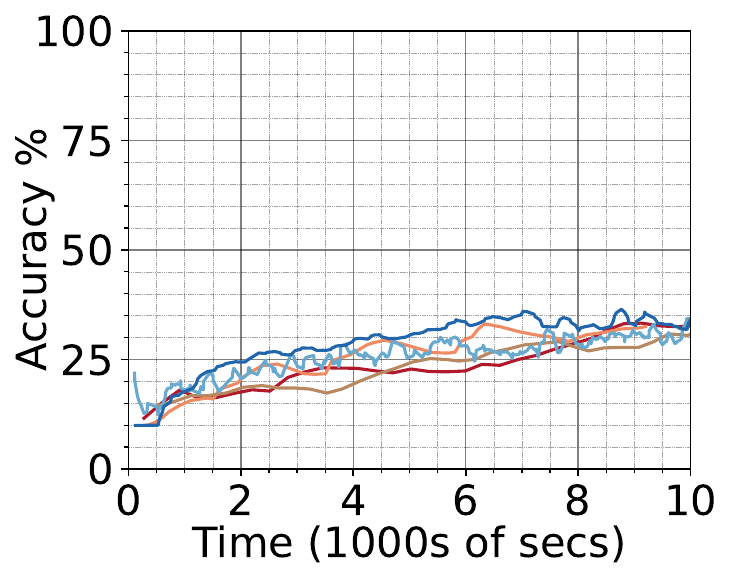}\quad\quad\quad
    }~~
    \hspace{-0.12in}
    \subfloat[MobileNet / NIID / Acc. vs Round]{
    \quad\quad\quad\includegraphics[width=0.28\linewidth]{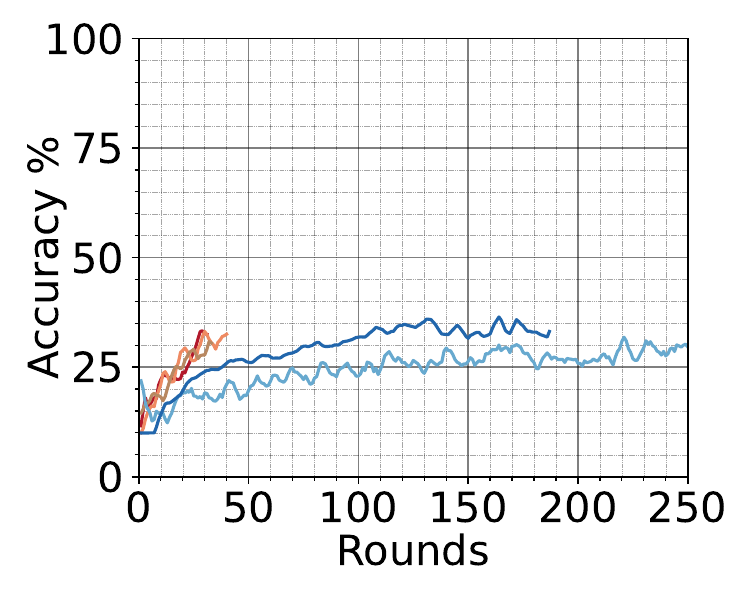}\quad\quad\quad
    }\\
    \vspace{-0.13in}
    \subfloat[\addc{MobileNet/Dir-NIID/Acc. vs Time}]{
    \quad\quad\quad\includegraphics[width=0.28\linewidth]{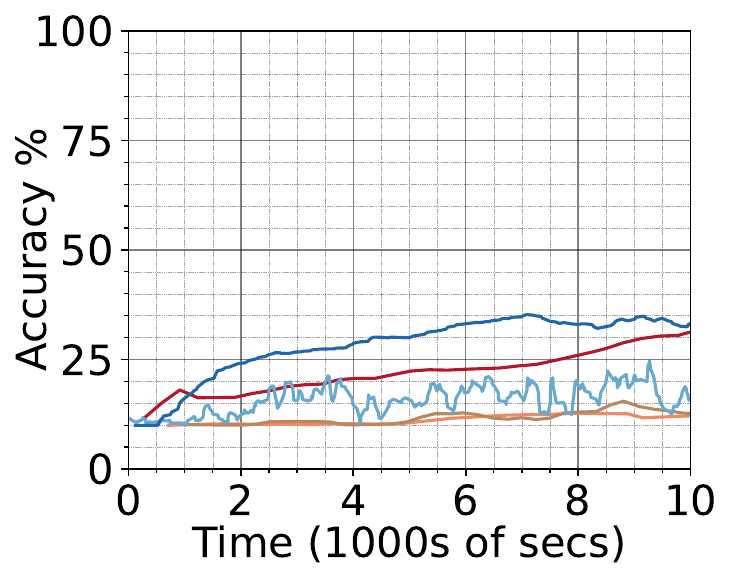}\quad\quad\quad
    }
    \subfloat[\addc{MobileNet/Dir-NIID/Acc. vs Round}]{
    \quad\quad\quad\includegraphics[width=0.28\linewidth]{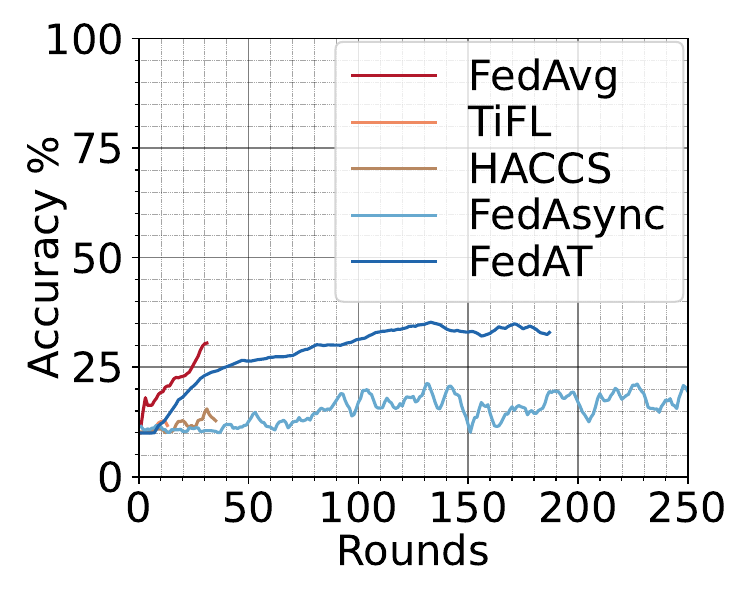}\quad\quad\quad
    }\\
    \vspace{-0.13in}
    
    \subfloat[CCNN / IID / Acc. vs Time\label{subfig: pi accuracy ccnn iid avg5}]{
    \quad\quad\quad\includegraphics[width=0.28\linewidth]{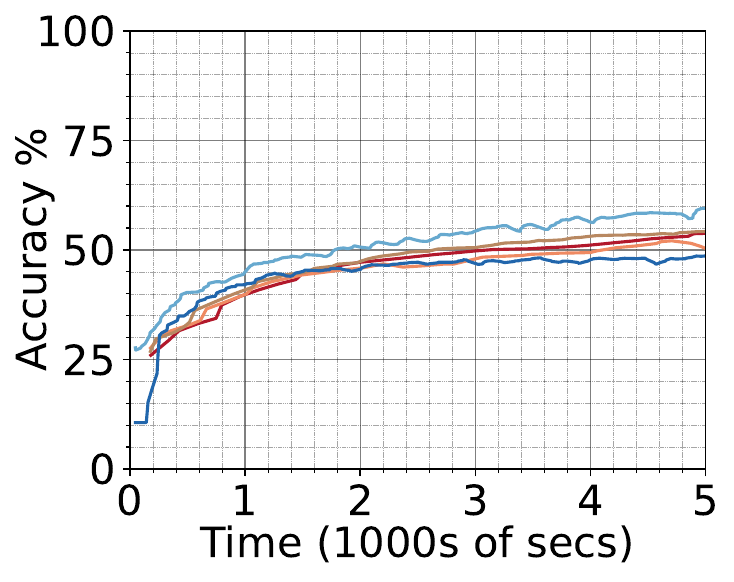}\quad\quad\quad
    }
    \subfloat[CCNN / IID / Acc. vs Round]{
    \quad\quad\quad\includegraphics[width=0.28\linewidth]{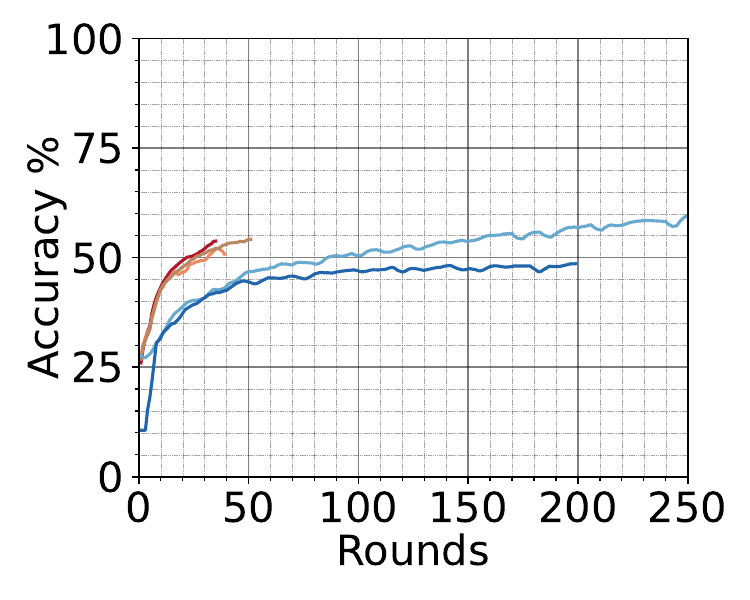}\quad\quad\quad
    }\\
    \vspace{-0.13in}
    \subfloat[CCNN / NIID / Acc. vs Time]{
    \quad\quad\quad\includegraphics[width=0.28\linewidth]{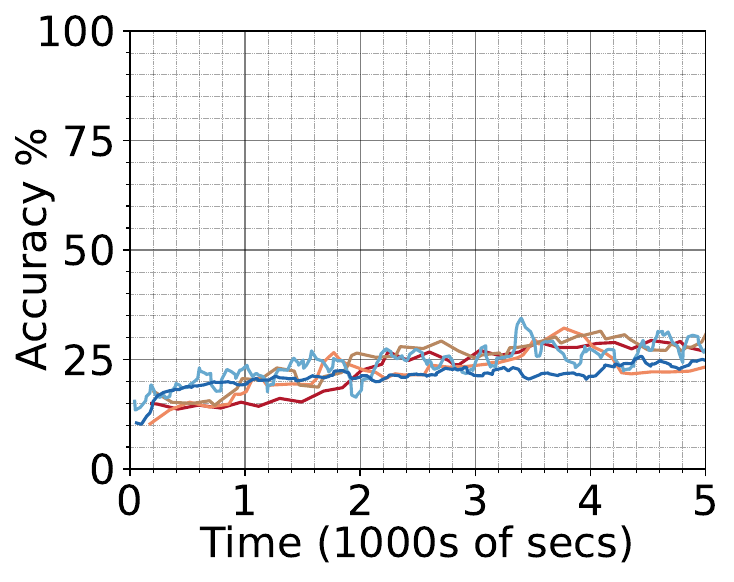}\quad\quad\quad
    }
    \subfloat[CCNN / NIID / Acc. vs Round]{
    \quad\quad\quad\includegraphics[width=0.28\linewidth]{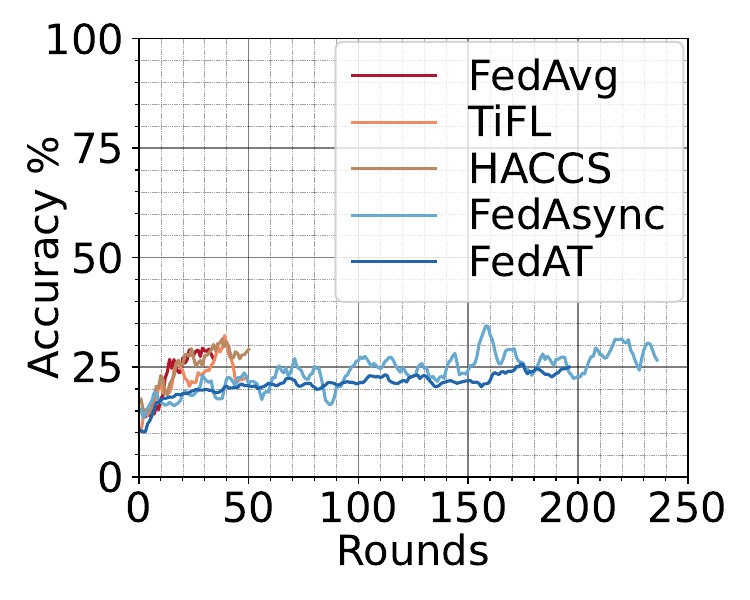}\quad\quad\quad
    }
    \caption{Changes in accuracy with \textit{wallclock time} and \textit{FL rounds} for \modc{\textit{MobileNet on Jetson cluster} and \textit{CCNN on Pi cluster} for \textit{IID (rows 1 and 4)} and \textit{non-IID (rows 2, 3 and 5)}} data distribution using five FL strategies implemented in \FL. The accuracies are averaged over 5 rounds for smoothing the noise.}
    
    \vspace{-0.1in}
    \label{fig: jetson-accuracy-plots}
\end{figure}

\begin{figure}[t!]
    \centering
    \includegraphics[width=0.45\linewidth]{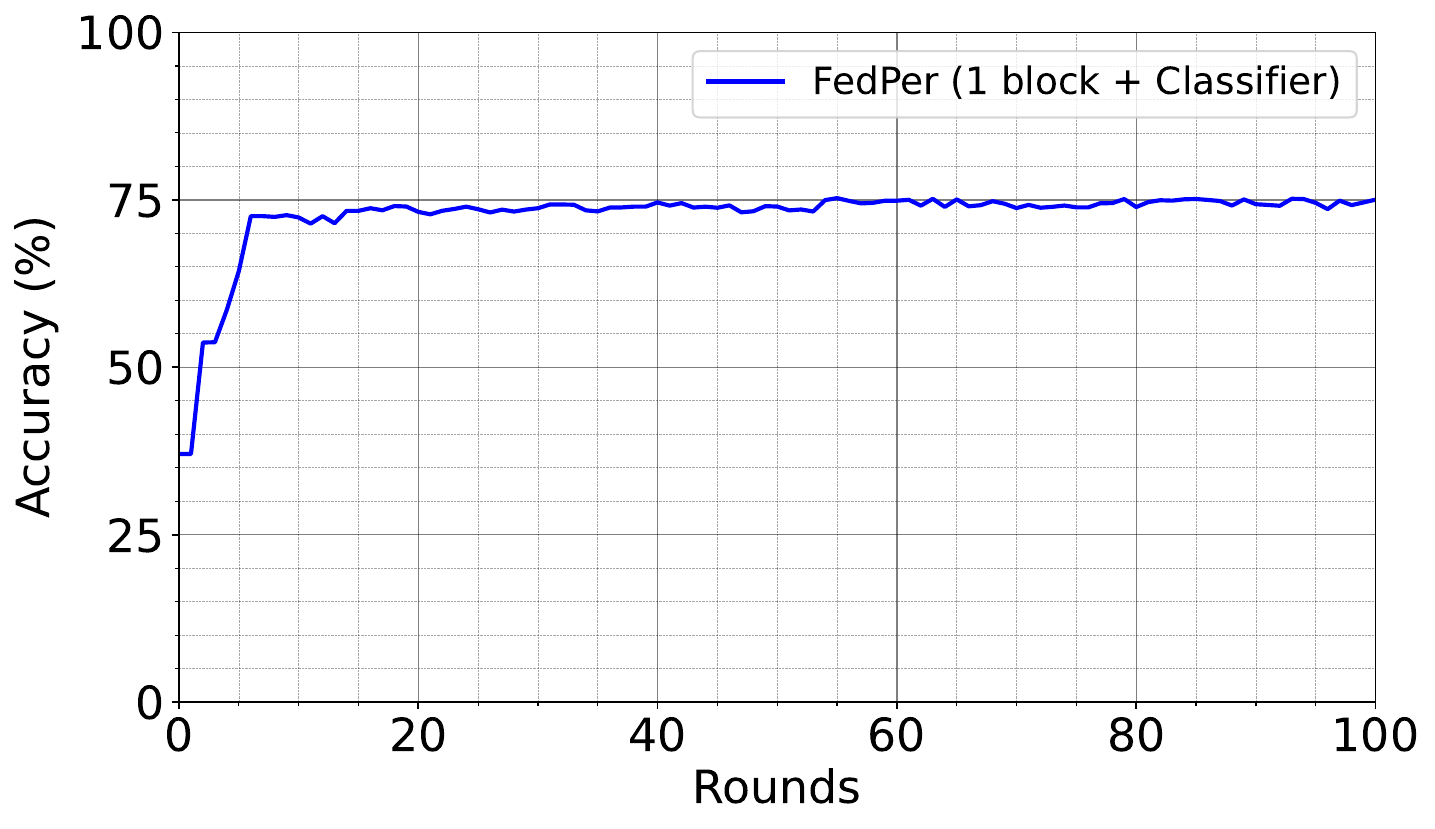}
    \caption{\addc{Changes in accuracy with \textit{FL rounds} for MobileNet on Dirichlet-based non-IID data using \textit{FedPer strategy}, using \FL in simulation mode.}}
    \label{fig:fedper_accuracy}
\end{figure}

\addc{To evaluate this, we perform a \textit{simulation-based FL training} of MobileNetV2 model on CIFAR10 partitioned with a Dirichlet distribution across 12 clients with $50\%$ client selection using the FedPer strategy for 100 rounds. \FL runs on a GPU workstation with a $12$-core AMD Ryzen $9$ $7900$X CPU~($3.7$GHz) with $128$GiB RAM, having an NVIDIA RTX $4090$ GPU card with $24$GiB GPU memory.
In this architecture, the last layer is a personalized layer while the rest are shared base layers.
The results are shown in Fig.~\ref{fig:fedper_accuracy}. As reported in the FedPer~\cite{arivazhagan2019federated} we observe a similar accuracy curve where accuracy starts from $\approx 35\%$ and shows rapid improvement within eight rounds, after which it continues to stabilize around  $\approx 75\%$. This also highlights the ability of \FL to perform a simulation-based study, besides real hardware and emulation-based runs. We do not discuss FedPer or the simulation based approach further.}


\subsection{Evaluating the Performance of FL Strategies}
\label{sec:exp:fl-perf}
Being able to rapidly implement new FL strategies in \FL allows for an apples-to-apples comparison of the training quality and runtime of these strategies on real devices. We put this to test by comparing the five FL strategies listed above. Interestingly, as we report below, the FL strategies do not always exhibit convergence performance that match the claims reported in their papers when evaluated on real hardware, despite careful hyper-parameter tuning to match the configurations in the papers.

\begin{figure}[h]
\centering
\subfloat[Accuracy achieved within Fixed Time]{
\includegraphics[width=0.85\textwidth]{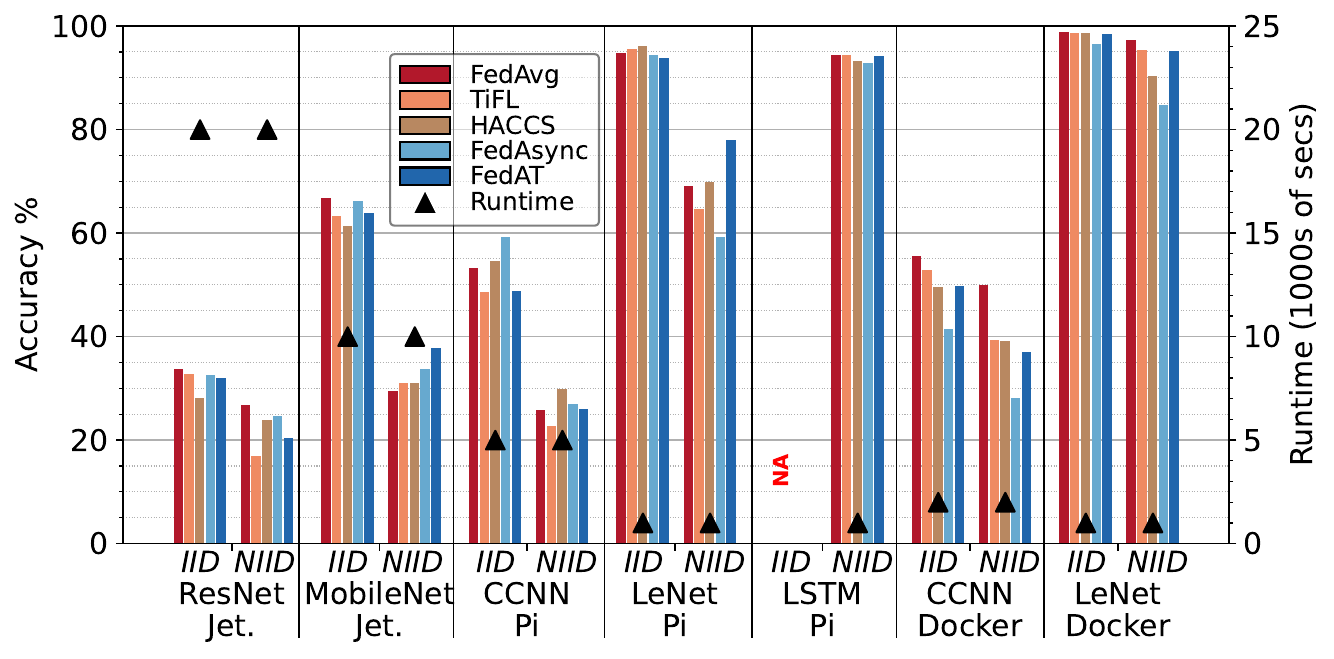}
\label{subfig:AAT}
} \\
\subfloat[Time to Fixed Accuracy (TTA)]{
\includegraphics[width=0.85\textwidth]{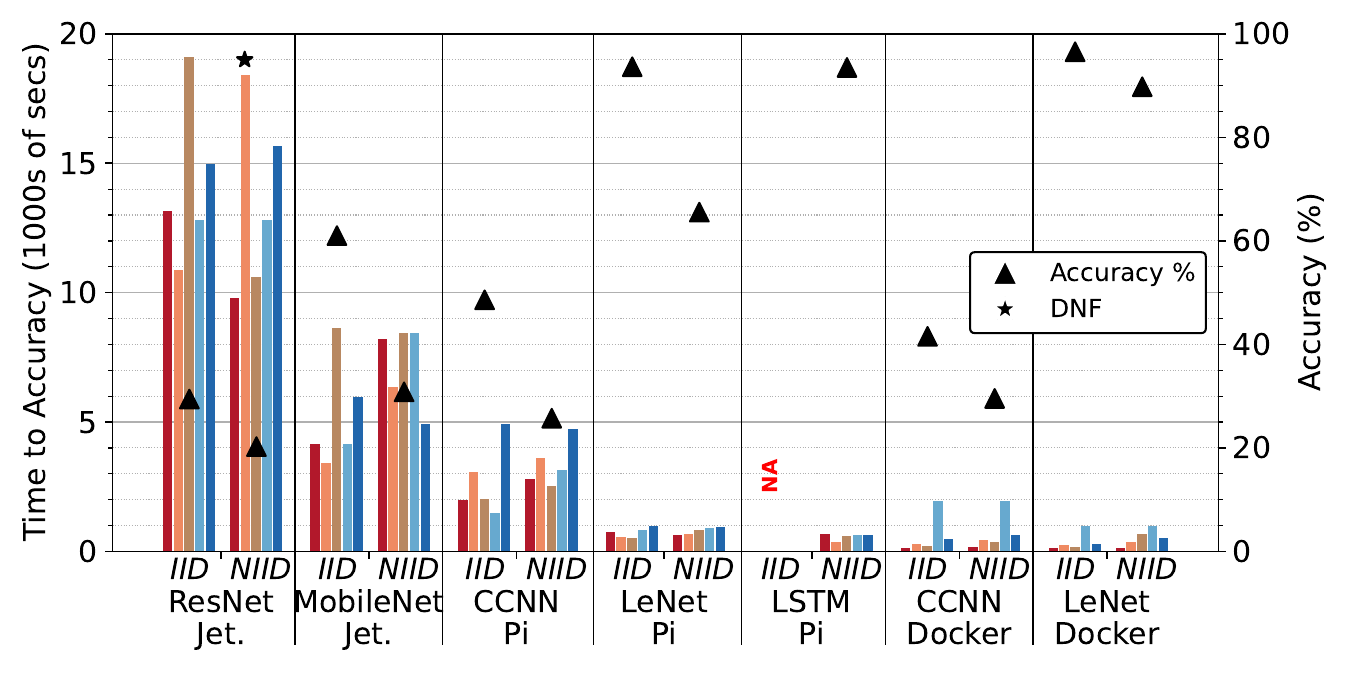}
\label{subfig:TTA}
}
\vspace{-0.1in}
\caption{
Time and accuracy performance for all FL strategies, models and clusters trained using \FL. Right Y axes $\blacktriangle$ show the fixed time or accuracy, as applicable.}

\label{fig:TTAandAATplots}
\end{figure}

In Figs.~\ref{fig: jetson-accuracy-plots}, we report the change in accuracy of the global model for these FL strategies, with the number of rounds and with the wallclock time. We train the MobileNet model on the Jetson cluster and the CCNN model on the Pi cluster for 10k and 5k seconds, respectively, using IID and non-IID data. For the asynchronous strategies, each global model update forms a round. All reported accuracies are smoothed over 5 rounds.
We also report the \textit{final accuracy achieved}
for all 7 model--cluster combinations for these FL strategies in Fig.~\ref{subfig:AAT}.

Our results show that the true benefits of complex FL strategies are modest.
\textit{For IID data} (rows 1 and 3 of Fig.~\ref{fig: jetson-accuracy-plots}), even baseline strategies like FedAvg and FedAsync offer similar or better results than sophisticated SOTA ones like TiFL, HAACS and FedAT. While FedAvg has a slower initial growth for MobileNet on the Jetson cluster, both FedAvg and its asynchronous variant, FedAsync, outperform all other models beyond 6k~seconds to reach a better final accuracy at 10k~seconds (Fig.~\ref{subfig: jetson accuracy mobnet iid avg5}). For the CCNN model trained on the Pi cluster, FedAsync is better than FedAvg, HACCS matches FedAvg, while TiFL and FedAT are worse. This is contrary to results reported in a simulated setting for FedAT~\cite{chai2021fedat}. TiFL claims to converge faster than FedAvg when using heterogeneous devices due to its performance-based clustering~\cite{chai2020tifl} but we do not observe this.
 Many frameworks do not report results on standard models like MobileNet and ResNet, instead using custom~\cite{chai2020tifl,chai2021fedat} or trivial ones like LeNet~\cite{wolfrath2022haccs}.

With \textit{non-IID data distributions} (rows \modc{2, 3 and 5}  of Fig.~\ref{fig: jetson-accuracy-plots}), FL optimizations meant to handle data and performance diversity fail to offer faster convergence due to unrealistic assumptions. E.g., FedAT is reported to outperform FedAvg and TiFL~\cite{chai2021fedat} through better client clustering. But in a real-world setting, the actual training time for a local model on a device type is not deterministic (unlike in simulation). So the clustering done by FedAT assuming static client latencies per device type does not hold, e.g., with the per-round training time on JXNX for MobileNet varying from 
262s (Q1)
to 311s (Q3). \addc{FedAT however does do better on Dir-NIID, reaching a higher accuracy earlier than others and ultimately achieving a marginally better accuracy than FedAvg.} 

HACCS is designed for non-IID but does not benefit from a majority-labelled distribution assumed by them, compared to \delc{our non-IID our}\addc{either of our }non-IID \modc{setups}\delc{that is more uniformly randomized [18]}. 
\modc{TiFL performs better than FedAvg in the case of NIID}\addc{, but poorly for Dir-NIID, since clients with very little data are clustered with clients faster clients, and hence compromise the cluster probabilities}. 
\addc{FedAsync also suffers for Dir-NIID as compared to NIID since the updates from clients with little data derail the convergence of the global model.}

More generally, when we compare the \textit{accuracy achieved within a fixed time} for 14 configurations of models, clusters and data distributions (Fig.~\ref{subfig:AAT}), FedAvg and/or FedAsync baselines are the best or comparable to the best overall strategy in all but a few exceptions -- FedAT is better for MobileNet non-IID and LeNet non-IID on Pi; and HACCS is better for CCNN non-IID on Pi. When we see the complementary \textit{time-to-accuracy (TTA)} plots in Fig.~\ref{subfig:TTA}
TiFL is occasionally faster than FedAvg, but only because it shows a better accuracy improvement early on but saturates more quickly. 
\textit{Such critical analysis is possible only because of the ability to rapidly implement and run the strategies on real hardware under realistic conditions using the same \FL framework.}
We also see the impact of long training runs on real hardware. TiFL picks clients from a tier that has the highest client validation loss in the prior round. 

This causes a single tier to be picked repeatedly, which leads to the \textit{devices overheating and failing} if the tier has few devices.
Such a device skew and overheating is also seen in FedAT, where clients with a similar performance tier are often picked.
As a result, the ResNet18 training on non-IID data using FedAT on the Jetsons could not run beyond 16k seconds despite numerous tries (DNF, `$\star$').
\textit{These again indicate the limitations of performing just a simulation based study and the need for real hardware experiments.} It also open up the possibility of designing reliability or thermal aware client selection algorithms.


\subsection{Training Resilience of \FL}
\label{subsec: server_resilience_exp}

Next, we demonstrate \FL's reliability during FL training by evaluating both server and client failures.

\subsubsection{Resilience to Server Failures}
\label{subsubsec: server failure}
Robustness to server failures is a novel feature of \FL and under-explored elsewhere.
Here, we train the CCNN/IID/FedAvg model on the Pi Cluster using \FL, enabling the external Redis key-value store to maintain the live session states.
A 3080 GPU \textit{primary} workstation has the Leader service (\texttt{GPU1}) with another identical \textit{secondary} machine on standby (\texttt{GPU2}) to start a new Leader service upon failure~\footnote{In production scenarios, we anticipate a single VM for the \FL Leader service, a coordination service like Zookeeper to detect failures and initiate failover to an alternate VM instantiated on-demand,

and Redis (or an equivalent like Amazon ElastiCache) being a reliable cloud-hosted service.}.

In the \textit{first failure mode setup (single machine)}, we simulate server failures by killing the \FL Leader Service on the primary GPU1 after every 5 rounds of training and restarting the Leader service on the same workstation, with the flag set to resume the prior session from the Redis state store. This is repeated several times.
In \textit{the second setup (failover)}, we kill the Leader Service on the primary \texttt{GPU1} after 5 rounds but restore the session on the secondary \texttt{GPU2} to resume training; and again kill the service on \texttt{GPU2} after 5 rounds and restore it on \texttt{GPU1}, and so on.

\begin{figure}[t]
\vspace{-0.1in}
    \centering
    \subfloat[Server Failure every 5 rounds, for CCNN/CIFAR10-IID/Fed\-Avg on Pi cluster.]{%
    
    \includegraphics[width=0.32\columnwidth]{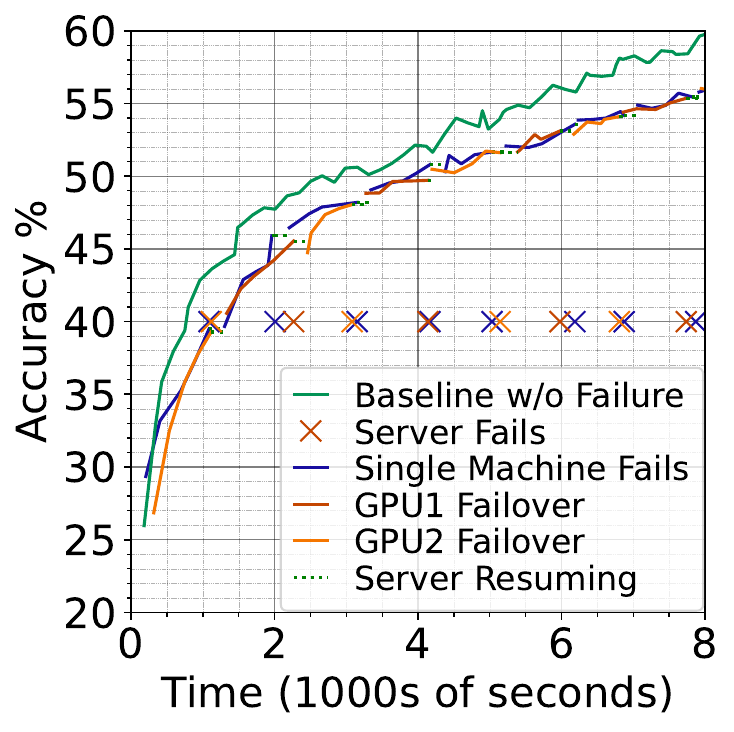}%
    \label{subfig: server_reliability_failure}%
    }~
        \subfloat[Checkpointing Time on disk (bar, left Y) and Size (marker, right Y) for models on Jetson/Pi cluster.]{%
    \includegraphics[width=0.35\columnwidth]{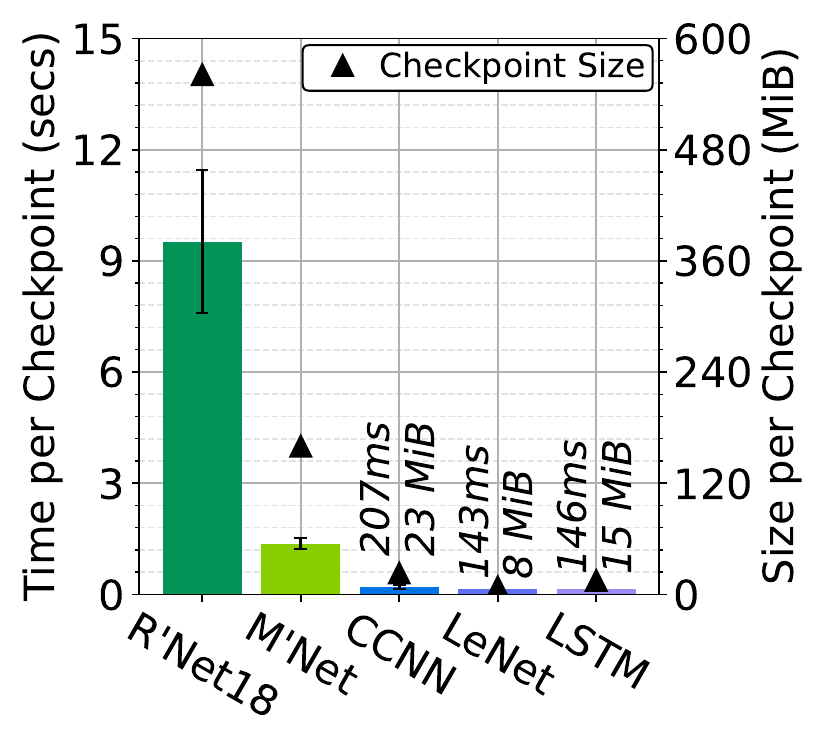}%
    
    \label{fig:reliability:checkpoint}%
    }~
    \subfloat[Size of Redis External State over time for CCNN/CIFAR10-IID/Fed\-Avg on Pi cluster.]{%
    \includegraphics[width=0.32\columnwidth]{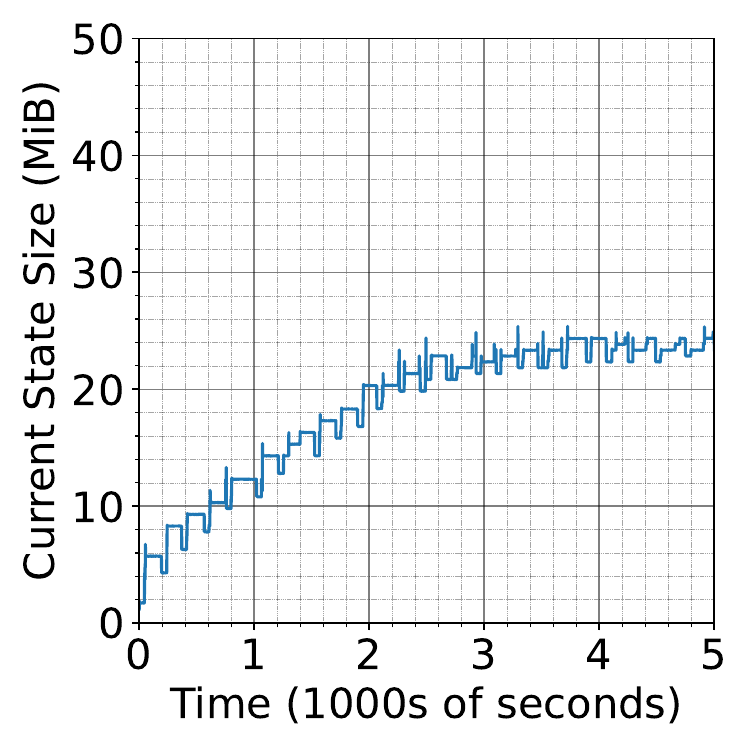}%
    
    \label{fig:reliability:session}
    }    
    \caption{\FL resilience with server failure and its overheads.}
    
    \label{fig: server and client failure plots}
    \vspace{-0.1in}
\end{figure}

Fig.~\ref{subfig: server_reliability_failure} shows these two failure setups and a baseline without failures. `$\times$' marks the time points at which the servers are killed. \textit{Single Machine Fail} is for setup 1, while \textit{GPU1} and\textit{ GPU2 Failover} are for setup 2; solid line indicate the server is training and a dotted line shows the restoration and resumption period, between the last completed round and the next successful round. We do not smoothen the accuracies here to clearly show the impact.

Both of the failure mode experiments achieve an accuracy trend comparable to the baseline run without server failure. 
Of the time spent in recovery after a failure,
only $\approx750$ms is spent in starting the \FL Leader Service and a further $\approx 75$ms in restoring the state from Redis, on average. Additional time is spent in completing the partial round after the last failure to aggregate and validate a new global model ($\approx 171$s).
These \textit{sub-second resumption overheads} are negligible compared to the mean round train time of $\approx 178$s, and more so given the adversarial failure rates we use here.
This trend is consistent for both the single and dual-server setups. 
In contrast, the time to restore a periodic round checkpoint from disk is marginally faster at $\approx 51$ms.

\subsubsection{Server Resilience Overheads}

Next, we report the overheads for achieving server resilience using both these approaches -- periodically checkpointing rounds to disk
and incrementally maintaining session state on the Redis store.

Fig.~\ref{fig:reliability:checkpoint} 
(bars on the left Y axis) reports the time to perform the \textit{disk checkpoint} after 5 rounds for given different model--dataset--cluster configurations, while the markers on the right Y axis report the size on disk of each checkpoint.
As expected, the size of the state checkpoint and the time taken increases with the size of the model parameters, e.g., with ResNet taking 560MiB and 9.26s, while LeNet takes 8MiB and 143ms per checkpoint to local disk. The model weights account for the bulk of the saved session state, with the global model using 43MiB and the local models stashed for 12 clients consuming 517MiB, while the non-model state maintained per round, client, etc., adds only 0.1MiB when training ResNet.
These overheads of a few seconds per checkpoint are modest compared to the per-round training time that runs into 100s of seconds and form 0.007--0.32\% of the overall FL time in our runs.

Fig.~\ref{fig:reliability:session} shows the incremental growth in the \textit{external session state size} on Redis over time when training CCNN. As expected, the session state grows as more clients and their states are included over multiple rounds. 
We saturate at 2500s when all have clients participated and are part of the cumulative state stored in Redis of 24.8MiB. The eventual size of the disk checkpoint and the external state store are similar, as expected, except that checkpointing is done in bulk after a round boundary while the external state is updated incrementally on each state operation.
The time overhead to externalize states to Redis is negligible.
E.g., the time to run 25 FL rounds for CCNN/CIFAR10-IID/FedAvg on the Pi cluster, with in-memory vs. externalized states are $2813.9$s vs. $2815.5$s -- within $2$s or $0.07\%$ of each other.

\subsubsection{Client Failures}
We now evaluate the resilience of \FL to client failures. Here, we use the Docker-208 cluster to train the CCNN/IID/FedAvg model. We configure the clients to have a failure rate that matches a Poisson distribution, as is common in literature~\cite{lemoine1985failure}, with the Mean Time To Failure~(MTTF, $\mu$) set to $\mu=10$~days -- chosen to mimic an \textit{adversarial scenario} where about $40\%$ of all clients fail within the 2000~second run. Failures are triggered every $5$ seconds, with each client killed at time $t$ with a probability: $1-e^{(-\frac{t}{\mu})}$, where $\mu$ is in seconds.
Failed clients do not recover.

\begin{figure}[t]
    \centering
    \includegraphics[width=0.45\columnwidth]{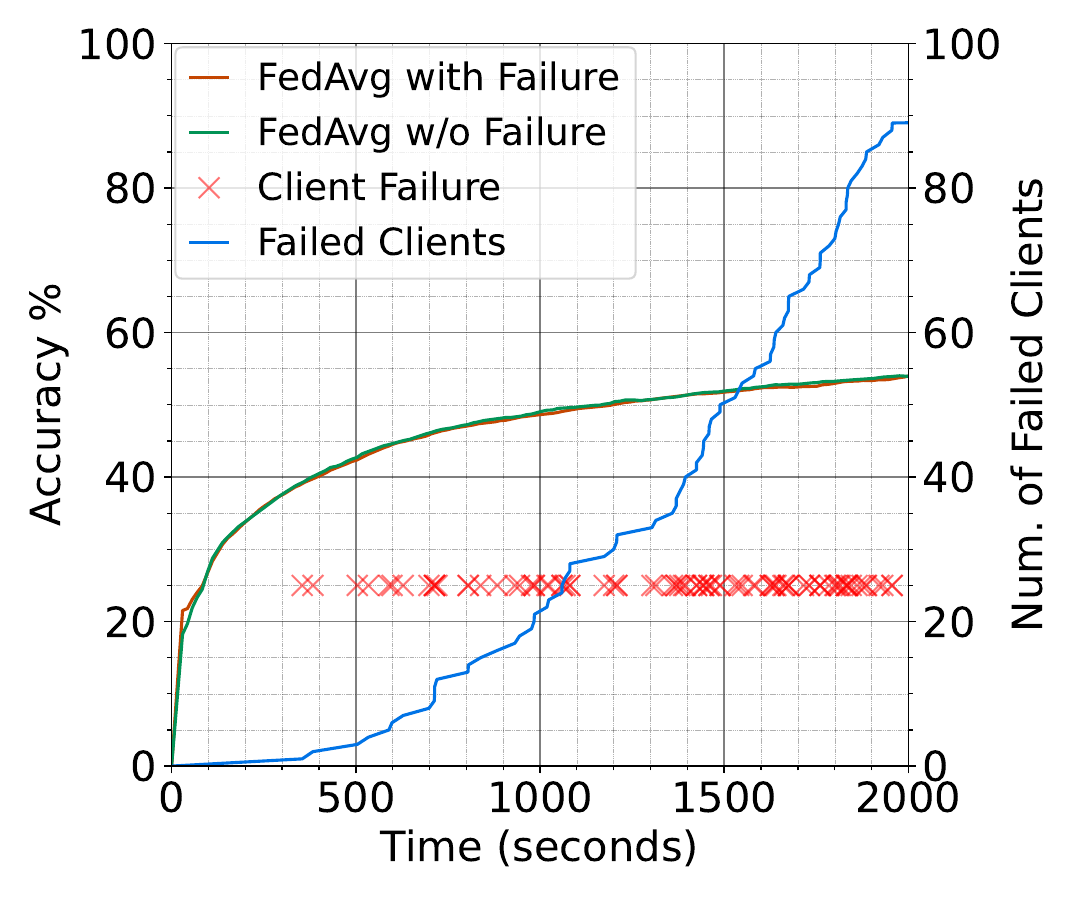}
   
    \caption{Training accuracy achieved with clients failing with Poisson distribution, for CCNN/CIFAR10-IID/FedAvg on Docker-208 cluster.}
    \label{subfig: client failure}%
    
\end{figure}

Fig.~\ref{subfig: client failure} shows the accuracies with and without client failures on the left Y axis over time, with the right Y axis reporting the cumulative number of client failures and `$\times$' markers indicating the time at which a client fails. \FL tolerates client failures without any significant drop in training accuracy. E.g., even with 89 of the 208 clients being killed over the 2000~second run 
the final accuracy with and without client failures is near-identical, $57.4\%$ vs. $57.1\%$, and within the margins of variability across rounds.
This is as expected. Since the data distribution among the clients is IID, even a decreasing pool of clients to select from in subsequent rounds does not have a significant effect on the accuracy of the global model. \FL's heartbeat mechanism detects a client failure within $\approx 30$s, when it misses 5 consecutive heartbeats sent every 5s. Further, the Session Manager skips waiting for a client response beyond the configured timeout of 18s for the round. These allow for quick detection and mitigation of client failures on a best-effort basis. We see a mean per-round training of $\approx10.4$s with failures, comparable to a session without failures of $\approx10.2$s.


\subsection{Scalability}
\label{subsec: scalability and edge deployment}

We next evaluate the scalability of \FL using \addc{a} containerized \addc{setup that uses} Docker \addc{to emulate edge} clients.
 
Such a container-based setup is helpful when a large number of edge hardware devices are not available at testing time and yet we wish to test the effectiveness of the FL strategy at scale.
Here, we also contrast the scalability of \FL with \textit{Flower}, among the most popular FL frameworks~\cite{lai2022fedscale}. 

We first measure their \textit{weak-scaling performance} by training the CCNN/ CIFAR10-IID/FedAvg model for 100 rounds on a subset of the Docker-208 cluster with 56, 112, 160 and 208 clients enabled, while pinning each client container on a vCPU core, and select 10\% of clients in each round for training.

We then expand the study to the Docker-1080 cluster with 1080 clients, training the CCNN/CIFAR100-NIID/FedAvg model for 300 rounds using 100 clients for training per round, to assess the \textit{scaling bottlenecks}.

\begin{figure}[t!]
    \centering
    \subfloat[Total FL time with weak-scaling, using clusters with different \# of client in Docker-208.]{
    \includegraphics[height=5cm]{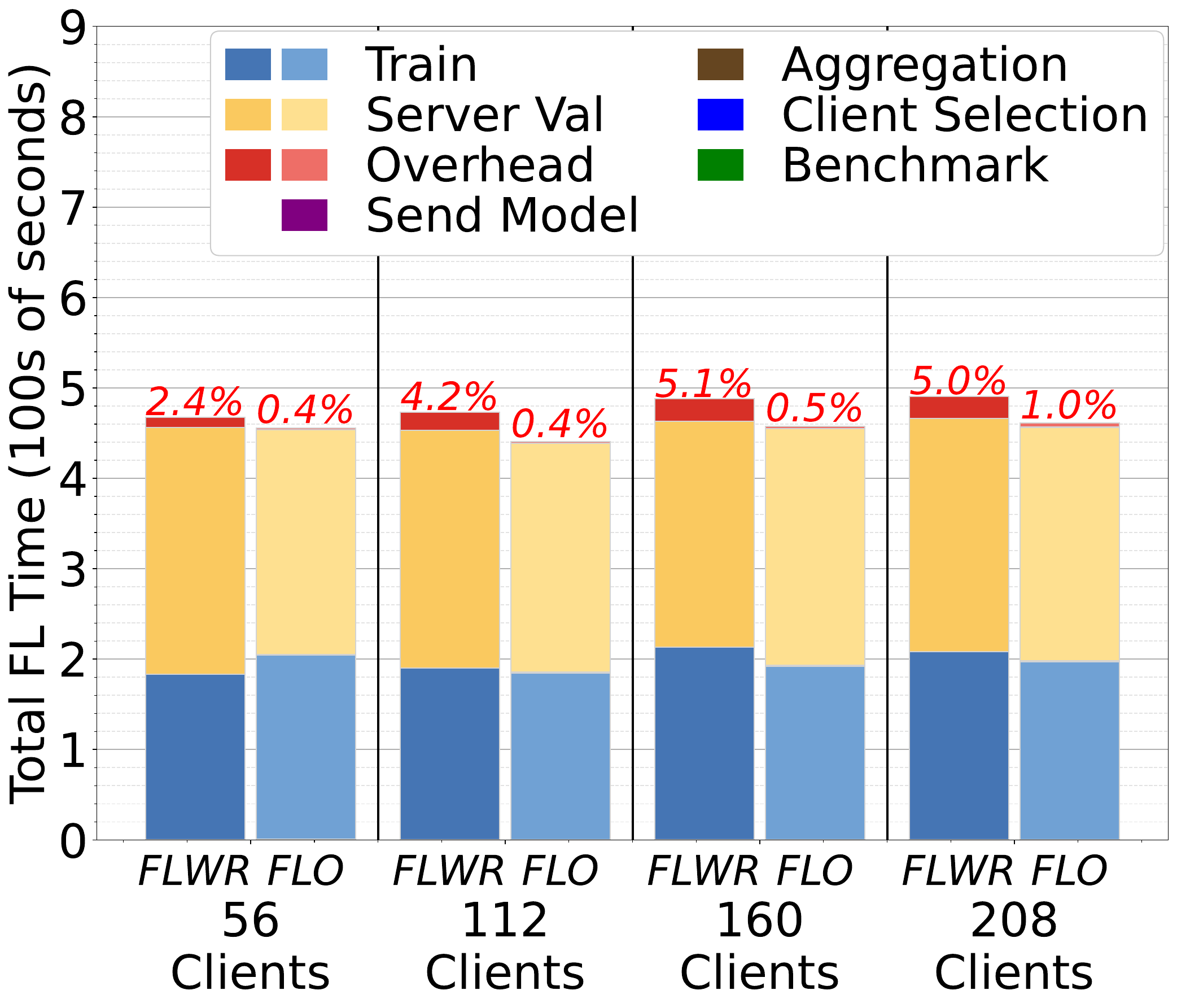}~
    \label{fig: scalability-sirius}
    }~
    \subfloat[Total FL time and client invocation time on all clients of Docker-1080.]{
    \includegraphics[height=5cm]{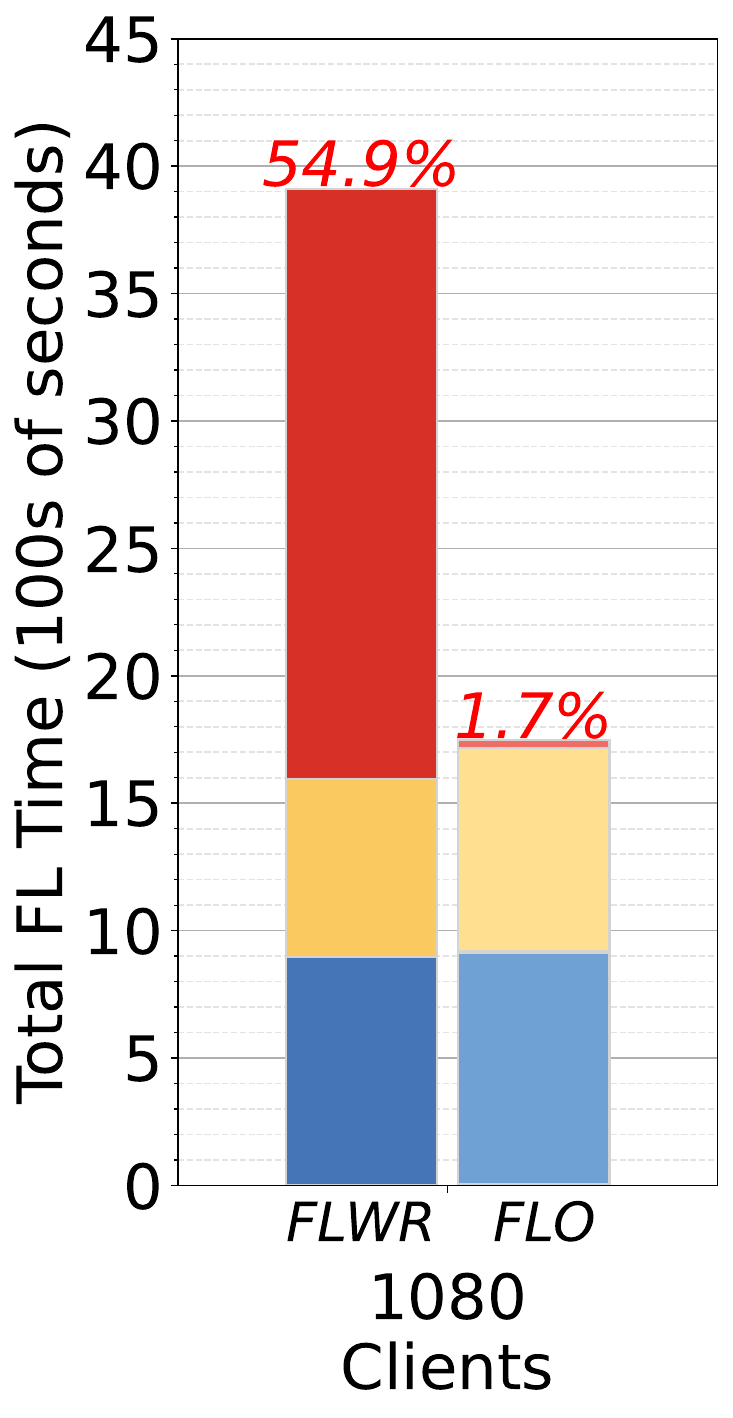}
    \label{fig: scalability-aws}
    \includegraphics[height=5cm]{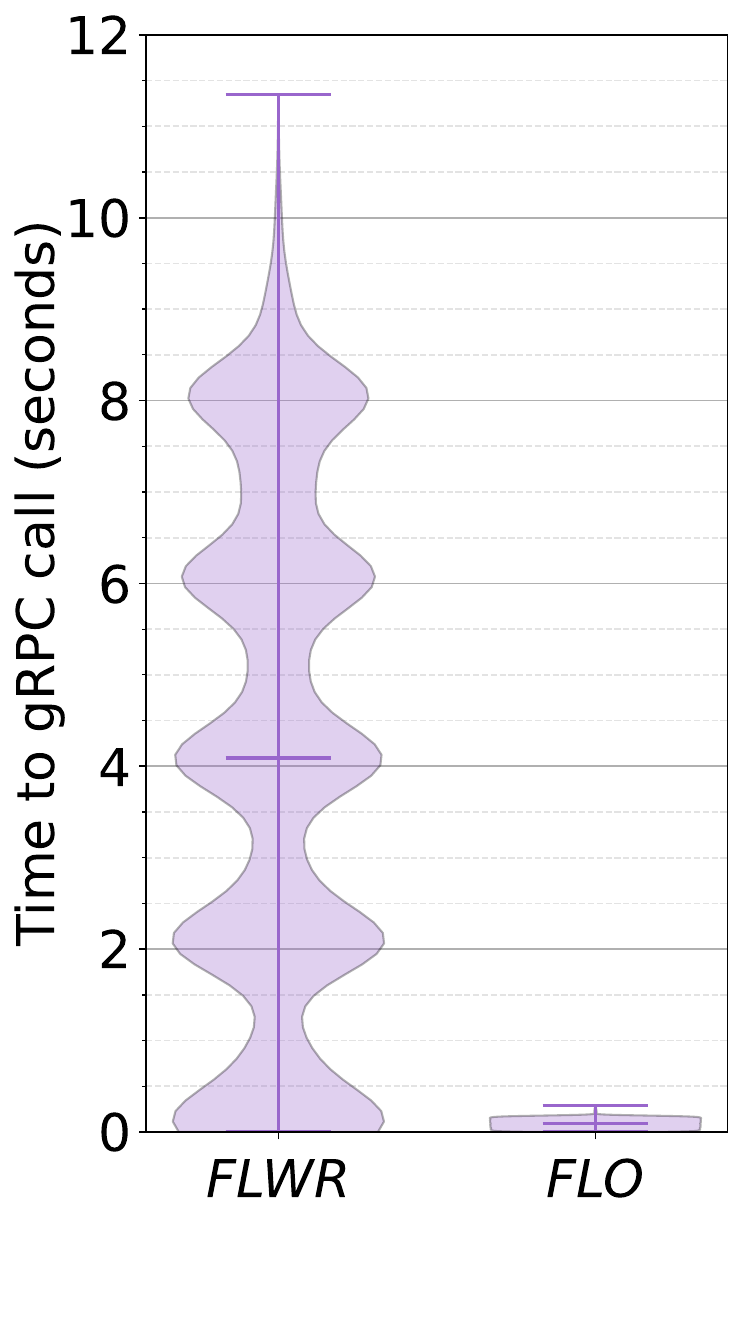}
    \label{fig: launch-overhead}
    }
    
    \caption{
    Scaling experiments comparing Flower (FLWR) and \FL (FLO) for training CCNN/CIFAR10/FedAvg on the Docker clusters. \% overheads are shown above the bars.}

    \label{fig: Flotilla-vs-FlWR-scalability}
\end{figure}

We report the end-to-end FL training time and the component times as stacked bars in Fig.~\ref{fig: Flotilla-vs-FlWR-scalability}. In the \textit{weak scaling results} in Fig.\ref{fig: scalability-sirius}, the training times are similar as the number of clients increase and comparable for both Flower and \FL. This indicates that both frameworks exhibit weak scaling at these scales. Here, the dominant times are spent on the client training time (blue stack) and the server validation time (yellow) that are the compute-intensive parts. 

Since these clients have homogeneous resources, they take similar times to complete their local training. The time taken for initial model deployment, client benchmarking, client selection and aggregation are negligible.

However, the framework overheads, which excludes these other productive steps, gently increase with the number of clients, reaching 5\% of total FL time for Flower and 1\% for \FL with 208 clients. This overhead sharply grows to 54\% for Flower when we scale this to the \textit{Docker-1080 setup} (Fig.~\ref{fig: scalability-aws}, left), while \FL retains a modest 1.7\% overhead, with the other components of FL time being comparable. This limits the scaling of Flower to 1000+ clients.

The sources of overheads vary between the two frameworks.
Due to the event-driven asynchronous training loop of \FL, each client update triggers a set of client selection and aggregation executions, even if they are often no-ops in FedAvg. This, coupled with Python's limitation on concurrency due to the Global Interpreter Lock (GIL), causes some overheads as more client are active per round causing more callbacks to happen. Still these costs are low, with just a 990ms (1.7\%) overhead per round for the Docker-1080 run with 100 client callbacks in each round.

In contrast, Flower has a $5\%$ overhead even for 208 client, training over 20 clients per round, and
sharply grows to 
$54.9\%$ when training 100 clients per round in Docker-1080.
This is due to Flower's use of ThreadPoolExecutor to concurrently issue client training requests. It initiates client training in waves for each round, waiting for the earlier asynchronous requests to be acknowledged before initiating the next set of requests. 
The gRPC request overheads per client request per round is shown in Fig.~\ref{fig: scalability-aws} (right). As can be seen, \FL shows a tight distribution of $190$ms for all requests, i.e., from the client training round starting till the request being acknowledged by the client, while Flower has diverse request durations grouped by the different batched requests.
This introduces an average delay of $8.96$s between the first and the last training request in a round for Flower, of which $7.72$s is visible as part of training overhead and is cumulatively $2316$s over 300 rounds.


\begin{figure}[h!]
    \includegraphics[width=\columnwidth]{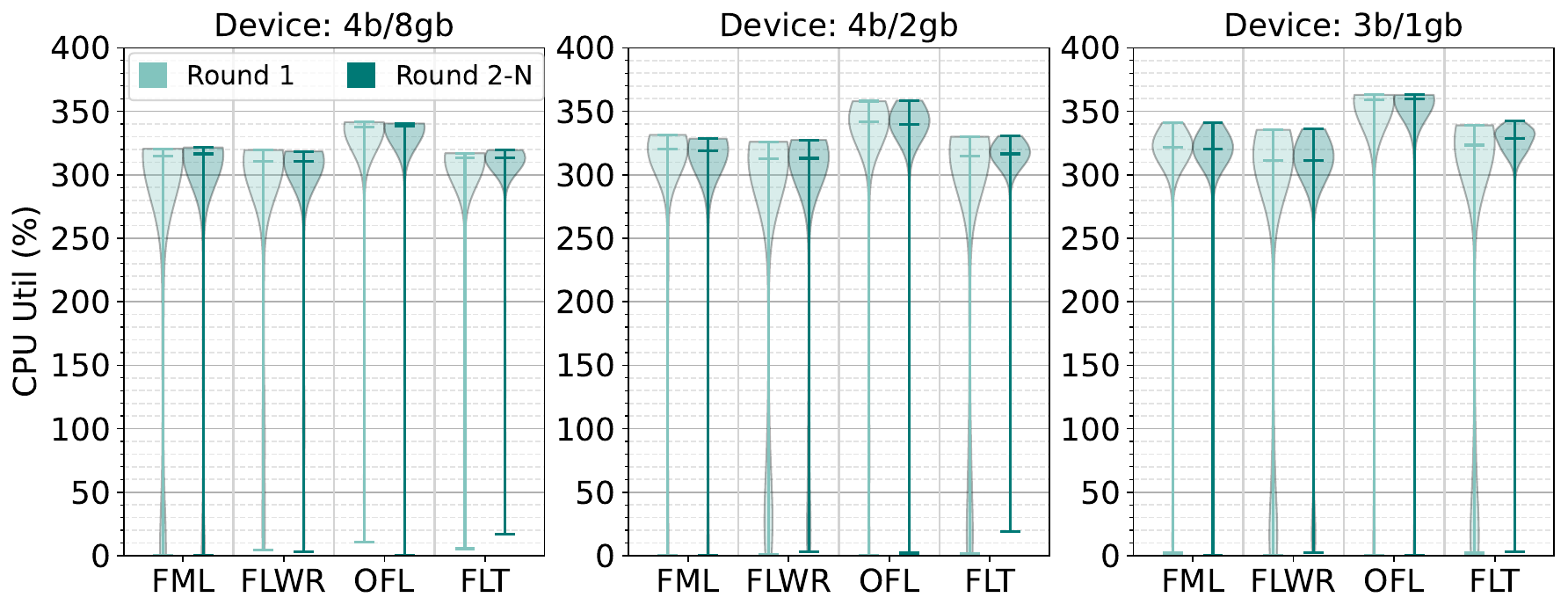}
    \includegraphics[width=\columnwidth]{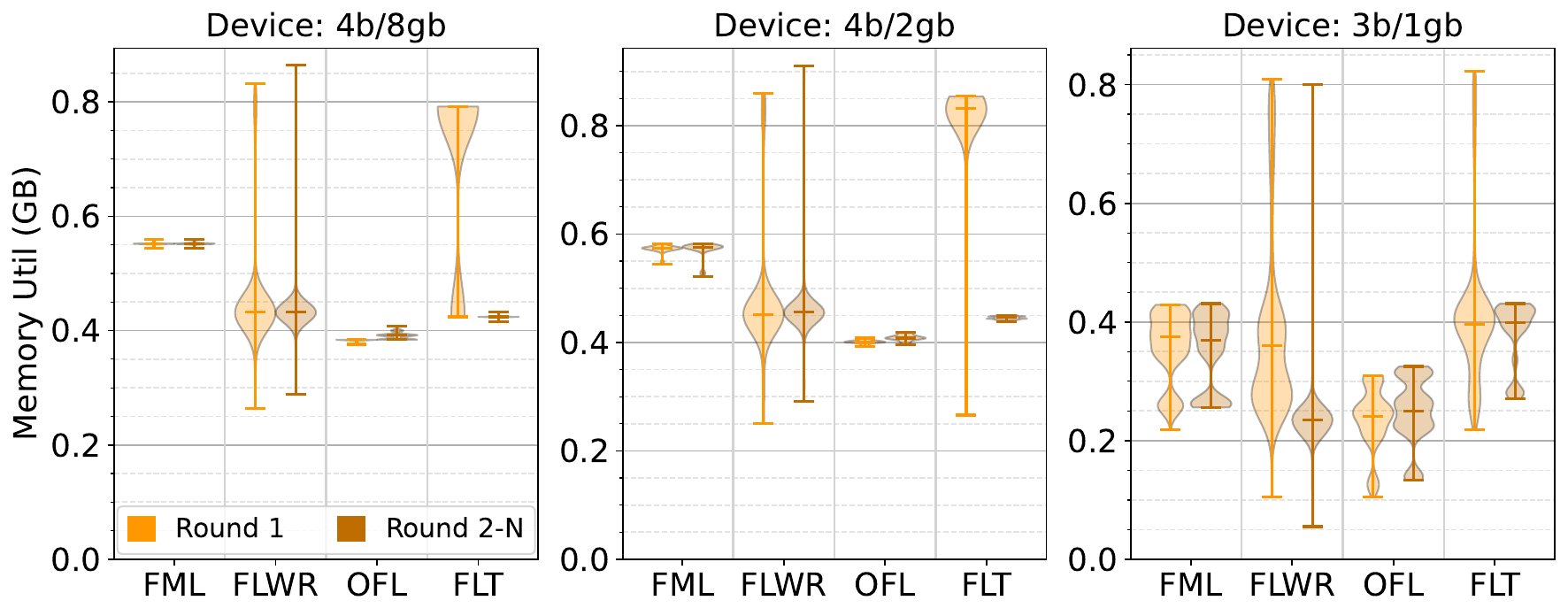}
    \includegraphics[width=\columnwidth]{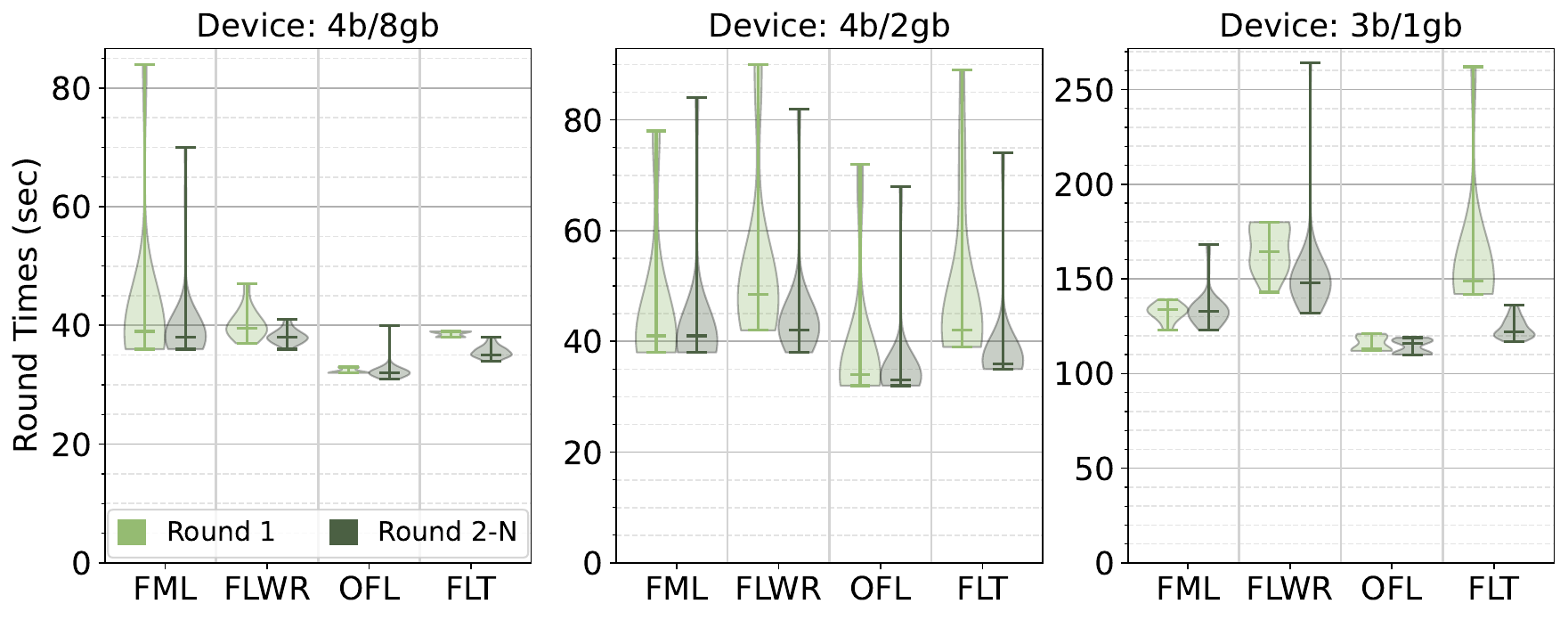}
    \vspace{-0.2in}
    \caption{\textit{CPU (top row)} and \textit{Memory (middle row) usage} and \textit{per-round times~(bottom row)} for clients while performing local training, during FL of CCNN/CIFAR10-IID/FedAvg using FedML (FLM), Flower (FLWR), OpenFL (OFL) and \FL (FLT). Columns indicate the 3 device types in the Pi cluster. The left violin in each plot has metrics for the first round in which a client participates and the right violin shows the same for later rounds.}

    \label{fig: Framework Comparison CPU and Mem}
    
\end{figure}

\subsection{Comparison with Other Frameworks}
\label{sec:exp:compare}

Finally, we compare \FL with three other open-source FL frameworks: \textit{Flower~v1.6.0}, \textit{OpenFL~v1.5.1} and \textit{FedML~v0.8.28}.

\subsubsection{Functional Features}

The features of these framework were already compared with \FL in Table~\ref{tbl:framework_comparison}.
\textit{Flower} is one of the earliest and widely used frameworks in the FL community~\cite{beutel2020flower} and provides an easy interface to implement FL strategies and ML models, but unlike \FL is limited to synchronous strategies. It is light-weight, claiming to run even on a Raspberry Pi Zero. \textit{OpenFL} is aimed at data scientists examining secure FL. It offers long-lived server and client components, allowing for realistic deployments.
However, it lacks support for asynchronous strategies, server-side global model validation and failure detection of clients.

\textit{FedML} is also limited in modularity, as it hard-codes the list of clients per session and also lacks support for asynchronous strategies.

\subsubsection{Resource Performance Comparison}
\label{subsubsec: resource}

We next compare the resource footprint for FedML~(FML), Flower~(FLWR), OpenFL~(OFL) and \FL~(FLT) on heterogeneous Pi devices. We train the CCNN/CIFAR10-IID model using each framework's native FedAvg strategy for 30 rounds using the Pi cluster, retaining the defaults and hyper-parameters in Table \ref{tab:model_datatset_table}. 
We use the in-built trainer and dataloaders for all frameworks except Flower, for which an equivalent dataloader was not available at the time of of this experiment. Hence, we incorporated Flotilla's dataloader into Flower. 
We pre-compute a random static sampling of clients for each round and use these same set of client for all frameworks to ensure identical behavior.

We record the CPU and memory utilization every 2~s and the per-round time on each Pi device when they are selected for training by these frameworks, and report them as violin plots in Fig.~\ref{fig: Framework Comparison CPU and Mem}, grouped by the three Pi device types. We notice that the first round may include bootstrapping for the FL session by some of the frameworks. So we report a pair of violins for each device and framework, the first violin (lighter color) showing the metric for Round 1 and the second (darker color) indicating the metric over the remaining rounds.

OpenFL has the least memory usage followed by \FL and Flower, while FedML has a relatively higher memory usage (row 2 of Fig.~\ref{fig: Framework Comparison CPU and Mem}).
OpenFL uses 250--400MiB of RAM when training, depending on the device type,
due to the use of a custom dataloader that uses their native batch generator instead of PyTorch's default dataloader. However, it loads the local training data into memory immediately after the client starts-up rather than at the start of a round, preventing it from being dynamically configured for different FL sessions.
Flower and \FL load the data on-demand when a request to train is received for a round, giving them flexibility but consuming more time. Since Flower uses \FL's dataloader, their memory usage is comparable. However, \FL further optimizes this by optionally caching the dataloader on a client from the first round, and reusing it in later rounds. Hence, the memory usage in the initial round for a client is higher for \FL, while in later rounds, it drops to a median of 400--450MiB.
The FedML framework natively has a higher memory footprint of $360$MiB compared to the others, and takes $370$--$570$MiB of RAM while training.

Another consequence of OpenFL's native dataloader and trainer is a higher CPU utilization, thanks to which it also has the lowest median per-round time, ranging from 32--115 seconds on the Pi 3B--4B. The other frameworks have similar CPU utilization. \FL has the second best median round times of 36--125 seconds on the three Pi types, other than in the initial round, and this is marginally better than Flower and FedML. 

In summary, despite the more complex set of features supported by \FL, we are competitive or better than these other FL platforms on resource usage and training time even on low-end Pi-class devices, allowing \FL to be deployed on heterogeneous edge devices.


\section{Conclusions and \addc{Future Work}}\label{sec:conclusion}

In this article, we present \FL, a novel, modular, scalable and resilient federated learning framework to develop, test and deploy FL strategies. \FL is built to scale real FL workloads on diverse edge clusters, and is resilient to client failures during training. Further, its unique external state store alongside periodic checkpointing allows \FL to be resilient to server failures too.
 Thanks to its modular design, access to observability metrics, and asynchronous training lifecycle, \FL naturally supports a much wider range of FL strategies than contemporary frameworks like Flower and FedScale. This is validated through implementing 5 baseline and SOTA FL strategies and comaparing their performance.
We also demonstrate the scalable design of \FL that allows it to operate on 1000+ clients and its low resource footprint even on devices like Raspberry Pi.

As future work, we plan to support concurrent FL sessions \addc{that can help efficiently utilize the client devices and improve training throughput across multiple users and models, without affecting the training latency per session}, and incorporate security and privacy techniques such as secure aggregation~\cite{Bonawitz2017} and differential privacy~\cite{dwork2006differential, MaoqiangWu2021} into \FL. Our design allows for \modc{these extensibilities}. 
We will also examine hierarchical tiering over multiple levels of servers~\cite{liu2020client}, along with vertical FL~\cite{10.1145/3543433}. There are opportunities to offer strong consistency for the external state store to help design more robust and correct FL strategies in the presence of failures.

\addc{The \textit{extensibility} of \FL goes beyond the strategies mentioned and includes other contemporary ones that are emerging as well. \textit{Prototype-based FL} like FedProto~\cite{Tan2022fedproto} and FedTGP~\cite{zhang2024} leverage class prototypes -- representative feature vectors of data classes -- to improve model convergence and generalization across distributed clients. Here, the client interfaces can be used to train a local model and calculate class-wise prototypes as the mean feature vector per class, and these can be sent to the leader for aggregation through a custom aggregator that regularizes their training.}

\addc{Further, as we design support for hierarchical federated learning into \FL, we can also design multi-layer aggregation and \textit{hierarchical FL strategies} such as NebulaFL~\cite{Lian2024HFL} and HeirFAVG~\cite{Liu2020HFL}. This primarilly involves introducing a new ClusterSessionManager that runs on the leaders for a cluster of clients in the hierarchy. These will play the role of a client to the primary leader, receiving local model training requests, and the role of a leader to clients it manages, passing these requests to the clients and aggregating their local models as its response to the primary leader. This also requires introducing a TopologyConstructor phase before each round that can support dynamic hierarchies.}
\addcrtwo{While decentralized FL is less common due to communication overheads~\cite{yuan2024decentralized} and convergence instabilities~\cite{zhou2024accelerate}, similar design approaches can be take to extend \FL to support this paradigm.}

\addc{Operating in semi-trusted/untrusted environments requires additional security and privacy capabilities that we plan to explore in Flotilla.
Techniques such as secure aggregation and differential privacy can help protect against external threats, e.g., when the communication channel is not reliable. If the clients themselves can be malicious, we need to include approaches to detect and correct for such adverse behavior, including poisoning attacks~\cite{kabir2023flshield,Zhu2023FedValidate}. The validation, aggregation and client selection phases can help score the credibility of clients, adaptively adjust the client-weights during aggregation, and reduce the probability of selection of clients based on these scores. Our architecture can be extended to accommodate features.}

\addc{We also plan to use \FL to examine the impact of different non-IID data distributions, including data count and feature heterogeneity, and additional system diversity, such as network heterogeneity and device reliability, on various FL strategies to evaluate their efficacy.}

\section*{Acknowledgement}
This first author was supported by a Prime Minister's Research Fellowship (PMRF). This work was partly supported by a research grant from the Cabinet Secretariat, Government of India. The authors thank Pranjal Naman and Prashanti S.K. from the DREAM:Lab, Indian Institute of Science, for their assistance.


 \bibliographystyle{elsarticle-num} 
 \bibliography{references}

\begin{thebibliography}{10}
\expandafter\ifx\csname url\endcsname\relax
  \def\url#1{\texttt{#1}}\fi
\expandafter\ifx\csname urlprefix\endcsname\relax\def\urlprefix{URL }\fi
\expandafter\ifx\csname href\endcsname\relax
  \def\href#1#2{#2} \def\path#1{#1}\fi

\bibitem{iot-smart-city-1}
P.~Williams, I.~K. Dutta, H.~Daoud, M.~Bayoumi, A survey on security in internet of things with a focus on the impact of emerging technologies, IEEE IoT-J (2022).

\bibitem{kumar2019internet}
S.~Kumar, P.~Tiwari, M.~Zymbler, Internet of things is a revolutionary approach for future technology enhancement: a review, J. Big Data (2019).

\bibitem{iot-science-anl}
J.~R. Elias, R.~Chard, J.~A. Libera, I.~Foster, S.~Chaudhuri, The manufacturing data and machine learning platform: Enabling real-time monitoring and control of scientific experiments via iot, in: IEEE WF-IoT-J, 2020.

\bibitem{iot-science-parashar}
V.~J. Aski, V.~S. Dhaka, A.~Parashar, I.~Rida, et~al., Internet of things in healthcare: A survey on protocol standards, enabling technologies, wban architectures and open issues, PHYCOM (2023).

\bibitem{ml-traffic}
M.~Ali, G.~Lavanya~Devi, R.~Neelapu, Intelligent traffic signal control system using machine learning techniques, in: ICMEET, 2021.

\bibitem{ml-science-field}
P.~Jain, S.~C. Coogan, S.~G. Subramanian, M.~Crowley, S.~Taylor, M.~D. Flannigan, A review of machine learning applications in wildfire science and management, Environ. Rev. (2020).

\bibitem{ml-medical}
I.~Kononenko, Machine learning for medical diagnosis: history, state of the art and perspective, Artif. Intell. Med. (2001).

\bibitem{liu2020client}
L.~Liu, J.~Zhang, S.~Song, K.~B. Letaief, Client-edge-cloud hierarchical federated learning, in: ICC, 2020.

\bibitem{Nguyen2024preservingprivacy}
T.~Nguyen, M.~T. Thai, Preserving privacy and security in federated learning, IEEE/ACM Trans. Netw. (2024).

\bibitem{ml-health-regulatory}
D.~C. Nguyen, Q.~Pham, P.~N. Pathirana, M.~Ding, A.~Seneviratne, Z.~Lin, O.~A. Dobre, W.~Hwang, Federated learning for smart healthcare: {A} survey, {ACM} Comput. Surv. (2023).

\bibitem{fintech-health-regulatory}
G.~Long, Y.~Tan, J.~Jiang, C.~Zhang, Federated learning for open banking, in: Federated Learning: Privacy and Incentive, 2020.

\bibitem{McMahan2016CommunicationEfficientLO}
H.~B. McMahan, E.~Moore, D.~Ramage, S.~Hampson, B.~A. y~Arcas, Communication-efficient learning of deep networks from decentralized data, in: AISTATS, 2016.

\bibitem{9252927}
H.~T. Nguyen, V.~Sehwag, S.~Hosseinalipour, C.~G. Brinton, M.~Chiang, H.~Vincent~Poor, Fast-convergent federated learning, IEEE JSAC (2021).

\bibitem{refl}
A.~M. Abdelmoniem, A.~N. Sahu, M.~Canini, S.~A. Fahmy, Refl: Resource-efficient federated learning, in: EuroSys, 2023.

\bibitem{ching2024totoro}
C.-W. Ching, X.~Chen, T.~Kim, B.~Ji, Q.~Wang, D.~Da~Silva, L.~Hu, Totoro: A scalable federated learning engine for the edge, in: EuroSys, 2024.

\bibitem{khan2024float}
A.~F. Khan, A.~A. Khan, A.~M. Abdelmoniem, S.~Fountain, A.~R. Butt, A.~Anwar, Float: Federated learning optimizations with automated tuning, in: EuroSys, 2024.

\bibitem{chai2020tifl}
Z.~Chai, A.~Ali, S.~Zawad, S.~Truex, A.~Anwar, N.~Baracaldo, Y.~Zhou, H.~Ludwig, F.~Yan, Y.~Cheng, Tifl: A tier-based federated learning system, in: HPDC, 2020.

\bibitem{wolfrath2022haccs}
J.~Wolfrath, N.~Sreekumar, D.~Kumar, Y.~Wang, A.~Chandra, Haccs: heterogeneity-aware clustered client selection for accelerated federated learning, in: IPDPS, 2022.

\bibitem{xie2019asynchronous}
C.~Xie, S.~Koyejo, I.~Gupta, Asynchronous federated optimization, arXiv preprint arXiv:1903.03934 (2019).

\bibitem{chai2021fedat}
Z.~Chai, Y.~Chen, A.~Anwar, L.~Zhao, Y.~Cheng, H.~Rangwala, Fedat: A high-performance and communication-efficient federated learning system with asynchronous tiers, in: SC, 2021.

\bibitem{zhang2020federated}
X.~Zhang, X.~Zhu, J.~Wang, H.~Yan, H.~Chen, W.~Bao, Federated learning with adaptive communication compression under dynamic bandwidth and unreliable networks, Inf. Sci. (2020).

\bibitem{lai2022fedscale}
F.~Lai, Y.~Dai, S.~Singapuram, J.~Liu, X.~Zhu, H.~Madhyastha, M.~Chowdhury, Fedscale: Benchmarking model and system performance of federated learning at scale, in: ICML, 2022.

\bibitem{tff}
Google, Tensorflow—federated learning, \url{https://www.tensorflow.org/federated/federated_learning} (2020).

\bibitem{caldas2018leaf}
S.~Caldas, S.~M.~K. Duddu, P.~Wu, T.~Li, J.~Kone{\v{c}}n{\`y}, H.~B. McMahan, V.~Smith, A.~Talwalkar, Leaf: A benchmark for federated settings, arXiv preprint arXiv:1812.01097 (2018).

\bibitem{beutel2020flower}
D.~J. Beutel, T.~Topal, A.~Mathur, X.~Qiu, J.~Fernandez-Marques, Y.~Gao, L.~Sani, K.~H. Li, T.~Parcollet, P.~P.~B. de~Gusm{\~a}o, et~al., Flower: A friendly federated learning research framework, arXiv preprint arXiv:2007.14390 (2020).

\bibitem{openfl_citation}
P.~Foley, M.~J. Sheller, B.~Edwards, S.~Pati, W.~Riviera, M.~Sharma, P.~N. Moorthy, S.-h. Wang, J.~Martin, P.~Mirhaji, P.~Shah, S.~Bakas, Openfl: the open federated learning library, Phys. Med. Biol. (2022).

\bibitem{he2020fedml}
C.~He, S.~Li, J.~So, X.~Zeng, M.~Zhang, H.~Wang, X.~Wang, P.~Vepakomma, A.~Singh, H.~Qiu, et~al., Fedml: A research library and benchmark for federated machine learning, arXiv preprint arXiv:2007.13518 (2020).

\bibitem{flatscale}
K.~Bonawitz, H.~Eichner, W.~Grieskamp, D.~Huba, A.~Ingerman, V.~Ivanov, C.~Kiddon, J.~Kone\v{c}n\'{y}, S.~Mazzocchi, B.~McMahan, T.~Van~Overveldt, D.~Petrou, D.~Ramage, J.~Roselander, Towards federated learning at scale: System design, in: Proceedings of Machine Learning and Systems, 2019.

\bibitem{vahidian2023rethinkingdatahetero}
S.~Vahidian, M.~Morafah, M.~Shah, B.~Lin, Rethinking data heterogeneity in federated learning: Introducing a new notion and standard benchmarks, IEEE TAI (2023).

\bibitem{fl-on-hetero-dev-survey}
K.~Pfeiffer, M.~Rapp, R.~Khalili, J.~Henkel, Federated learning for computationally constrained heterogeneous devices: A survey, ACM Comput. Surv. (2023).

\bibitem{emperical-hetero-fl}
A.~M. Abdelmoniem, C.-Y. Ho, P.~Papageorgiou, M.~Canini, A comprehensive empirical study of heterogeneity in federated learning, IEEE IoT-J (2023).

\bibitem{fedcav}
H.~Zeng, T.~Zhou, Y.~Guo, Z.~Cai, F.~Liu, Fedcav: Contribution-aware model aggregation on distributed heterogeneous data in federated learning, in: ICPP, 2021.

\bibitem{li2019convergence}
X.~Li, K.~Huang, W.~Yang, S.~Wang, Z.~Zhang, On the convergence of fedavg on non-iid data, in: ICLR, 2019.

\bibitem{AQFL}
A.~M. Abdelmoniem, M.~Canini, Towards mitigating device heterogeneity in federated learning via adaptive model quantization, in: EuroMLSys, 2021.

\bibitem{hetero-fl-survey}
M.~Ye, X.~Fang, B.~Du, P.~C. Yuen, D.~Tao, Heterogeneous federated learning: State-of-the-art and research challenges, ACM Comput. Surv. (2023).

\bibitem{Wu2019SAFAAS}
W.~Wu, L.~He, W.~Lin, R.~Mao, C.~Maple, S.~A. Jarvis, Safa: A semi-asynchronous protocol for fast federated learning with low overhead, IEEE Trans. Comput. (2019).

\bibitem{fedPEC}
Y.~Cai, W.~Xi, Y.~Shen, Y.~Peng, S.~Song, J.~Zhao, High-efficient hierarchical federated learning on non-iid data with progressive collaboration, Future Gener. Comput. Syst. (2022).

\bibitem{feddyn}
W.~Zhang, Y.~Zhao, F.~Li, H.~Zhu, A hierarchical federated learning algorithm based on time aggregation in edge computing environment, Applied Sciences (2023).

\bibitem{Roy2019BrainTorrentAP}
A.~G. Roy, S.~Siddiqui, S.~P{\"o}lsterl, N.~Navab, C.~Wachinger, Braintorrent: A peer-to-peer environment for decentralized federated learning, arXiv preprint arXiv:1905.06731 (2019).

\bibitem{pysyft}
T.~Ryffel, A.~Trask, M.~Dahl, B.~Wagner, J.~Mancuso, D.~Rueckert, J.~Passerat-Palmbach, A generic framework for privacy preserving deep learning, arXiv preprint arXiv:1811.04017 (2018).

\bibitem{xu2023asynchronous}
C.~Xu, Y.~Qu, Y.~Xiang, L.~Gao, Asynchronous federated learning on heterogeneous devices: A survey, Comput. Sci. Rev. (2023).

\bibitem{wolf2020transformers}
T.~Wolf, L.~Debut, V.~Sanh, J.~Chaumond, C.~Delangue, A.~Moi, P.~Cistac, T.~Rault, R.~Louf, M.~Funtowicz, J.~Davison, S.~Shleifer, P.~von Platen, C.~Ma, Y.~Jernite, J.~Plu, C.~Xu, T.~L. Scao, S.~Gugger, M.~Drame, Q.~Lhoest, A.~M. Rush, Transformers: State-of-the-art natural language processing, in: EMNLP, 2020.

\bibitem{wang2019deep}
M.~Y. Wang, Deep graph library: Towards efficient and scalable deep learning on graphs, in: ICLR, 2019.

\bibitem{naman2024optimizing}
P.~Naman, Y.~Simmhan, Optimizing federated learning using remote embeddings for graph neural networks, in: Euro-Par, 2024.

\bibitem{violet}
S.~Baheti, S.~Badiger, Y.~Simmhan, Violet: An emulation environment for validating iot deployments at large scales, ACM Trans. Cyber-Phys. Syst. (2021).

\bibitem{Bonawitz2017}
K.~Bonawitz, V.~Ivanov, B.~Kreuter, A.~Marcedone, H.~B. McMahan, S.~Patel, D.~Ramage, A.~Segal, K.~Seth, Practical secure aggregation for privacy-preserving machine learning, in: ACM CCS, 2017.

\bibitem{BatchCrypt}
C.~Zhang, S.~Li, J.~Xia, W.~Wang, F.~Yan, Y.~Liu, {BatchCrypt}: Efficient homomorphic encryption for {Cross-Silo} federated learning, in: USENIX ATC, 2020.

\bibitem{MaoqiangWu2021}
M.~Wu, D.~Ye, J.~Ding, Y.~Guo, R.~Yu, M.~Pan, Incentivizing differentially private federated learning: A multidimensional contract approach, IEEE IoT-J (2021).

\bibitem{cao2019understanding}
D.~Cao, S.~Chang, Z.~Lin, G.~Liu, D.~Sun, Understanding distributed poisoning attack in federated learning, in: 2019 IEEE 25th International Conference on Parallel and Distributed Systems (ICPADS), 2019.

\bibitem{liu2021privacy}
X.~Liu, H.~Li, G.~Xu, Z.~Chen, X.~Huang, R.~Lu, Privacy-enhanced federated learning against poisoning adversaries, IEEE Transactions on Information Forensics and Security (2021).

\bibitem{Wang2020model}
Y.~Wang, T.~Zhu, W.~Chang, S.~Shen, W.~Ren, Model poisoning defense on federated learning: A validation based approach, in: Network and System Security, 2020.

\bibitem{zhuang2022mufl}
W.~Zhuang, Y.~Wen, S.~Zhang, Smart multi-tenant federated learning, arXiv preprint arXiv:2207.04202 (2022).

\bibitem{bhuyan2022multi}
N.~Bhuyan, S.~Moharir, Multi-model federated learning, in: 2022 14th International Conference on COMmunication Systems \& NETworkS (COMSNETS), 2022.

\bibitem{kabir2023flshield}
E.~Kabir, Z.~Song, M.~R. Ur~Rashid, S.~Mehnaz, Flshield: A validation based federated learning framework to defend against poisoning attacks, in: 2024 IEEE Symposium on Security and Privacy (SP), 2024.

\bibitem{frey2002condor}
J.~Frey, T.~Tannenbaum, M.~Livny, I.~Foster, S.~Tuecke, Condor-g: a computation management agent for multi-institutional grids, in: HPDC, 2001.

\bibitem{Hu2022incentive}
M.~Hu, D.~Wu, Y.~Zhou, X.~Chen, M.~Chen, Incentive-aware autonomous client participation in federated learning, TPDS (2022).

\bibitem{10.1049/iet-sen.2017.0251}
A.~Luoto, K.~Systä, Fighting network restrictions of request-response pattern with mqtt, IET Software (2018).

\bibitem{lecun1998gradient}
Y.~LeCun, L.~Bottou, Y.~Bengio, P.~Haffner, Gradient-based learning applied to document recognition, Proc. IEEE (1998).

\bibitem{sandler2018mobilenetv2}
M.~Sandler, A.~Howard, M.~Zhu, A.~Zhmoginov, L.-C. Chen, Mobilenetv2: Inverted residuals and linear bottlenecks, in: CVPR, 2018.

\bibitem{he2016deep}
K.~He, X.~Zhang, S.~Ren, J.~Sun, Deep residual learning for image recognition, CVPR (2016).

\bibitem{yao2024fedgcn}
Y.~Yao, W.~Jin, S.~Ravi, C.~Joe-Wong, Fedgcn: Convergence-communication tradeoffs in federated training of graph convolutional networks, NeurIPS (2024).

\bibitem{hamilton2017inductive}
W.~Hamilton, Z.~Ying, J.~Leskovec, Inductive representation learning on large graphs, NeurIPS (2017).

\bibitem{deng2009imagenet}
J.~Deng, W.~Dong, R.~Socher, L.-J. Li, K.~Li, L.~Fei-Fei, Imagenet: A large-scale hierarchical image database, CVPR (2009).

\bibitem{yurochkin2019bayesian}
M.~Yurochkin, M.~Agarwal, S.~Ghosh, K.~Greenewald, N.~Hoang, Y.~Khazaeni, Bayesian nonparametric federated learning of neural networks, in: ICML, 2019.

\bibitem{tan2023towards}
A.~Z. Tan, H.~Yu, L.~Cui, Q.~Yang, Towards personalized federated learning, IEEE Transactions on Neural Networks and Learning Systems (2023).

\bibitem{arivazhagan2019federated}
M.~G. Arivazhagan, V.~Aggarwal, A.~K. Singh, S.~Choudhary, Federated learning with personalization layers, arXiv preprint arXiv:1912.00818 (2019).

\bibitem{lemoine1985failure}
A.~J. Lemoine, M.~L. Wenocur, On failure modeling, Naval research logistics quarterly (1985).

\bibitem{dwork2006differential}
C.~Dwork, Differential privacy, in: International colloquium on automata, languages, and programming, Springer, 2006.

\bibitem{10.1145/3543433}
A.~Das, T.~Castiglia, S.~Wang, S.~Patterson, Cross-silo federated learning for multi-tier networks with vertical and horizontal data partitioning, ACM Trans. Intell. Syst. Technol. (2022).

\bibitem{Tan2022fedproto}
Y.~Tan, G.~Long, L.~LIU, T.~Zhou, Q.~Lu, J.~Jiang, C.~Zhang, Fedproto: Federated prototype learning across heterogeneous clients, AAAI (2022).

\bibitem{zhang2024}
J.~Zhang, Y.~Liu, Y.~Hua, J.~Cao, Fedtgp: Trainable global prototypes with adaptive-margin-enhanced contrastive learning for data and model heterogeneity in federated learning, AAAI (2024).

\bibitem{Lian2024HFL}
Z.~Lian, J.~Cao, Q.~Cao, W.~Liu, Z.~Zhu, X.~Zhou, Nebulafl: Self-organizing efficient multilayer federated learning framework with adaptive load tuning in heterogeneous edge systems, IEEE Transactions on Computer-Aided Design of Integrated Circuits and Systems (2024).

\bibitem{Liu2020HFL}
L.~Liu, J.~Zhang, S.~Song, K.~B. Letaief, Client-edge-cloud hierarchical federated learning, in: ICC 2020 - 2020 IEEE International Conference on Communications (ICC), 2020.

\bibitem{yuan2024decentralized}
L.~Yuan, Z.~Wang, L.~Sun, P.~S. Yu, C.~G. Brinton, Decentralized federated learning: A survey and perspective, IEEE Internet of Things Journal (2024).

\bibitem{zhou2024accelerate}
M.~Zhou, G.~Liu, K.~Lu, R.~Mao, H.~Liao, Accelerating the decentralized federated learning via manipulating edges, in: Proceedings of the ACM Web Conference 2024, 2024.

\bibitem{Zhu2023FedValidate}
W.~Zhu, Z.~Liu, Z.~Chen, C.~Shi, X.~Zhang, S.~Guo, Fedvalidate: A robust federated learning framework based on client-side validation, 2023 8th International Conference on Data Science in Cyberspace (DSC) (2023).

\end{thebibliography}

\clearpage
\appendix

\section{Psuedocode for FedAt and TiFL in \FL}
\label{appendix:psudocode}

\subsection{FedAT Pseudo-code for Client Selection and Aggregation}
FedAT is implemented in \FL using the following simple pseudo-code, using the state objects for coordination:
\begin{lstlisting}[basicstyle=\small\ttfamily, language=python]
def |\textbf{clientSelectFedAT}(\textit{sessionID, availableClients, client\-Sel\-StateRW, aggStateRO, clientTrainStateRO, client\-Info\-StateRO, trainSessionStateRO, clientSelUserConfig}):|

    |\em Get round\_no from clientSelUserConfig.|
    clientsPerTier = clientSelUserConfig.get(clientsPerTier)
    if round_no ==0 and aggStateRO.is_empty():
        numTiers = clientSelUserConfig.get(numTiers)
        clientLatencies = 
            clientInfoStateRO[availableClients.benchmark]
        clientTiers = |\em Agglomerative clustering using clientLatencies. |
        Put clientTiers in clientSelStateRW.
        for tierID, tier in clientTiers
            selClients = |\em Randomly pick clientsPerTier \# of clients from tier.|
            clientSelStateRW[tierID.selected] = selClients
            allSelClients += selClients
            clientSelStateRW[tierID.tierAggNum] = 0
        return allSelClients, None
    else
        |\em Get clientTiers from clientSelStateRW.|
        for tierID, tier in clientTiers:
            cs_tier_agg_num = clientSelStateRW[tier.tierAggNum]
            agg_tier_agg_num = aggStateRO[tier.tierAggNum]
            if cs_tier_agg_num < agg_tier_agg_num:
                clientSelStateRW[tierID.tierAggNum+= 1
                selClients = |\em Randomly pick clientsPerTier \# of clients from tier.|
                clientSelStateRW[tierID.selected] = selClients
                return selClients, None
        return None, None

def |\textbf{aggregateFedAT}(\textit{sessionID, clientID, localModel, aggStateRW, client\-Sel\-StateRO, clientTrainStateRO,client\-Info\-StateRO, trainSessionStateRO, aggUserConfig}|):
    if aggStateRW is empty:
        |\em Put| aggStateRW[tierID.tierAggNum]=0 |\em for all tierIDs|
        globalModel = trainSessionStateRO[sessionID.globalModel]
        |\em Intialize | aggStateRW[tierID.tierModel] = globalModel|\em for all tierIDs|
        
    |\em Put localModel and clientID in aggStateRW.|
    |\em Get tierID of clientID from clientSelStateRW.|
    selClients = |\em Get selected clients for tierID from clientSelStateRO.|

    if all selClients in aggStateRW:
        aggStateRW[tierID.tierAggNum+= 1
        tierModel = |\em Aggregate tier model from localModel for clients in selClients and amount of data per client from clientInfoStateRO.|
        aggStateRW[tierID.tierModel] = tierModel
        newModel = |\em Aggregate tierModel for all tierIDs into a single model using tierAggNum from aggStateRW.|
        |\em Delete clientID for all clients in selClients from aggStateRW.|
        return newModel
    else
        return None
\end{lstlisting}

\subsection{TiFL Pseudo-code for Client Selection and Aggregation}
TiFL is implemented in \FL using the following simple pseudo-code, using the state objects for coordination:
\begin{lstlisting}[basicstyle=\small\ttfamily, language=python]
def |\textbf{clientSelectTiFL}(\textit{sessionID, availableClients, client\-Sel\-StateRW, aggStateRO, clientTrainStateRO, client\-Info\-StateRO, trainSessionStateRO, clientSelUserConfig}):|
    |\em Get round from clientSelUserConfig.|
    numClients = |\em Get \# clients to select per tier from clientSelUserConfig.|

    if aggStateRO is Empty:
        |\em Get valRoundInterval from clientSelUserConfig.|
        if round == 0:
            num_tiers = |\em Get \# tiers  from clientSelUserConfig.|
            creditsPerTier = |\em Get credits per tier from clientSelUserConfig.|
            tierProbs = |\em Equal for all tiers.|
            clientLatencies = |\em Get benchmark data for availableClients from clientInfoStateRO.|
            clientTiers = |\em Cluster clients using Agglomerative clustering.|
            |\em Put clientTiers in clientSelStateRW.|
            |\em Put valOngoing to be False in  clientSelStateRW.|
            |\em Put tierCredits to be 0 for all tiers in clientSelStateRW|
        else if round % valRoundInterval:
            |\em Get ValOngoing from clientSelStateRW.|
            if ValOngoing if False:
                |\em Put valOngoing to be True in clientSelStateRW.|
                return None, availableClients
            else: 
                |\em Put clientID, ValidationMetrics in clientSelStateRW.|
                if |\em all clientIDs in availableClients are in clientSelStateRW|:
                    |\em Put valOngoing to be False in clientSelStateRW.|
                    tierProbs = |\em Calculate tier probabilities using val metrics tierCredits.|
                    |\em Clear ClientIDs and metrics from clientSelStateRW.|
        |\em Get ValOngoing from clientSelStateRW.|
        if ValOngoing is False:
            chosenTier = |\em Choose tier using tierProbs.|
            clientTiers = |\em Get clientTiers from clientSelStateRW.|
            |\em Decrement tierCredits of chosenTier by 1.|
            selClients = |\em Randomly select numClients number of clients from chosenTier of clientTiers.|
            |\em Put selClients in clientSelStateRW.|

            return selClients, None
    return None, None

def |\textbf{aggregateTiFL}(\textit{sessionID, clientID, localModel, aggStateRW, client\-Sel\-StateRO, clientTrainStateRO,client\-Info\-StateRO, trainSessionStateRO, aggUserConfig}|):
    |\em Put localModel and clientID in aggStateRW.|
    |\em Get selClients from clientSelStateRO.|
    if all selClients are in aggStateRW: 
        globalModel = |\em Aggregate global model from localModels and amount of data per client from clientInfoStateRO.|
        |\em Clear aggStateRW.|
    return globalModel
  else:
    return None
\end{lstlisting}

\clearpage
\section{Sample YAML Configuration File for FL Training}
\label{appendix: training-config}

\FL offers a declarative means to configure a FL training session using a YAML file that does not require code to be written. The following is an example of configuring \FL to train a LeNet-5 model on the EMNIST (Non-IID)
dataset using the predefined FedAT client selector and aggregator. It should run for 100 rounds where each client trains 1 epoch using a batch size of 16 and a learning rate of 0.00005. Session checkpointing, which includes saving the model and session states, is configured to occur every 5 rounds   \

\begin{lstlisting}[style=yaml]
session_config:
  session_id: lenet_fedat_noniid
  use_gpu: False
  aggregator: fedat
  aggregator_args: None
  client_selection: fedat
  client_selection_args:
    num_tiers: 3
    num_clients_selected_per_tier: 2
  checkpoint_interval: 5
  validation_round_interval: 1
  generate_plots: False

benchmark_config:
  skip_benchmark: True

server_training_config:
  model_dir: ./models/LeNet5
  global_model_validation_batch_size: 100
  num_training_rounds: 100

client_training_config:
  model_id: LeNet5
  model_class: LeNet5_class
  dataset: EMNIST_NONIID3
  epochs: 1
  batch_size: 16
  learning_rate: 0.00005
  train_timeout_duration_s: 300
  loss_function: crossentropy
  loss_function_custom: False
  optimizer: adam
  optimizer_custom: False

model_config:
  use_custom_dataloader: False
  custom_loader_args: None
  use_custom_trainer: False
  custom_trainer_args: None
  use_custom_validator: False
  custom_validator_args: None
  
  model_args:
    num_classes: 10 
\end{lstlisting}

\clearpage
\section{\FL Pre-defined State Entries}

\label{appendix:System States}
Overview of session state objects and pre-defined set of keys used by \FL to orchestrate the FL session. \addc{The framework specifies two types of states: Read-only~(Table~\ref{tab:flotilla-state-detials}) and Read-Write States (Table~\ref{tab:client-selection-aggregation}). This allows phases to be loosely coupled when maintaining/sharing information and coordinating with each other when designing flexible FL strategies.
\begin{itemize}
    \item \textbf{Read-only States} These are predefined by the framework and are essential for coordination among phases and tracking the FL lifecycle.
    \item \textbf{Read-write} Entries in this state are defined by the user and created/updated by the Client Selection phase (Client\_Selection state) and Aggregation phase (Aggregation state). Their entries depend on the chosen strategy.
\end{itemize}  

}

\begin{table}[h]
\centering
\resizebox{\textwidth}{!}{%
\begin{tabular}{p{3.5cm}|l|lp{8cm}}
\hline
\textbf{State Object \textit{(Scope)}} &  \textbf{Primary Key} & \textbf{Secondary Key} & \textbf{Description} \\ \hline\hline
\multirow{7}{3.5cm}{Client Training State \textit{(Training session)}} & \multirow{7}{*}{\tt Client ID} & \tt missed\_deadline & Indicates whether the client has missed the deadline. \\
 &   & \multirow{1}{3cm}{\tt last\_round\_ participated} & Tracks the last round in which the client  participated.\\
 &   & \tt current\_model\_id & Represents the model on which the client is training.\\
 &  & \tt current\_dataset & Indicates the dataset for the client's training.\\
 &   & \tt model\_weights & Stores the pickled model weights for the client.\\
 &  & \tt training\_metrics & Contains metrics for the training on the client.\\
 &  & \tt validation\_metrics & Stores validation metrics from the client, providing insights into model performance.\\ 
 
\hline
\multirow{8}{3.5cm}{Client Info State \textit{(Application lifecycle)}} & \multirow{8}{*}{\tt Client ID} & \tt hardware\_information & Provides details about the client's hardware. \\
 &   & \tt role & Specifies the role of the client in the system.\\
 &    & \tt dataset\_details & Contains details about the datasets used by the client. \\
 &    & \tt models & Lists the models associated with the client. \\
 &    & \tt heartbeat\_timestamp & Records the last heartbeat timestamp from the client. \\
 &    & \tt heartbeat\_interval & Specifies the interval for client heartbeats. \\
 &    & \tt join\_timestamp & Records when the client joined the system. \\
 &    & \tt benchmark & Contains benchmarking details from the client. \\ 
 &    & \tt is\_training & Stores True if Client is currently training, stores False otherwise. \\
  &   & \tt failed\_rounds & Stores the rounds in which the client has failed to return within the gRPC timeout or has disconnected. \\
    &   & \tt \addc{is\_active} & \addc{Boolean indicating a client's liveliness based on heartbeats} \\
 \hline
\multirow{7}{3.5cm}{Training Session State \textit{(Across training sessions)}} & \multirow{7}{*}{\tt Session ID} & \tt global\_model & Represents the global model being used in the training session. \\
 &   & \tt training\_config & Contains configuration settings for the training session. \\
 &    & \tt last\_round\_number & Tracks the number of the last round that has been completed. \\
 &    & \tt training\_state\_id & Indicates the current state of the training session. \\
 &    & \tt agg\_state\_id & Represents the state of the aggregation process. \\
 &   & \tt cs\_state\_id & Tracks the state of the client in the training session. \\
 &  & \tt status & Indicates whether the training session is currently running or has completed. \\ 
\hline
\end{tabular}%
}
\caption{Table detailing the data structure and the corresponding information maintained by the three internal states of the \FL Server.}
\label{tab:flotilla-state-detials}
\end{table}

\begin{table}[t]
\centering
\resizebox{\textwidth}{!}{%
\begin{tabular}{p{4cm}|l|l|p{8cm}}
\hline
\textbf{State Object \textit{(Scope)}} & \textbf{Strategy} & \textbf{Primary Key} & \textbf{Description} \\ \hline\hline

\multirow{3}{4cm}{Client Selection State \textit{(Training session)}} 
& FedAvg  & \tt selected\_clients  & List of clients selected for the current round. \\ 

&  &  & \\ 

& \multirow{2}{*}{FedAT} 
& \tt selected\_clients\_tier\_\{i\} & Selected client IDs from the i-th Tier. \\  

&  & \tt client\_to\_tier\_id\_dict & Mapping of client IDs to their respective Tiers. \\  

\hline

\multirow{4}{4cm}{Aggregation State \textit{(Training sessions)}}  
& FedAvg  & \tt clientweights\_\{i\} & Mapping of client IDs to their respective weights. \\  

&  &  & \\ 

& \multirow{3}{*}{FedAT} 
& \tt clientweights\_\{i\} & Mapping of client IDs to their respective weights. \\  

&  & \tt update\_count\_tier\_\{i\} & Number of updates received for the i-th Tier. \\  

&  & \tt tier\_model\_tier\_\{i\} & Global model associated with the i-th Tier. \\  

\hline

\end{tabular}%
}
\caption{\addc{Table detailing the data structure and keys maintained by the Client Selection and Aggregation states. These are defined for specific FL strategies by the CS and Agg user logic, and shown here for FedAvg and FedAT strategies.}}
\label{tab:client-selection-aggregation}
\end{table}

\end{document}